\documentclass[11pt]{article}

 \usepackage{mathrsfs}
 \usepackage[utf8]{inputenc}
 \usepackage{epsfig,graphicx,color}
 \usepackage{subfig}
 \usepackage{latexsym}                 % extension symboles
 \usepackage{amsmath,amsfonts,verbatim,xkeyval,bm, amsthm, upgreek, amscd, wasysym, amssymb}
 \usepackage{cases}
\usepackage{subfig}

 \usepackage[english]{babel}
 \usepackage{listings}
 \usepackage{multicol}
 \usepackage[colorlinks]{hyperref}  %serve per fare collegamenti a pagine web
 \hypersetup{hidelinks}  %serve per fare i collegamenti in nero
 \usepackage{tikz-cd}
 \usepackage{booktabs}
 \usepackage[numbers, sort&compress]{natbib}
 \usepackage{sidecap}
 \usepackage[font=small,labelfont=bf]{caption}
 \usepackage{float}
\usepackage{titlesec}

\titleformat{\paragraph}
{\normalfont\normalsize\bfseries}{\theparagraph}{1em}{}
\titlespacing*{\paragraph}
{0pt}{3.25ex plus 1ex minus .2ex}{1.5ex plus .2ex}
 
\newtheorem{Theorem}{Theorem}[section]

\newtheorem{Proposition}[Theorem]{Proposition}

\def\vet#1{{\bm #1}}
\def\build#1_#2^#3{\mathrel{
\mathop{\kern 0pt#1}\limits_{#2}^{#3}}}
\def\reali{\mathbb{R}}

\def\naturali{\mathbb{N}}
\def\interi{\mathbb{Z}}

\def\e{\mathrm{e}}
\def\i{\mathrm{i}}
\def\AMD{\mathrm{AMD}}

\def\Ascr{\mathcal{A}}
\def\Bscr{\mathcal{B}}
\def\Cscr{\mathcal{C}}

\def\Escr{\mathcal{E}}
\def\Fscr{\mathcal{F}}
\def\Gscr{\mathcal{G}}
\def\Hscr{\mathcal{H}}

\def\Oscr{\mathcal{O}}

\def\rho{\varrho}

\def\poisson#1#2{\lbrace #1,#2 \rbrace}

\def\cgrav{\Gscr}
\def\t#1{\tilde{#1}}
\def\h#1{\hat{#1}}

\begin{document}

\pagenumbering{arabic}

\begin{center}
{\Large\bf The phase-space architecture in extrasolar systems with two planets in orbits of high mutual inclination}
\vskip 1cm
{\it R. Mastroianni\footnote{rita.mastroianni@math.unipd.it} and C. Efthymiopoulos\footnote{cefthym@math.unipd.it}\\
Dipartimento di Matematica Tullio Levi-Civita, Universit\`{a} degli 
Studi di Padova}\\
via Trieste 63, 35121 Padova
\end{center}

\begin{abstract}
{\small We revisit the secular 3D planetary three-body problem with the aim to provide a unified formalism allowing for a convenient parametric representation of all basic phenomena sculpting the form of the phase space as the mutual inclination between the two planets' orbits is set to gradually larger values, starting from zero (planar case) and ending to a value beyond the limit of the Lidov-Kozai instability. Our main findings are: i) on methodological ground, we propose an algebraic `book-keeping' technique, incorporated to the treatment of the secular Hamiltonian already at the level of the Jacobi reduction of the nodes, which allows to decompose the secular Hamiltonian in the form $\Hscr_{sec}=\Hscr_{planar}+\Hscr_{space}$, with $\Hscr_{space}$ collecting all terms depending on the planets mutual inclination $\i_{mut}$. This technique can be implemented both in expansions of the Hamiltonian in multipoles (Legendre polynomials) or in power of the planetary eccentricities (Laplace-Lagrange). We numerically compare several such models, aiming to establish an appropriate value of the minimum multipole truncation order $N_P$, as well as of the minimum book-keeping truncation order $N_{bk}$ (accounting simultaneously for powers of the planetary eccentricities, inclinations, as well as the system's Angular Momentum Deficit ($\AMD$)), such that the form of obtained phase portrait stabilizes for truncations at orders beyond the minimum ones. This analysis is made in a regime far from any hierarchical limit either in the planetary mass ratio or distance ratio. ii) On dynamical ground, we explore the transition, as the level of mutual inclination increases, from a `planar-like' to a `Lidov-Kozai' regime, both characterized by the kind of periodic orbits which dominate the structure of the phase space at the corresponding values of the energy. Using a typical case of non-hierarchical parameters (pertinent to the well known $\upsilon$-Andromedae system), we find that the structure of the phase portraits of the integrable secular dynamics of the planar case is reproduced to a large extent and for a wide range of energies also in the 3D case. We provide a semi-analytical criterion allowing to estimate the extent in energies and level of mutual inclination up to which dynamics remains nearly-integrable. In this regime, we propose a normal form method by which the basic periodic orbits of the nearly-integrable regime (modes A and B), can be computed semi-analytically. These modes generalize the apsidal corotation orbits (A: anti-aligned, B: aligned) of the planar case. On the other hand, we show how, as the energy increases, the system gradually moves from the nearly-integrable `planar-like' to the `Lidov-Kozai' regime. The latter is dominated by two different families of nearly circular and highly inclined periodic orbits ($C_1$ and $C_2$), of which $C_2$ becomes unstable via the usual Lidov-Kozai mechanism. We demonstrate how the $C_1$ and $C_2$ families are connected to the $A$ and $B$ families via a chain of saddle-node and pitchfork bifurcations. We provide an analytical criterion for the destabilization of the family $C_2$ obtained in terms of the system's orbital parameters (value of the eccentricity of the outer planet as well as value of the $\AMD$), obtained in the framework of the quadrupolar approximation. Finally, we perform a numerical study on the form of the phase portraits for different mass ratios as well as semi-major axis ratios of the two planets, aiming to establish how generic are the phenomena reported above as the systems parameters are chosen close to one or more hierarchical limits.}           
\end{abstract}

%%%%%%%%%%%%%%%%%%%%%%%%%%%%%%%%%%%%%%%%%%%%%%%%%%%%%%%%%%%%%%%%%%%%%%%%
\section{Introduction}
\label{Intro}
%%%%%%%%%%%%%%%%%%%%%%%%%%%%%%%%%%%%%%%%%%%%%%%%%%%%%%%%%%%%%%%%%%%%%%%%
In the present paper we revisit the planetary three body problem in Poincar\'{e} heliocentric canonical variables governed by the Hamiltonian
\begin{equation}
\label{ham.3BP}
\Hscr(\vet{r_2},\vet{r_3},\vet{p_2},\vet{p_3})=\frac{\vet{p_2}^2}{2\, m_2}-\frac{\mathcal{G}\, m_0\, m_2}{r_2}+\frac{\vet{p_3}^2}{2 \,m_3}-\frac{\mathcal{G}\, m_0\, m_3}{r_3}+\frac{(\vet{p_2}+\vet{p_3})^2}{2\, m_0}-\frac{\mathcal{G}\, m_2\, m_3}{|\vet{r_2}-\vet{r_3}|}\, ,
\end{equation}
where $m_0 \gg m_2,\, m_3\, $. Thus $m_0$ represents the mass of a star, and $m_{i}\,$, $\vet{p_i}\,$, $\vet{r_{i}}\,$, $i=2,3$ the masses, barycentric momenta and heliocentric position vectors of two planets orbiting the star, at distances $r_i=|\vet{r_i}|\,$.

Our focus below will be on a systematic study of how the structure of the phase space of the above Hamiltonian system is altered as the planetary masses and distances and, most importantly, the \textit{mutual inclination} between the two planets' trajectories is varied. Answering this question in the framework of the three-body problem is a key step towards understanding the phase-space architecture in planetary systems with two or more planets in orbits with high mutual inclination. A well-studied case of the latter is the $\upsilon$-Andromedae system. This is a double star system with four planets ({\bf{b}}, {\bf{c}}, {\bf{d}} and {\bf{e}}) orbiting one of the stars. Since the mass $m_b$ is much smaller than $m_c\, , m_d$ the motion of the innermost planet {\bf{b}} can be modelel to a good approximation via a restricted four-body problem, with planets {\bf{c}}, {\bf{d}} providing the main perturbations (this motivates the use of the indices 2,3 in the Hamiltonian (\ref{ham.3BP}) for the masses and canonical variables of the planets); the long-term motion of the innermost planet $(m_1,\vet{r}_1,\vet{p}_1)$ is considered in a separate work (Mastroianni and Locatelli, in preparation). More detailed models can include planet {\bf{e}} as well as the second star. At any rate, all the above models require providing first a good analytical model for the orbits of the giant planets of the system, say, for example, the planets {\bf{c}} and {\bf{d}}, whose masses, as estimated by observations, are larger than $10\,M_{J}\,$ (see e.g~\cite{deietal2015} and~\cite{mcaetal2010}). 

Now, the available data for the orbital parameters of an extrasolar system are typically affected by wide observational error bars. As will be discussed in detail below, one important problem with the uncertainties in the observations stems from the fact that even small changes in a system's estimated parameters, consistent with the observations, may imply a drastic change in the type of orbital state in which the observed system is assumed to have been settled. We will argue below that this sensitive dependence on  available parameter estimates affects mostly those predictions referring to the \textit{secular} (long-term) evolution of the orbital state, i.e., the variations of the planets' eccentricity and inclination vectors which take place in timescales of the order of $\sim 10^2-10^4$ orbital periods. Notwithstanding these long timescales, characterizing the whole variety of possible stable secular orbital states of exoplanetary systems can be  relevant also in the interpretation of short-in-time observations: most importantly, it can serve the purpose of constraining observational uncertainties on the basis of stability considerations, i.e., indicating which subdomains in parameter space favor the long-term stability of the planetary orbits \cite{petetal2018}. 

Our study in the present paper focuses on one phenomenon, whose role appears central to the aim of classifying and characterizing the variety of possible secular orbital states in 3D planetary systems: this is the chain of bifurcations of periodic orbits which mark the transition, as the planets' mutual inclination increases, from the \textit{apsidal corotation} to the \textit{Lidov-Kozai} regime. 

As is well known, the apsidal corotation (AC) states (see, for example,~\citep{lauetal2002},~\cite{beaetal2003},~\cite{leepea2003}) provide the backbone of the phase space of secular motions for two planets in coplanar orbits. Formally, the apsidal corotations are the continuation, in the nonlinear regime, of the linear (Laplace-Lagrange) normal mode solutions of the planar two-planet secular Hamiltonian. Physically, they represent periodic orbits along which the pericenters of the two planets constantly precess by remaining either always anti-aligned (state A) or always aligned (state B). Periodic orbits having the same property in the exact (non-averaged with respect to fast angles) Hamiltonian can also be found \cite{had1975},~\cite{had2006} once the dynamics is regarded in a frame rotating with angular velocity equal to the (common) precession rate of the apsides. 

Given their importance in the planar case, a natural question regards how the family of apsidal corotation orbits is continued, as well as what is the dynamical role played by the AC periodic, and nearby quasi-periodic, orbits, when we pass from the coplanar to the 3D planetary orbital configuration. Both the above questions have been addressed in the literature using diverse formalisms (see \cite{beaetal2012} and references therein, as well as the discussion below). Owing to reasons explained in detail in section \ref{sec:hampc} below, in our present study we employ our own-proposed formalism, in which, by applying a so-called \textit{book-keeping} technique already at the level of Jacobi's reduction of the nodes \cite{jac1842}), we arrive at a natural decomposition of the 3D secular Hamiltonian of a system with fixed Angular Momentum Deficit (AMD; see \cite{lasrob1995}) as 
\begin{equation}\label{h3ddecompo}
\Hscr_{sec}=\Hscr_{planar}(\mathbf{X},\mathbf{Y})+\Hscr_{space}(\mathbf{X},\mathbf{Y};\AMD)~~.
\end{equation}
In Eq.(\ref{h3ddecompo}), $(\mathbf{X},\mathbf{Y})$ are Poincar\'{e} canonical variables for the two planets ($(X_2,Y_2)$, $(X_3,Y_3)$ are approximately proportional to the planets' eccentricity vectors). The Angular Momentum Deficit is defined by 
$$
\AMD=L_2+L_3-L_z 
$$
where $L_2$, $L_3$ are the angular momenta of the circular orbits at semi-major axes $a_2,a_3$ equal to those of the two planets, and $L_z$ the modulus of the total angular momentum normal to the system's Laplace plane. 

The methodological benefits from working with a decomposition of the secular Hamiltonian as in Eq.(\ref{h3ddecompo}) stem from the following properties (see sections 2 and 3 for details):

i) $\Hscr_{planar}(\mathbf{X},\mathbf{Y})$ is an integrable Hamiltonian, the quantity $J=(X_2^2+Y_2^2+X_3^2+Y_3^2)/4$ being a second integral independent of the energy.  

ii) For every permissible value of the energy $\Escr$, all the orbits $(\mathbf{X}(t),\mathbf{Y}(t))$ under Hamilton's equations with the Hamiltonian $\Hscr_{planar}$ are confined to the Laplace plane, i.e., they have zero mutual inclination at the given (and fixed in advance) level of AMD. 

iii) The Hamiltonian $\Hscr_{planar}$ can be explicitly shown to admit two periodic orbits (called below the `modes' A and B) which correspond to the anti-aligned and aligned apsidal corotation states for the two planets in the planar case. 

iv) The Hamiltonian $\Hscr_{space}$ can be further decomposed as $\Hscr_{space}=\Hscr_{0,space}+\Hscr_{1,space}$, where $\Hscr_{0,space}$ also admits $J=(X_2^2+Y_2^2+X_3^2+Y_3^2)/4$ as a second integral. Then, the Hamiltonian
\begin{equation}\label{def.h0}
\Hscr_{int}=\Hscr_{planar}+\Hscr_{0,space} 
\end{equation}
is integrable, and it has a similar formal structure as the Hamiltonian $\Hscr_{planar}$. In particular, the existence of periodic orbits of the type A and B can be explicitly demonstrated for the Hamiltonian $\Hscr_{int}$ using the integrability property, in the same way as for the Hamiltonian $\Hscr_{planar}$. 

v) Focusing, now, on the full secular Hamiltonian
\begin{equation}\label{hamful}
\Hscr_{sec}=\Hscr_{int}+\Hscr_{1,space}
\end{equation}
a Birkhoff-like normal form construction (section \ref{sec:dynamics}) allows to demonstrate (up to an exponentially small error) that the periodic orbits A,B (accompanied by neighboring quasi-periodic orbits) continue to exist in the full 3D regime. Hence, we call these orbits the \textit{3D apsidal corotation states}. 

vi) We parameterize the level of non-coplanarity (mutual inclination $\i_2+\i_3$) of the system in terms of the (constant) energy level of the full Hamiltonian $\Escr=\Hscr_{sec}$. One way to regard the connection between the value of the energy $\Hscr_{sec}$ and the level of non-coplanarity of the orbits is by noting that the energy grows in absolute value nearly as a quadratic function of the planetary eccentricities, i.e., nearly proportionally to a linear combination of $\e_2^2$ and $\e_3^2$. Thus, restricted to the chosen Poincar\'e surface of section (see below), the constant energy condition $\Escr=\Hscr_{sec}$ yields an ellipsoid-like surface. Also (see section \ref{sec:dynamics}), for fixed $\AMD$, the condition of constant mutual inclination yields the contour of a function also quadratic in the orbital eccentricities. Then, as shown in detail in section \ref{sec:dynamics}, for any fixed value of the energy $\Escr$, there are two values of the mutual inclination $\i_{mut}=\i_2+\i_3$, namely $\i_{mut}^{min}(\Escr)$, $i_{mut}^{max}(\Escr)$ such that, the corresponding contours of constant mutual inclination come tangent to the ellipsoidal surface of constant energy $\Escr=\Hscr_{sec}$. As a consequence, for any value of the mutual inclination in the interval $\i_{mut}^{min}(\Escr)\leq \i_{mut}\leq \i_{mut}^{max}(\Escr)$ there is an energetically permissible domain of motions. By the geometric properties of the above considerations, we find that both $\i_{mut}^{min}(\Escr)$ and $\i_{mut}^{max}(\Escr)$ are monotonically increasing functions of the energy. Thus, selecting a particular level value of the energy $\Escr$ fixes the overall level of mutual inclinations allowed at the energy $\Escr$. The minimum possible energy $\Escr_{min}$ is defined by the condition that the level surface $\Hscr_{sec}=\Escr_{min}$ is tangent to the level surface of minimum possible mutual inclination $\i_{mut}=0$ (planar case), while all other points of the surface $\i_{mut}=0\,$, except for the points of tangency, are in the interior of the surface $\Hscr=\Escr_{min}$. Solving for these conditions allows to specify the value of $\Escr_{min}$.    

In conclusion, the level of mutual inclination for all orbits increases in general with the quantity $\delta \Escr=\Escr-\Escr_{min}$, where $\Escr_{min}\leq \Escr\leq 0$. Thus, the lowermost value of $\delta\Escr$ is $\delta\Escr_{min}=0$, while the highermost limit is $\delta\Escr_{max}=-\Escr_{min}$, i.e., $\Escr=0$. At this latter limit, the available phase space shrinks to a point with $\e_2(t)=\e_3(t)=0$. This means a unique possible orbital configuration of two mutually-inclined circular planetary orbits. We call this the \textit{limiting trajectory of the Lidov-Kozai regime}. Actually, we will see in section \ref{sec:dynamics} that, for negative energies $\Escr$ close to zero, the phase space acquires a structure reminiscent to the one of the non-integrable Lidov-Kozai case of the restricted three-body problem (i.e. non intersecting trajectories examined in a higher than quadrupolar development of the disturbing function, see \cite{libetal2011}). The typical behavior of the trajectories in the Lidov-Kozai regime is to (quasi-)periodically exchange eccentricity for mutual inclination. However, we will see that such a coarse illustration of the dynamics is rather simplistic in the case of non-hierarchical (in distance ratios or masses) two-planet systems; in reality, the dynamics around the central Lidov-Kozai periodic orbit is highly unstable and the corresponding phase space turns to exhibit strong chaos. 

In summary, our focus in the present paper will be on describing, with sufficient detail, the observed transition in the structure of the phase space, as the parameter $\delta \Escr$ increases from its lowermost limit, corresponding to a nearly planar orbital configuration, to the highermost limit, corresponding to nearly circular orbits with a high degree of mutual inclination. 

As expected in the study of any dynamical system, structural changes in the phase space are associated with the birth, bifurcations and stability evolution of the most important periodic orbits of the system. We already mentioned that these are the ACs, in the nearly planar limit, and the Lidov-Kozai orbits, in the maximum mutual inclination limit. We follow the evolution and the connections between these periodic orbits, as well as other ones emerging in half the way between the two limiting regimes. In our study we work with analytical estimates, as well as with a numerical example inspired by the $\upsilon$-Andromedae system. This is an example of system being reasonably far from any hierarchical limit: the estimated mass ratio of the two planets is $m_2/m_3\simeq 1.3$, while the estimated semi-major axis ratio is $a_2/a_3\simeq 0.3$. On the other hand, section 4 contains results from a numerical investigation referring to different choices for the mass and semi-major axis ratios, representing every one of all possible cases of hierarchical models that could arise in the problem under consideration.  
As a rough guide, the phenomena discussed below should cover most cases of interest in the range of mass ratios $1/10\leq m_2/m_3 \leq 10\,$, distance ratios $ 1/7\leq r_2/r_3\leq 1/3 $ and mutual inclination $0\leq \i_{mut}\leq 45^\circ\, $.\\
\\
Few more words about our hereby proposed formalism and method of study: the spectrum of methods and formalisms proposed for the study of the secular planetary three body problem is nearly as wide as the literature itself on the subject. Different proposals distinguish between cases in which the three body problem dealt with is considered hierarchical (e.g. in the distances $r_2/r_3 \ll 1\,$; see~\cite{bro1959},~\cite{foretal2000},~\cite{miggoz2011}), or non-hierarchical (see~\cite{henlib2004},~\cite{naoetal2013a}). The choice of variables and/or proposed representation of the phase space of the system has often been motivated by whether the focus of a particular study is on phenomena related to apsidal-corotations (see~\cite{beaetal2003},~\citep{lauetal2002},~\cite{leepea2003}), or the Kozai instability (see~\cite{koz1962},~\cite{libdel2012},~\cite{libtsi2009},~\cite{litnao2011}). 
In addition, different formalisms stem from choices related to:\\

i) the type of \textit{Hamiltonian expansion}: this can be performed as the usual (Laplace-Lagrange) series expansion in powers of the planets' eccentricities and inclinations (see for example~\cite{murder1999} and~\cite{libhen2007}) or as a Legendre multipolar expansion (see~\cite{nao2016},~\cite{libdel2012} and references therein). One can note that, although the two types of expansions are equivalent \textit{in the limit} of infinite order of the expansion, different truncations (even with the same type of series) can lead to quantitatively, and even qualitatively, different results. Libert \& Henrard in~\cite{libhen2007} have discussed the question of the correct order of truncation in the framework of the Laplace-Lagrange expansion. Mogaszewki \& Go\'{z}dziewki and Naoz (see~\cite{miggoz2011},~\cite{naoetal2013a} and~\cite{nao2016}) discuss various truncated multipolar models for spatial hierarchical systems. To our knowledge, there is no literature on a comparison between the results obtained by truncated models with the two types of expansion for non-hierarchial spatial systems.\\

ii) \textit{Method of averaging}: different secular models are obtained by a different choice of method for averaging the Hamiltonian with respect to the fast angles. Such methods include: a) averaging ``by scissors'',  i.e., by just dropping-off the Hamiltonian all fast periodic terms (see, for example,~\cite{libhen2007},~\cite{henlib2004}). b) ``Closed-form'' averaging (e.g.~\cite{miggoz2011}). This method has the benefit of avoiding expansions in the orbital eccentricities (whose convergence becomes limited due to the limit in the series inversion of Kepler's equation, see, for example~\cite{szelam1982}). However, closed-form averaging can only be performed after a multipolar expansion of the Hamiltonian. Thus the method is particularly suited for systems hierachical with respect to the planetary distances, while its precision in the case of non-hierarchical systems is an open issue (see some results in section~\ref{sec:parametric}). c) Numerical computation (e.g. by Gauss' method) of the quadratures involved in the averaging (see~\cite{thomor1996},~\cite{miggoz2008}). This method is particularly suited for systems with intersecting trajectories, owing to theorems (see, for example, \cite{gromil1998}) establishing the continuity of the secular equations of motion at the points of intersection, which are singularities of the integrand functions appearing in the quadratures. d) Elimination of the mean anomalies from the Hamiltonian via a canonical transformation (see~\cite{bro1959} or~\cite{naoetal2013a}). From the theoretical point of view, the use of a canonical transformation is imperative when precision of second order in the planetary masses is sought for in the secular model. It should be stressed that while $\Oscr(m^2)$ expansions are straightforward to obtain in the framework of the Laplace-Lagrange series (see~\cite{locgio2000}), their counterpart in the form of closed-form series is an open question. In fact, closed-form averaging in the framework of the three-body problem requires the use of some `relegation' technique (see~\cite{paletal2006}), or alternatives as those recently proposed in~\cite{caveft2022}).\\

iii) \textit{Choice of coordinates}. Several sets of variables have been proposed as convenient for the  visualization and study of the phase space of secular motions. Examples are: 
$(\e_2\,\sin(\omega_2),\,\e_3\,\sin(\omega_3))$ (see for istance~\cite{libhen2007},~\cite{miggoz2011},~\cite{libtsi2009});
$(\e_j\cos(\Delta\varpi),\,\e_j\sin(\Delta\varpi))\,$, $j=2,\,3\,$ (see~\cite{micmal2004});
$(\e_2\,\cos(2\omega_2),\,\e_3\cos(\Delta\omega))$ (see~\cite{libdel2012});
$(\e_2\,\cos(\Delta\varpi),\,\e_3\,\cos(2\,\omega_2)\,)$ (see~\cite{miggoz2011},~\cite{micetal2006});
$(\e_j\,\cos(\omega_j),\,\e_j\,\sin(\omega_j))\,$ (see~\cite{thomor1996}).
Among the motivations behind the choice of a particular set of variables are: a) the treatment of singular cases (e.g. $\e_2=0$, $\e_3=0$, in which the longitude of the perihelium is no longer defined), b) the possibility to include all the main families of possible orbital states in one plot. As a characteristic example of the latter, see figure 3 of~\cite{micetal2006}; one can remark, however, the complexity involved in properly deciphering the information given in that figure, which is evident from the accompanying caption.

Our own choice regarding the points (i) to (iii) above is: (i) and (ii) we base most of our results on a closed-form averaging of a multipolar expansion (in powers of the distance ratio $r_2/r_3<1$) of the Hamiltonian~\eqref{ham.3BP}. Besides its compact form, convenient for analytic studies (as, for example, in subsection~\ref{sub:Kozai}), such a model, truncated at a sufficiently high multipole order, circumvents the problem of slow convergence of the Laplace-Lagrange series for highly eccentric orbits, without compromising precision even far from the hierarchical limit (e.g. for $r_2/r_3\sim 0.3\,$). In order to specify the suitable order of multipole  truncation, in section \ref{sec:hampc} we make a comparison of the phase portraits obtained via the secular Hamiltonian  arrived at by the above method versus those obtained by a scissors' averaging of the Laplace-Lagrange series truncated at order 10 in the eccentricities. Also, in obtaining the final Hamiltonian we introduce a book-keeping procedure for the Jacobi reduction of the nodes, which, as mentioned already, leads to a convenient decomposition of the Hamiltonian as in Eq.(\ref{h3ddecompo}). (iii) We illustrate all phase portraits using the usual Poincar\'e surface of section $(\e_2\cos\omega_2,\e_2\sin\omega_2)$ every time when $\omega_3=\pi$, $\dot{\omega_3}\geq 0$. The sequence of canonical transformations leading to such a representation of phase portraits is described in section \ref{sec:hampc}. Owing to the conservation of angular momentum, one can easily see (\cite{cusbat1997}) that the phase space of the integrable models $\Hscr_{planar}$ and $\Hscr_{int}$ is the sphere $S^2$ (instead of $\mathbb{R}^4$, as generically true for Hamiltonian systems with two degrees of freedom). Furthermore, as analyzed in section \ref{sec:hampc}, some points of the sphere `inflate' to curves in the Poincar\'{e} surface of section defined as above. Finally, using an appropriate set of good variables whose Poisson algebra admits the angular momentum as a Casimir, the two modes A and B are seen to be separated by a meridian circle in the sphere $S^2$, which however does not correspond to a dynamical separatrix (since the integrable model contains no unstable periodic orbits). We devote some effort to carefully describe these phenomena, which are often found to generate some confusion in the literature when use is made of some of the components of the planets' eccentricity vectors to describe the structure of the problem's phase space. 

The paper is organized as follows. Section~\ref{sec:hampc} describes all the steps followed in order to arrive at the finally adopted secular Hamiltonian model. This includes the introduction of an appropriate `book-keeping', the way this is used in the Jacobi reduction of the Hamiltonian, the chain of canonical transformations leading the Hamiltonian to its final form in Poincar\'{e} canonical variables, as well as a number of precision tests about the order of the multipole truncation, comparison with the Laplace-Lagrange series, etc.  Section~\ref{sec:dynamics} analyzes the dynamics and the phase portraits under the secular Hamiltonian model computed as in section~\ref{sec:hampc}. In particular, subsections~\ref{subsub:integrable} and~\ref{subsub:hopf} discuss the phase-space properties of the integrable approximations $\Hscr_{planar}$ and $\Hscr_{int}$, and introduces the basic terminology for the orbits of the apsidal corotation type (modes A and B). Subsection~\ref{subsub:normalforms} deals, instead,  with an analysis of the phase-space structure under the full (non-integrable) Hamiltonian (\ref{h3ddecompo}).  This includes the semi-analytical method (via a normal form) used to demonstrate the continuation of the A and B modes as well as to produce the time series yielding the evolution of the eccentricity vectors for both planets along these modes. Subsection~\ref{sub:bifurcation} discusses numerical results obtained by computing the surfaces of section as the energy $\Escr$ increases (or the parameter $\delta \Escr$ decreases), showing the sequence of bifurcations that connect the apsidal corotation with the Lidov-Kozai regime. Subsection~\ref{sub:Kozai} contains some results related to the `Kozai mechanism', i.e.,  the transition from linear stability to instability for the inclined circular orbit of one of the two planets. These results are compared to an analytical approximation obtained in the framework of the quadrupolar approximation. Section~\ref{sec:parametric} contains the investigation of how the phase portraits change in cases with different distance or mass ratios of the planets, covering various hierachical models ($r_2/r_3=1/7,\,1/3,\,$ $m_2/m_3=1/10,\,1/3,\,1,\,3,\,10\,$). Section~\ref{sec:conclusions} provides a summary of the main conclusions from the present study.   

%%%%%%%%%%%%%%%%%%%%%%%%%%%%%%%%%%%%%%%%%%%%%%%%%%%%%%%%%%%
\section{Hamiltonian model}
\label{sec:hampc}
%%%%%%%%%%%%%%%%%%%%%%%%%%%%%%%%%%%%%%%%%%%%%%%%%%%%%%%%%%%

%-----------------------------------------------------
\subsection{Averaged Hamiltonian}
\label{sub:Closed}
%-----------------------------------------------------
We will focus below mostly on the properties of a secular model $\Hscr_{sec}$ for the Hamiltonian~\eqref{ham.3BP} obtained by performing averaging with respect to the fast angles just `by scissors'. We denote by $(a,\e,\i,M,\omega,\Omega)$ the Keplerian elements of a body (semi-major axis, eccentricity, inclination, mean anomaly, argument of the periastron, argument of the nodes), and by $\lambda=M+\omega+\Omega$, $\varpi=\omega+\Omega$ the mean longitude and longitude of the periastron respectively. We then have the following properties of the secular model obtained after averaging `by scissors': 
\begin{eqnarray}\label{hamscis}
\Hscr_{sec}&=&{1\over 4\pi^2}
\int_{0}^{2\pi}\int_{0}^{2\pi} H(\vet{r_2},\vet{r_3},\vet{p_2},\vet{p_3}) \,dM_{2}\,dM_{3} \nonumber\\
&=&
{1\over 4\pi^2}\int_{0}^{2\pi}\int_{0}^{2\pi} 
\left(\left({1\over m_0}+{1\over m_2}\right)\frac{\vet{p_2}^2}{2}-\frac{\mathcal{G}\, m_0\, m_2}{r_2}
+
\left({1\over m_0}+{1\over m_3}\right)\frac{\vet{p_3}^2}{2}-\frac{\mathcal{G}\, m_0\, m_3}{r_3} 
\right)\,dM_{2}\,dM_{3} \nonumber\\
&+&{1\over 4\pi^2}\int_{0}^{2\pi}\int_{0}^{2\pi}
\frac{\vet{p_2}\cdot\vet{p_3}}{ m_0}dM_{2}\,dM_{3}
-{1\over 4\pi^2}\int_{0}^{2\pi}\int_{0}^{2\pi}\frac{\mathcal{G}\, m_2\, m_3}{|\vet{r_2}-\vet{r_3}|}\,dM_{2}\,dM_{3}~~.
\end{eqnarray}
We have
\begin{eqnarray}\label{hkep0sec}
~&~& {1\over 4\pi^2}\int_{0}^{2\pi}\int_{0}^{2\pi} 
\left(\left({1\over m_0}+{1\over m_2}\right)\frac{\vet{p_2}^2}{2}-\frac{\mathcal{G}\, m_0\, m_2}{r_2}
+
\left({1\over m_0}+{1\over m_3}\right)\frac{\vet{p_3}^2}{2}-\frac{\mathcal{G}\, m_0\, m_3}{r_3} 
\right)\,dM_{2}\,dM_{3}  \nonumber\\
&=&-{\mathcal{G} m_0 m_2\over 2a_2}-{\mathcal{G} m_0 m_3\over 2a_3}+
{\mathcal{G}  m_2^2\over 2a_2}+{\mathcal{G} m_3^2\over 2a_3}~~.
\end{eqnarray}
The indirect part of the disturbing function, depending on the product $\vet{p_2}\cdot\vet{p_3}$, yields 
zero average
\begin{equation}\label{indmomenta}
{1\over 4\pi^2}\int_{0}^{2\pi}\int_{0}^{2\pi}\frac{\vet{p_2}\cdot\vet{p_3}}{ m_0}\,dM_{2}\,dM_{3}
={1\over 4\pi^2}
{\Gscr \,m_2 m_3\over \sqrt{a_2^3\,a_3^3}}
\left( \int_{0}^{2\pi}{d\vet{r_2}\over dM_2}  \,dM_{2}
\int_{0}^{2\pi}{d\vet{r_3}\over dM_3}  \,dM_{3}\right)=0~~.
\end{equation}
To compute the average of the direct part $\mathcal{G}m_2m_3/|\vet{r_2}-\vet{r_3}|$, we assume a dynamical regime of the planetary system in which the distance ratio $r_2/r_3$ (where $r_{i}=|\vet{r}_{i}|\,$, $i=2,3$) remains always smaller than unity. Then, the quantity $\mathcal{G}m_2m_3/|\vet{r_2}-\vet{r_3}|$ admits a convergent Legendre multipolar expansion in powers of the quantity $r_2/r_3<1$:
\begin{align*}
-{1\over 4\pi^2}\int_{0}^{2\pi}\int_{0}^{2\pi}\frac{\mathcal{G}\, m_2\, m_3}{|\vet{r_2}-\vet{r_3}|} \,&dM_{2}\,dM_{3}\, =\\
&-{\mathcal{G}\, m_2\, m_3\over 4\pi^2}
\int_{0}^{2\pi}\int_{0}^{2\pi}\left(\frac{1}{r_{3}} +\frac{\vet{r}_{2}\cdot\vet{r}_{3}}{r_{3}^3}-\frac{1}{2}\frac{r_2^2}{r_3^3}+\frac{3}{2}\frac{(\vet{r}_{2}\cdot\vet{r}_{3})^2}{r_{3}^5} +\ldots\right) 
\,dM_{2}\,dM_{3}\, .
\end{align*}
We have:
\begin{align*}
&-{\mathcal{G}\, m_2\, m_3\over 4\pi^2}
\int_{0}^{2\pi}\int_{0}^{2\pi}\frac{1}{r_{3}} dM_1dM_2=-\frac{\mathcal{G}m_2 m_3}{a_3}\, , &
&\int_{0}^{2\pi}\int_{0}^{2\pi}\frac{\vet{r}_{2}\cdot\vet{r}_{3}}{r_{3}^3}\,dM_{2}\,dM_{3}=0~~.
\end{align*}  
We then need to compute
\begin{equation}\label{remainder}
\mathcal{R}_{sec}= {1\over 4\pi^2}
\int_{0}^{2\pi}\int_{0}^{2\pi}-\frac{\mathcal{G}\, m_2\, m_3}{r_3}\left(-\frac{1}{2}\frac{r_2^2}{r_3^2}+\frac{3}{2}\frac{(\vet{r}_{2}\cdot\vet{r}_{3})^2}{r_{3}^4} +\ldots\right) \,dM_{2}\,dM_{3}\, .
\end{equation}
Keeping only the lowest order term in the integrand of (\ref{remainder}), proportional to $(r_2/r_3)^2\,$, leads to the so-called \textit{quadrupole} approximation; the next truncation (up to terms proportional to $(r_2/r_3)^3\,$) is the \textit{octupole} approximation, etc. The integrals of any multipole approximation  can be computed in so-called \textit{closed form} (i.e. without expansions in the eccentricities), by avoiding completely the series reversion of Kepler's equation, using, instead, the change of variables $M_2\rightarrow u_2$ (eccentric anomaly), $M_3\rightarrow f_3\,$ (true anomaly). We have
\begin{equation}\label{trick.closed}
dM_2=(1-\e_2\cos u_2)\,du_2,~dM_3={r_3^2\over a_3^2\sqrt{1-\e_3^2}} \, df_3,~
r_2=a_2(1-\e_2\cos(u_2)),~\frac{1}{r_3}=\frac{1+\e_3\cos(f_3)}{a_3\,(1-\e_3^2)}\,  .
\end{equation}
Replacing the above expressions in~\eqref{remainder} and performing all trigonometric reductions, we find a trigonometric polynomial series containing only terms of the form $\cos(K_2u_2+K_3f_3+...)$, with $K_2,K_3$ integers. This implies that the average can be computed by just scissor-cutting all the terms in the integrand of ~\eqref{remainder} for which $|K_2|+|K_3|\neq 0$. This leads to a closed-form expression for the secular Hamiltonian 
\begin{equation}\label{H.closed}
\Hscr_{sec}=-{\mathcal{G} m_0 m_2\over 2a_2}-{\mathcal{G} m_0 m_3\over 2a_3}+{\mathcal{G}  m_2^2\over 2a_2}+{\mathcal{G} m_3^2\over 2a_3}-{\mathcal{G} m_2 m_3\over a_3}+\mathcal{R}_{sec}(a_2,a_3,\e_2,\e_3,\i_2,\i_3,\omega_2,\omega_3,\Omega_2-\Omega_3)~~.
\end{equation} 

The following are some relevant remarks regarding the Hamiltonian~\eqref{H.closed}:\\
\\
\textit{Remark 1:} the averaging~\eqref{remainder} yields a valid secular model only when the system is assumed to be \textit{far from any low-order mean-motion resonance}. By `low' it is implied that no resonance condition of the form 
$$
|K_2n_2+K_3n_3|<\mathcal{O}\left((\mathcal{G}m_0\mu)^{1/2}/a_2^{3/2}\right)~~,
$$
should be satisfied, with $K_2,K_3$ integers, $\mu=\max(m_2/m_0,m_3/m_0)$, $n_2=\left(\mathcal{G}(m_0+m_2)/a_2^3\right)^{1/2}$, $n_3=\left(\mathcal{G}(m_0+m_3)/a_3^3\right)^{1/2}$, for an order $|K_2|+|K_3|$ inferior or equal to the order $N_P$ of the multipole expansion in Eq.(\ref{remainder}). Formally, such a requirement reflects the fact that the averaging by scissors serves to substitute the complete procedure of first-order averaging, which involves a canonical transformation to properly eliminate from the Hamiltonian the fast angles $M_2,M_3$. Such a transformation can be defined in closed form (see, for example, \cite{caveft2022}), but it involves divisors of the form $K_2n_2+K_3n_3$ which become very small near any mean motion resonance of order smaller or equal to $N_P$.\\
\\
\textit{Remark 2}: the function $\mathcal{R}_{sec}$ is trigonometric polynomial in the angles $\omega_2,\omega_3$, as well as $\Omega_2-\Omega_3$. The dependence on $\Omega_2,\Omega_3$ only by the difference $\Omega_2-\Omega_3$ is a consequence of the fact that the sum of the angular momenta $H_2+H_3$ perpendicularly to the system's Laplace plane is an invariant of the motion of the complete Hamiltonian~\eqref{ham.3BP}, which is preserved also in the secular Hamiltonian~\eqref{H.closed}. \\
\\
\textit{Remark 3}: the function $\mathcal{R}_{sec}$ is polynomial in the quantities $a_2,\e_2,\e_3,\sin \i_2,\sin \i_3,\cos \i_2,\cos \i_3$, while it is rational in the quantities $a_3,\eta_3=\sqrt{1-\e_3^2}$. As regards the dependence on $\sin \i_2,\sin \i_3,\cos \i_2,\cos \i_3$, this follows the symmetry that $\mathcal{R}_{sec}$ can only depend on the quantity $\cos(\i_2+\i_3)$, where $\i_2+\i_3=\i_{mut}$ is the \textit{mutual inclination} of the planetary trajectories. This symmetry is exploited when performing Jacobi reduction of the nodes in the Hamiltonian $\Hscr_{sec}$ (see next subsection). On the other hand, the rational dependence on the quantity $\sqrt{1-\e_3^2}$ implies that any expansion in the eccentricities performed \textit{after} the closed-form averaging is convergent in the entire domain $|\e_2|<1,|\e_3|<1$. This allowance to perform an a posteriori expansion of the Hamiltonian $\Hscr_{sec}$ in the planetary eccentricities proves useful in the theoretical analysis of the dynamical properties of such a Hamiltonian (see section \ref{sec:dynamics}). 

%------------------------------------------------------------------------
\subsection{Book-keeping and Jacobi reduction of the nodes}
\label{sub:bookkeeping}
%------------------------------------------------------------------------
Consider the set of Delaunay canonical variables:
\begin{align}\label{Delau}
&L_j=m_j\sqrt{\mathcal{G}\,m_0\,a_j}\, , & &l_j=M_j\, ,\notag\\
&G_j=L_j\sqrt{1-\e^2_j}\, , & &g_j=\omega_j\, , \\
&H_j=G_j\cos(\i_j)\, , & &h_j=\Omega_j\, \notag
\end{align}
with $j=2,3$. Since $a_j=a_j(L_j)$, $\e_j=\e_j(L_j,G_j)$, $\i_j=\i_j(L_j,G_j,H_j)$, substituting the corresponding expressions in the Hamiltonian $\mathcal{H}_{sec}$ (Eq.~\eqref{H.closed}) leads to a Hamiltonian model with six degrees of freedom
\begin{equation}\label{hsecdelau}
\mathcal{H}_{\sec}\equiv\mathcal{H}_{sec}(L_2,L_3,G_2,G_3,H_2,H_3,\omega_2,\omega_3,\Omega_2-\Omega_3)~~. 
\end{equation}
Since the angles $M_2,M_3$ are ignorable, the Delaunay momenta $L_2,L_3$ are integrals of motion under the Hamiltonian flow of $\mathcal{H}_{sec}$. There are two more integrals of motion in involution with $L_2,L_3$, corresponding to the two components of the total angular momentum, normal and parallel to the Laplace plane. For the normal component we have $L_z=H_2+H_3=C=const$, while for the parallel component we have $L_{\parallel}=G_2\sin(\i_2(L_2,G_2,H_2))-G_3\sin(\i_3(L_3,G_3,H_3))=0$. Thus, with an appropriate reduction, we can reduce the system to a Hamiltonian with two degrees of freedom. This reduction, implemented on the invariant manifold $L_z=C$, $L_\parallel=0$ is called the Jacobi reduction of the nodes \cite{jac1842} and it exploits one more invariance of the system: since $L_\parallel=0$ we necessarily have $\Omega_2-\Omega_3=\pi=const$. 

In our analytical treatment of the Hamiltonian $\mathcal{H}_{sec}$ it turns convenient to perform the Jacobi reduction of the nodes by simultaneously introducing two `book-keeping symbols', $\varepsilon$, $\eta$ \cite{eft2012}, both with numerical values $\varepsilon=1$, $\eta=1$, whose role is the following: i) $\varepsilon$ keeps track of the order of a certain term in the planetary eccentricities and inclinations, ii) $\eta$ separates Hamiltonian terms which depend on powers of the quantity $\cos(\i_2+\i_3)$ from those terms which do not depend on the mutual inclination $\i_{mut}=\i_2+\i_3$. 

More specifically, we Jacobi-reduce the Hamiltonian $\mathcal{H}_{sec}$ by the following steps:\\
\\
\textit{Step 1: Canonical transformation}. Similarly as in~\citep{libhen2007}, we introduce the canonical transformation
\begin{align}\label{Delaujac}
&\Lambda_2=L_2\, , & &\lambda_2=M_2+\omega_2+\Omega_2\, ,\notag\\
&\Lambda_3=L_3\, , & &\lambda_3=M_3+\omega_3+\Omega_3\, ,\notag\\
&W_2=L_2-G_2\,   , & &w_2=-\omega_2\, , \\
&W_3=L_3-G_3\,   , & &w_3=-\omega_3\, , \notag\\
&R_2=L_2-H_2\,   , & &\theta_{r2}=\Omega_3-\Omega_2\, , \notag\\
&R_3=L_2+L_3-H_2-H_3=\AMD\,   , & &\theta_{r3}=-\Omega_3\, . \notag
\end{align}
Inverting the transformation~\eqref{Delaujac} and substituting the result into the Hamiltonian~\eqref{hsecdelau}, the secular Hamiltonian obtains the form:
\begin{equation}\label{hsecdelaujac}
\mathcal{H}_{\sec}\equiv
\mathcal{H}_{sec}(\Lambda_2,\Lambda_3,W_2,W_3,R_2,R_3,w_2,w_3,\theta_{r2})~~. 
\end{equation}
\\
\textit{Step 2: Jacobi reduction with book-keeping}. Hamilton's equations for the Hamiltonian~\eqref{hsecdelaujac} yield $\dot{\Lambda}_2=\dot{\Lambda}_3=0$ (constancy of the semi-major axes $a_2$, $a_3$), as well as $\dot{R}_3=0$ (constancy of $L_z$, implying the constancy of the AMD). The crucial point stems from the following invariance property of 
the Hamiltonian (\ref{hsecdelaujac}):
\begin{equation}\label{dhamdthr2}
\dot{\theta}_{r2}=
{\partial\mathcal{H}_{sec}(\Lambda_2,\Lambda_3,W_2,W_3,R_2,R_3=\AMD,w_2,w_3,\theta_{r2}=\pi)\over\partial R_2}=0
\end{equation}
corresponding to the invariance in time of the relation $\theta_{r2}=\Omega_3-\Omega_2=\pi$ for all trajectories. Equation~\eqref{dhamdthr2}, however, implies that when the substitution $\Omega_3=\Omega_2+\pi$ is made in the Hamiltonian~\eqref{hsecdelaujac}, the resulting expression becomes independent of $R_2$, and the reduced set of Hamilton's equations
\begin{align}\label{hameqred}
&\dot{w}_2={\partial\mathcal{H}_{sec}(\Lambda_2,\Lambda_3,W_2,W_3,R_3=\AMD,w_2,w_3,\theta_{r2}=\pi)
\over\partial W_2}
\, , \notag\\
&\dot{w}_3={\partial\mathcal{H}_{sec}(\Lambda_2,\Lambda_3,W_2,W_3,R_3=\AMD,w_2,w_3,\theta_{r2}=\pi)
\over\partial W_3}
\, , \\
&\dot{W}_2=-{\partial\mathcal{H}_{sec}(\Lambda_2,\Lambda_3,W_2,W_3,R_3=\AMD,w_2,w_3,\theta_{r2}=\pi)
\over\partial w_2}
\, , \notag\\
&\dot{W}_3=-{\partial\mathcal{H}_{sec}(\Lambda_2,\Lambda_3,W_2,W_3,R_3=\AMD,w_2,w_3,\theta_{r2}=\pi)
\over\partial w_3}
\, , \notag
\end{align}
remains valid. Note that $\Lambda_2,\,\Lambda_3$ in the above \textit{secular equations of motion} are constant, and can be effectively treated as parameters, depending on the (also constant) parameters $a_2,a_3$. Also, the terms 
$$
-{\mathcal{G} m_0 m_2\over 2a_2}-{\mathcal{G} m_0 m_3\over 2a_3}+{\mathcal{G}  m_2^2\over 2a_2}+{\mathcal{G} m_3^2\over 2a_3}-{\mathcal{G} m_2 m_3\over a_3}
$$
appearing in Eq.~\eqref{H.closed}, do not contribute to the secular equations of motion, and can be omitted from further analysis, by just renaming $\mathcal{R}_{sec}\rightarrow\mathcal{H}_{sec}$. Finally, Eqs.~\eqref{hameqred} allow to compute the secular evolution of only the eccentricity vectors of the two planets. To obtain the evolution in inclination, instead, we use the following relations,  obtained directly from the conservation of the angular momentum:
\begin{equation}\label{cosi23e23}
\cos(\i_2)={\Lambda_2^2(1-\e_2^2)-\Lambda_3^2(1-\e_3^2)+L_z^2\over 2L_z\Lambda_2\sqrt{1-\e_2^2}},~~~ 
\cos(\i_3)={L_z-\Lambda_2\sqrt{1-\e_2^2}\cos(\i_2)\over \Lambda_3\sqrt{1-\e_3^2}}~~. 
\end{equation}

In the practical implementation of step 2, we perform the substitutions of the angle variables $\Omega_3=\Omega_2+\pi$, as well as $\omega_2=-w_2$, $\omega_3=-w_3$ in the Hamiltonian, but leave implicit the latter's dependence on the action variables $(\Lambda_2,\Lambda_3,W_2,W_2,R_2,R_3)$ through the elements $(a_2,a_3,\e_2,\e_3,\i_2,\i_3)$. Owing, however, to the non-dependence of the Hamiltonian on $R_2$ \textit{after} the substitution $\Omega_3=\Omega_2+\pi$, we have that the Hamiltonian depends on the inclinations only through the trigonometric combination $\cos(\i_2+\i_3)=\cos(\i_2)\cos(\i_3)-\sin(\i_2)\sin(\i_3)\,$. Taking into account that 
$$
\left(1- \cos(\i_2)\cos(\i_3)\right)=\Oscr(\varepsilon^2)\, , \qquad\, 
\sin(\i_2)\sin(\i_3)=\Oscr(\varepsilon^2)~~,
$$
where the attribution of $\Oscr(\varepsilon^2)$ to an expression stands for `second order in the eccentricities and inclinations', we then introduce the following \textit{book-keeping control identities}:
\begin{align}\label{riduzione.cos.sen}
&\cos(\i_2)\cos(\i_3)=\varepsilon^2\eta\left(\cos(\i_2)\cos(\i_3)-1\right)+1\, , & 
&\sin(\i_2)\sin(\i_3)=\varepsilon^2\eta\sin(\i_2)\sin(\i_3)\, . 
\end{align}
We also use the substitution rule
\begin{equation}\label{seni2i3}
\sin(\i_2)\sin(\i_3)= \cos(\i_2)\cos(\i_3)+\frac{\Lambda_2\sqrt{1-\e_2^2}}{2\,\Lambda_3\sqrt{1-\e_3^2}}
+\frac{\Lambda_3\sqrt{1-\e_3^2}}{2\,\Lambda_2\sqrt{1-\e_2^2}}
-\frac{L_z^2}{2\,\Lambda_2\sqrt{1-\e_2^2}\,\Lambda_3\sqrt{1-\e_3^2}}\, .
\end{equation}
Substituting the above expressions into the Hamiltonian, and truncating the resulting expression up to a pre-selected \textit{maximum order in book-keeping} $N_{bk}$, by symmetry the terms in equal powers of the products $\sin(\i_2)\sin(\i_3)$ and $\cos(\i_2)\cos(\i_3)$ are opposite, and thus they are cancelled. Hence, the Hamiltonian resumes the form:
\begin{equation}\label{hambketa}
\mathcal{H}_{sec}=\sum_{s_1=0}^{N_{bk}/2}\eta^{s_1}\varepsilon^{2s_1}\mathcal{H}_{sec,s1}(\e_2,\e_3,w_2,w_3;a_2,a_3,L_z)~~. 
\end{equation}
We finally restore the values $\varepsilon=1$ and $\eta=1$ of the book-keeping parameters, and write the truncated (up to book-keeping order $N_{bk}$) Hamiltonian as:
\begin{equation}\label{hamplsp}
\mathcal{H}_{sec}=\mathcal{H}_{planar}+\mathcal{H}_{space} 
\end{equation}
where
\begin{equation}\label{hamplspanal}
\mathcal{H}_{planar}= \mathcal{H}_{sec,0}(\e_2,\e_3,w_2-w_3;a_2,a_3),~~~
\mathcal{H}_{space}= \sum_{s_1=1}^{N_{bk}/2}\mathcal{H}_{sec,s1}(\e_2,\e_3,w_2,w_3;a_2,a_3,L_z)~.
\end{equation}
Note that the term $\mathcal{H}_{planar}$ depends only on the difference $w_2-w_3$, since, in the planar case, the sum $W_2+W_3$, which represents the total angular momentum deficit for planar orbits, is a conserved quantity.  
\\
\textit{Step 3: Expansion in the orbital eccentricities}. We re-introduce the book-keeping:
\begin{equation}\label{ecclzbk}
\e_2\rightarrow \varepsilon \e_2,~~\e_3\rightarrow \varepsilon \e_3,~~ 
L_z \rightarrow \Lambda_2+\Lambda_3-\varepsilon^2\AMD\, .
\end{equation}
Substituting the above expressions in the Hamiltonian~\eqref{hamplspanal}, and expanding the resulting expressions in powers of the book-keeping parameter $\varepsilon$, apart from a term depending only on the constants $a_1,a_2$ and $\AMD$, we arrive at the truncated Hamiltonian:
\begin{eqnarray}\label{hamnostra}
\label{Ham.Lapl.nostra2}
\Hscr_{sec}&=&\Hscr_{planar}+\Hscr_{space} \\
&=&\sum_{s_2=1}^{N_{bk}/2}\varepsilon^{2 s_{2}} h_{s_{2}}(\e_2,\e_3, w_2-w_3;a_2,a_3) 
+\sum_{s_2=1}^{N_{bk}/2}\varepsilon^{2 s_{2}}\tilde{h}_{s_2}(\e_2,\e_3,w_2,w_3;a_2,a_3,\AMD)~~.\nonumber
\end{eqnarray}
\\
\textit{Step 4: introduction of Poincar\'e variables:} 
The Hamiltonian~\eqref{Ham.Lapl.nostra2} is polynomial in the eccentricities $\e_2,\,\e_3$. It was already stressed that the series expansion with respect to the quantities $\eta_2=\sqrt{1-\e_2^2}$ and $\eta_3=\sqrt{1-\e_3^2}$ introduces no divergence in the polydisc $|\e_2|<1$, $|\e_3|<1$. Besides, the Hamiltonian satisfies the D'Alembert rules in the eccentricities: every trigonometric term in it is of the form $h_{\ell_2,\ell_3,k_2,k_3}(a_2,a_3)(\AMD)^{\ell_1}\e_2^{\ell_2}\e_3^{\ell_3}{\cos}(k_2w_2+k_3w_3)$, with $\ell_1$ positive integer, and the integers $\ell_2,\ell_3,k_2,k_3$ satisfying i) $\ell_2,\ell_3>0$, ii) $\ell_2+\ell_3\geq |k_2|+|k_3|$, and iii) $mod(\ell_2,2)=mod(k_2,2)$, $mod(\ell_3,2)=mod(k_3,2)$. Rules (i) to (iii) imply, now, that the Hamiltonian is polynomial in the Poincar\'e canonical variables stemming from the variables $(w_j,W_j)$, $j=2,3$ through the canonical transformation
\begin{align}\label{Poincare1}
& X_j= -\sqrt{2 \,W_j}\cos(w_j)\, , & & Y_j=\sqrt{2\,W_j}\sin(w_j)\, , & &j=2,3~~.
\end{align} 
To obtain a truncated polynomial series of the Hamiltonian in the variables $(X_j,Y_j)$, we substitute into the Hamiltonian~\eqref{Ham.Lapl.nostra2} the expressions
$$
\sin(w_j)={Y_j\over\sqrt{2W_j}},~~~\cos(w_j)=-{X_j\over\sqrt{2W_j}},~~~\e_j=\sqrt{{2W_j\over\Lambda_j}-\varepsilon^2\left({W_j\over\Lambda_j}\right)^2},~~j=2,3~
$$
and expand the result in powers of the book-keeping parameter $\varepsilon$, up to the truncation order $N_{bk}$. This leads to an expression which no longer contains trigonometric functions of the angles $w_j$, while it still contains integer powers of the actions $W_j$. We then substitute $W_j\rightarrow(X_j^2+Y_j^2)/2$, and, finally, set back $\varepsilon=1$. In this way we arrive at the final secular model:
\begin{eqnarray}\label{Ham.final}
\Hscr_{sec}(X_2, X_3, Y_2, Y_3; \AMD)&=&
\Hscr_{planar}(X_2, X_3, Y_2, Y_3)+\Hscr_{space}(X_2, X_3, Y_2, Y_3;\AMD) \nonumber\\
&=&
\sum_{\ell\in\mathbb{N}^4,|\ell|=2}^{N_{bk}}
\mathcal{K}_{plane,\ell}X_2^{\ell_1}\,X_3^{\ell_2}\,Y_2^{\ell_3}\,Y_3^{\ell_4} \\
&+&
\sum_{\ell\in\mathbb{N}^4,|\ell|=2}^{N_{bk}}
\mathcal{K}_{space,\ell}(\AMD)X_2^{\ell_1}\,X_3^{\ell_2}\,Y_2^{\ell_3}\,Y_3^{\ell_4}~~.\nonumber
\end{eqnarray}
Note that, by symmetry, the value of the disturbing function (and hence of the secular Hamiltonian) remains invariant by the rotation of both planets' argument of the pericenter by $\pi$, hence the secular Hamiltonian is necessarily even in the planetary eccentricities. This implies that the Hamiltonian~\eqref{Ham.final} contains only even powers of the Poincar\'e variables $(X_j,Y_j)\,$, i.e. $\ell_1+\ell_2$ and $\ell_3+\ell_4$ are even. In particular, Hamilton's equations take the form: 
\begin{equation}\label{conseguenza.Lib-Henr}
\begin{split}
&\dot{X_i}=\frac{\partial \Hscr_{sec}}{\partial Y_i}
=Y_2\,F_{X_i,2}(X_2, X_3, Y_2, Y_3)+Y_3\,F_{X_i,3}(X_2, X_3, Y_2, Y_3)\, ,\\
&\dot{Y_i}=-\frac{\partial \Hscr_{sec}}{\partial X_i}
=X_2\,F_{Y_i,2}(X_2, X_3, Y_2, Y_3)+X_3\,F_{Y_i,3}(X_2, X_3, Y_2, Y_3)\, ,
\end{split}
\end{equation}
where the polynomials $F_{X_i,2}$, $F_{X_i,3}$, $F_{Y_i,2}$, $F_{Y_i,2}\,$, $ i=2,\,3\, ,$ are even, starting with constant terms. 

%--------------------------------------------------------------
\subsection{Poincar\'e surface of section. Precision tests}
\label{sub:precision.tests} 
%--------------------------------------------------------------
We will now discuss several precision tests, based on the method of comparison of phase portraits, which aim to establish which is the minimum multipole order, as well as minimum order in the eccentricities of the Hamiltonian $\Hscr_{sec}$, such that the Hamiltonian, truncated at the above orders, represents with sufficient precision the dynamics at the timescales of the secular system produced by averaging of the original Hamiltonian~\eqref{ham.3BP}. 

%--------------------------------------------------------------
\subsubsection{Poincar\'e surface of section: definitions}
\label{subsub:Poincare.section} 
%--------------------------------------------------------------
To visualize phase portraits, use is made below of a Poincar\'e surface of section $\mathcal{P}(\Escr;\AMD)$ defined by the relations:
\begin{equation}\label{pcsec}
\begin{aligned}
\mathcal{P}(\Escr;\AMD)=&
\bigg\{ (X_2,Y_2,X_3,Y_3)\in\mathbb{R}^4:~ 
Y_3=0, ~
\dot{Y}_3=-{\partial\Hscr_{sec}(X_2,Y_2,X_3,Y_3=0;\AMD)\over\partial X_3}\geq 0,~\\
&~~~
\cos(\i_{max})\leq\cos(\i_{mut})(X_2,Y_2,X_3,Y_3=0;\AMD)\leq 1\bigg\}~~,
\end{aligned}
\end{equation}
where the mutual inclination $\i_{mut}=\i_2+\i_3$ for fixed $\AMD$, or angular momentum $L_z=\Lambda_2+\Lambda_3-\AMD\,$, is given by 
\begin{equation}\label{cosimutua}
\cos(\i_{mut})=
\frac{L_z^2-\Lambda_2^2-\Lambda_3^2+\Lambda_2^2\,\e_2^2+\Lambda_3^2\,\e_3^2}{2\Lambda_2\Lambda_3\sqrt{1-\e_2^2}\sqrt{1-\e_3^2}}
=
\frac{L_z^2-\Lambda_2^2\left(1-{X_2^2+Y_2^2\over 2L_2}\right)^2
-\Lambda_3^2\left(1-{X_3^2+Y_3^2\over 2L_3}\right)^2}
{2\Lambda_2\Lambda_3\left(1-{X_2^2+Y_2^2\over 2L_2}\right)\left(1-{X_3^2+Y_3^2\over 2L_3}\right)}
\end{equation}
and the maximum possible mutual inclination consistent with the given $\AMD$ is
\begin{equation}\label{imutuamax}
\i_{max}=
\cos^{-1}\left(\frac{L_z^2-\Lambda_2^2-\Lambda_3^2}{2\Lambda_2\Lambda_3}\right)~~.
\end{equation}
The phase portrait corresponding to the Poincar\'e surface of section at a fixed level of energy $\Escr$ is obtained numerically, by choosing several initial conditions $(X_2,Y_2)\in\mathcal{D}(\Escr)\subset\mathbb{R}^2$, where $\mathcal{D}(\Escr)$ is the domain of permissible initial conditions consistent with the definition of the surface of section as in Eq.(\ref{pcsec}). For each initial condition, we then iterate the corresponding orbit under the Hamiltonian $\Hscr_{sec}$, and plot the consequent points $(X_2,Y_2)$, or, equivalently $(\e_2\cos\omega_2,\e_2\sin\omega_2)$, with 
\begin{equation}\label{xy2e2ome}
\e_2=\left(1-(1-(X_2^2+Y_2^2)/(2\Lambda_2))^2\right)^{1/2},~~~\omega_2=\mathrm{sgn}\left(\frac{-Y_2}{\sqrt{X_2^2+Y_2^2}}\right)\arccos\left(\frac{-X_2}{\sqrt{X_2^2+Y_2^2}}\right)
\end{equation}
every time when the orbit intersects the surface of section. Note that the conditions $Y_3=0,\dot{Y}_3\geq 0$ imply i) that the Poincar\'e mapping defined by the successive iterates is symplectic, and ii) that the section definition corresponds physically to instants when the orbit of the outer planet crosses the pericenter $\omega_3=\pi$. The symplecticity of the Poincar\'e mapping, along with the straightforward physical interpretation, are the main motives for the choice of variables and for the visualization of phase portraits via the definition of the Poincar\'e section as above. 

The domain $\mathcal{D}(\Escr)$ is non-null in a range of energies $\Escr_{min}\leq\Escr<0$. The energy $\Escr_{min}$ is computed as follows: consider the surface $\mathcal{I}_0$ of all possible points satisfying the section condition $Y_3=0$ as well as the lowermost limit of possible mutual inclination $\i_{mut}=0$. The surface  $\mathcal{I}_0$ is the sphere given by
\begin{equation}\label{imut0}
\mathcal{I}_0=\bigg\{X_3^2+X_2^2+Y_2^2=2\AMD\bigg\}~~.
\end{equation}
We then find the point $(X_{2,0},Y_{2,0})$ on the surface $\mathcal{I}_0$ where the energy is minimum by looking to the solutions of the system of equations
\begin{eqnarray}\label{eqenmin}
~&~&
{\partial\Hscr_{sec}(X_2,Y_2,X_3^2=2\AMD-X_2^2-Y_2^2,Y_3=0; \AMD)\over\partial X_2}=0 \\
~&~&
{\partial\Hscr_{sec}(X_2,Y_2,X_3^2=2\AMD-X_2^2-Y_2^2,Y_3=0; \AMD)\over\partial Y_2}=0~,\nonumber
\end{eqnarray}
satisfying also the section condition
\begin{equation}\label{y3dotsec}
\dot{Y}_3=\left({\partial\Hscr_{sec}(X_2,Y_2,X_3,Y_3=0;\AMD)\over\partial X_3}\right)_{X_3^2=2\AMD-X_2^2-Y_2^2}\geq 0~~. 
\end{equation}
We find two solutions $(X_{2,0},Y_{2,0}=0)\,$ and $(X_{2,1},Y_{2,1}=0)$ of the system of equations (\ref{eqenmin}) and (\ref{y3dotsec}); however, only $(X_{2,0},Y_{2,0}=0)\,$ corresponds to the minimum value of the energy, i.e. $\Hscr(X_2=X_{2,0}, Y_2=Y_{2,0}=0, X_3= X_{3,0}=\sqrt{2\AMD - X_{2,0}^2}, Y_3=0 )=\Escr_{min}$, as verified by the Hessian matrix 
$$
\begin{pmatrix}
&\displaystyle{\frac{\partial^2 \Hscr_{sec}(X_3=\sqrt{2\AMD - X_{2}^2},Y_3=0)}{\partial X_2^2}}
& \displaystyle{\frac{\partial^2 \Hscr_{sec}(X_3=\sqrt{2\AMD - X_{2}^2},Y_3=0)}{\partial Y_2\partial X_2}}
\\
&\displaystyle{\frac{\partial^2 \Hscr_{sec}(X_3=\sqrt{2\AMD - X_{2}^2},Y_3=0)}{\partial X_2\partial Y_2}} 
& \displaystyle{\frac{\partial^2 \Hscr_{sec}(X_3=\sqrt{2\AMD - X_{2}^2},Y_3=0)}{\partial Y_2^2}}
\end{pmatrix}_{(X_2=X_{2,0}, Y_2=Y_{2,0})}
$$
which is positive definite. On the other hand, the solution $(X_{2,1},\,Y_{2,1}=0,\,X_{3,1}=\sqrt{2\AMD - X_{2,1}^2})$ corresponds to the energy $\Escr_{2,3}$, and yield a negative definite Hessian matrix. Thus, $\Escr_{2,3}$ corresponds to the maximum energy for which we can have co-planar orbits (see discussion below).  

The uniqueness of the solution realizing the minimal value of the energy can be understood by the following argument: for every fixed value of the energy $\Escr$, sufficiently close to the origin the restriction of the manifold of constant energy to the section $Y_3=0$
\begin{equation}\label{manifene}
\mathcal{M}(\Escr)=\bigg\{(X_2,Y_2,X_3)\in\mathbb{R}^3: \Hscr_{sec}(X_2,X_3,Y_2,Y_3=0;\AMD)=\Escr\bigg\} 
\end{equation}
yields a surface. We will show in the next section that, for energies smaller than a suitably defined threshold,  the surface $\mathcal{M}(\Escr)$ is close to the surface of an integrable model
\begin{equation}\label{manifeneint}
\mathcal{M}_{int}(\Escr)=\bigg\{(X_2,Y_2,X_3)\in\mathbb{R}^3: \Hscr_{int}(X_2,X_3,Y_2,Y_3=0;\AMD)=\Escr\bigg\} 
\end{equation}
where $\Hscr_{int}$ contains $\Hscr_{planar}$ as well as a part of $\Hscr_{space}$. The manifolds $\mathcal{M}_{int}(\Escr)$, in turn, are ellipsoidal-like closed convex surfaces. The ellipsoidal form follows from the fact that the leading order terms of the restriction of $\Hscr_{int}$ to the section $Y_3=0$ yield a negative-definite quadratic form
$$
\Hscr_{int}(X_2,X_3,Y_2,Y_3=0)=-\nu_2\left({X_2^2+Y_2^2\over 2}\right)-\nu_3{X_3^2\over 2}+c_{23}X_2X_3+\ldots
$$
with $\nu_2,\nu_3>0$, $|c_{2,3}|<\min(\nu_2,\nu_3)$. 

\begin{figure}[h!]
\begin{center}
\includegraphics[scale=.7]{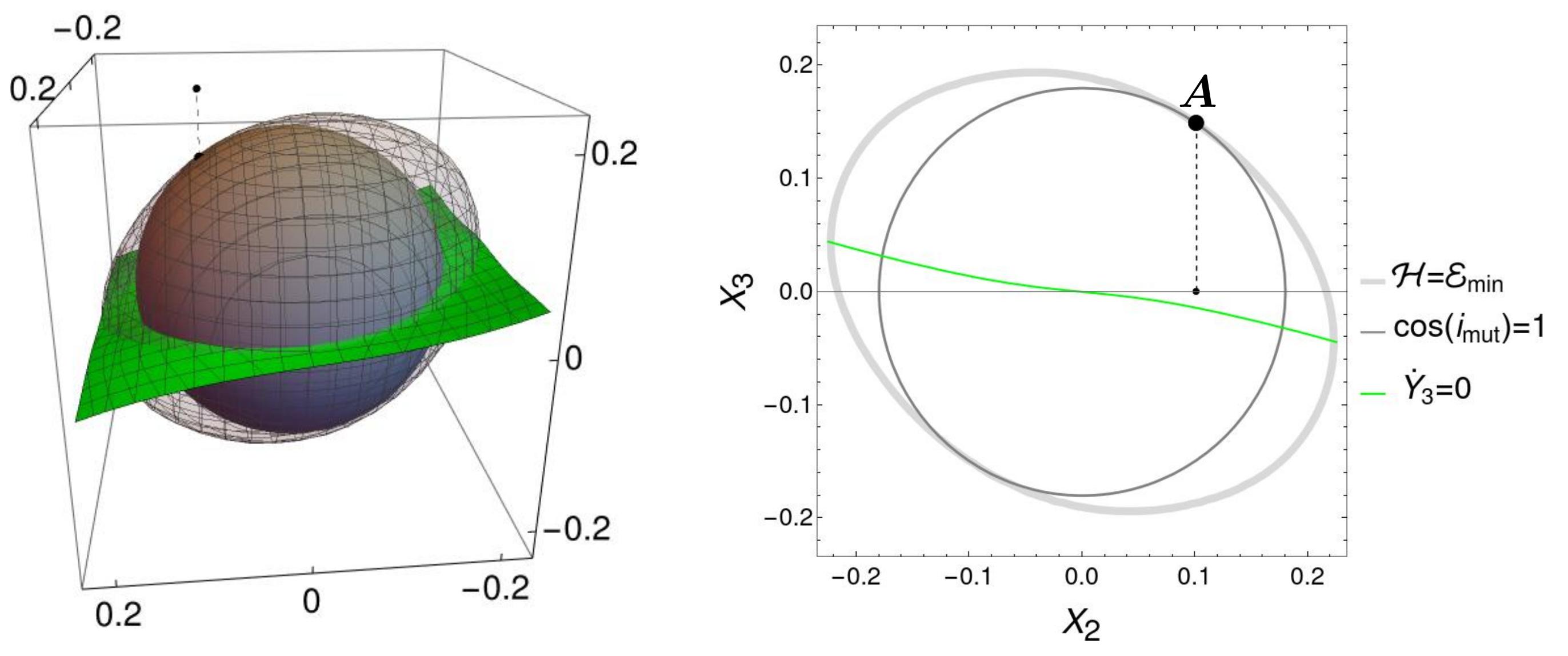}
\caption{Left: tangency between the ellipsoidal surfaces of minimum possible energy $\mathcal{M}(\Escr_{min})$, with $\Escr_{min}=-1.18237\cdot 10^{-4}$ (light gray) and the sphere $\mathcal{I}_0$ (gray). The only permissible initial condition on the surface of section corresponds to the point of tangency (black dot). The green slightly inclined surface  represents the condition $\dot{Y}_3=0$. Right: the tangency between the surfaces $\mathcal{M}(\Escr_{min})$  (thick light gray closed curves) and the sphere $\mathcal{I}_0$ (thin gray curve) as seen in the plane $(X_2,X_3)$ for $Y_2=0$. The green slightly inclined curve represents the condition $\dot{Y}_3=0$ and separates the plane $(X_2,X_3)$ in an upper domain, where $\dot{Y}_3>0$, and a lower domain ($\dot{Y}_3<0$). The point of tangency (thick dot) marks a fixed point corresponding to a planar periodic orbit of the apsidal corotation type called `mode A' (anti-aligned) in section \ref{sec:dynamics}, yielding $(X_2 = X_{2,0}=0.101237$, $X_3 = X_{3,0}=\sqrt{2\AMD-X_{2,0}^2}=0.148862$, $Y_2 = Y_{2,0}=0$, $Y_3=0\,)$, that correspond to $\e_2=0.337415$, $\e_3=0.433811$.}
\label{Fig.ellisphere.Enmin}
\end{center}
\end{figure}

Owing to their proximity to the manifolds $\mathcal{M}_{int}(\Escr)$, for energies low enough, the manifolds $\mathcal{M}(\Escr)$ are also ellipsoidal-like (the exact evolution of the form of $\mathcal{M}(\Escr)$ as the energy increases will be discussed in the next section). As a consequence, there is a negative value of the energy $\Escr=\Escr_{min}$ such that for $\Escr<\Escr_{min}$ the ellipsoidal manifold $\mathcal{M}(\Escr)$ surrounds the sphere $\mathcal{I}_0$, thus leading to the unphysical condition $\cos(\i_{mut})>1$. At the energy $\Escr=\Escr_{min}$ the ellipsoidal manifold $\mathcal{M}(\Escr_{min})$ has two points of tangency with the sphere $\mathcal{I}_0$ (see Fig.\ref{Fig.ellisphere.Enmin}). Both tangencies occur at the plane $Y_2=0$, but only one of them satisfies the condition $\dot{Y}_3\geq 0$. Then, we set $(X_{2,0},Y_{2,0}=0,X_{3,0}^2=2\AMD-X_{2,0}^2\,)$ equal to the coordinates of the corresponding point of tangency, and $\Escr_{min}=\Hscr_{sec}(X_2=X_{2,0},X_3=X_{3,0},Y_2=0,Y_3=0;\AMD)$. 
%--------------------------------------------------------------------------------
%----------------------------------------------------------------------------------------------------------

\begin{figure}[h!]
\begin{center}
\includegraphics[scale=.7]{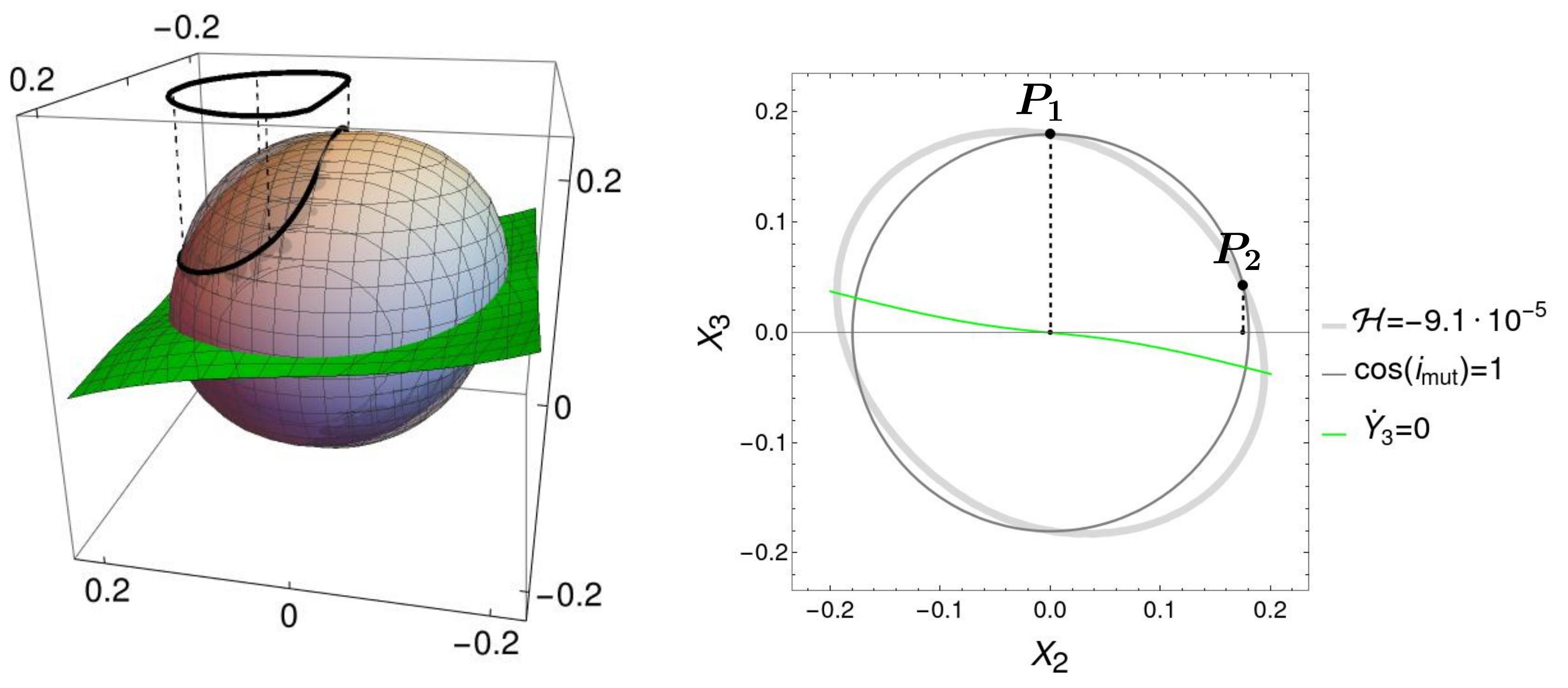}
\caption{Left: intersections between the surfaces $\mathcal{M}(\Escr)$ at the energy  $\Escr=-9.16\cdot 10^{-5}$ (light gray), the sphere $\mathcal{I}_0$ (thin gray), and the surface $\dot{Y}_3=0$ (green slightly inclined). The thick black curve projected on the plane $(X_2,Y_2)$ defines the permissible domain of initial conditions (see text). Right: the intersections between the surface $\mathcal{M}(\Escr)$ (thick light gray closed curve), the sphere $\mathcal{I}_0$ (thin gray curve and the condition $\dot{Y}_3=0$) (green slightly inclined curve) as viewed in the plane $(X_2,X_3)$ for $Y_2=0$.  The points of intersection mark the limits of possible initial conditions along $X_2$ for $X_3=Y_2=0$.}
\label{Fig.ellisphere.Enpanel2}
\end{center}
\end{figure}

In order to numerically specify, now, the limits of the domain $\mathcal{D}(\Escr)$ on the Poincar\'e section for any value of the energy in the range $\Escr_{min}<\Escr<0$ we work as follows: fixing any value of the angle $w_2=-\omega_2$ in the interval $0\leq w_2\leq\pi$, the line defined parametrically by the relations:
\begin{equation}\label{linexy2s2}
\mathcal{L}(w_2):\{X_2=s_2\cos(w_2),~Y_2=s_2\sin(w_2),s_2\in\mathbb{R}\}
\end{equation}
defines a plane $\mathcal{PL}(w_2):\{(X_2,Y_2)\in\mathcal{L}(w_2),X_3\in\mathbb{R}\}.$  The plane $\mathcal{PL}(w_2)$ intersects the ellipsoidal manifold $\mathcal{M}(\Escr)$ at a nearly-elliptic closed curve $\mathcal{C}_{PLM}(\Escr,w_2)$, while it intersects the sphere $\mathcal{I}_0$ at the circle $\mathcal{C}_{PLI_0}: s^2+X_3^2=2\AMD$. The curve $\mathcal{C}_{PLM}(\Escr,w_2)$ has central symmetry, and, owing to the fact that its quadratic-form approximation is negative definite, its overall size decreases as $\Escr$ increases from the most negative possible value $\Escr=\Escr_{min}$ towards the value $\Escr=0$, at which $\mathcal{C}_{PLM}(\Escr,w_2)$ reduces to a point at the origin of the plane $\mathcal{PL}(w_2)$. Finally, the surface $\dot{Y}_3=0$ also intersects the plane $\mathcal{PL}(w_2)$ at a curve $\mathcal{C}_{PL\dot{Y}_3=0}(w_2)$. As a result, fixing the value of $w_2$, there are three possibilities as regards the intersections of the curve $\mathcal{C}_{PLM}(\Escr,w_2)$, which varies with the energy, and the curves $\mathcal{C}_{PL\dot{Y}_3=0}(w_2)$ and $\mathcal{C}_{PLI_0}$:

\textit{Regime 1}: the curve $\mathcal{C}_{PLM}(\Escr,w_2)$ intersects the circle $\mathcal{C}_{PLI_0}$ at four points, two of which ($P_1,P_2$) are above the curve $\mathcal{C}_{PL\dot{Y}_3=0}(w_2)$, hence corresponding to $\dot{Y}_3>0$, while it intersects the curve $\mathcal{C}_{PL\dot{Y}_3=0}(w_2)$ itself at two points which are \textit{exterior} to the circle $\mathcal{C}_{PLI_0}$ (Fig.\ref{Fig.ellisphere.Enpanel2}). In this case, the permissible initial conditions in the Poincar\'{e} surface of section are given by $X_2=s_2\cos(w_2)$, $Y_2=s_2\sin(w_2)$ with 
$s_2$ in the interval $s_{P_1}\leq s_2\leq s_{P_2}$, where $s_{P_1}$, $s_{P_2}$ are the values of $s_2$ to which project the points of intersection $P_1,P_2$. At both these points we have $\dot{Y}_3>0$.
%----------------------------------------------------------------------------------------------------------
%----------------------------------------------------------------------------------------------------------

\begin{figure}[h!]
\begin{center}
\includegraphics[scale=.7]{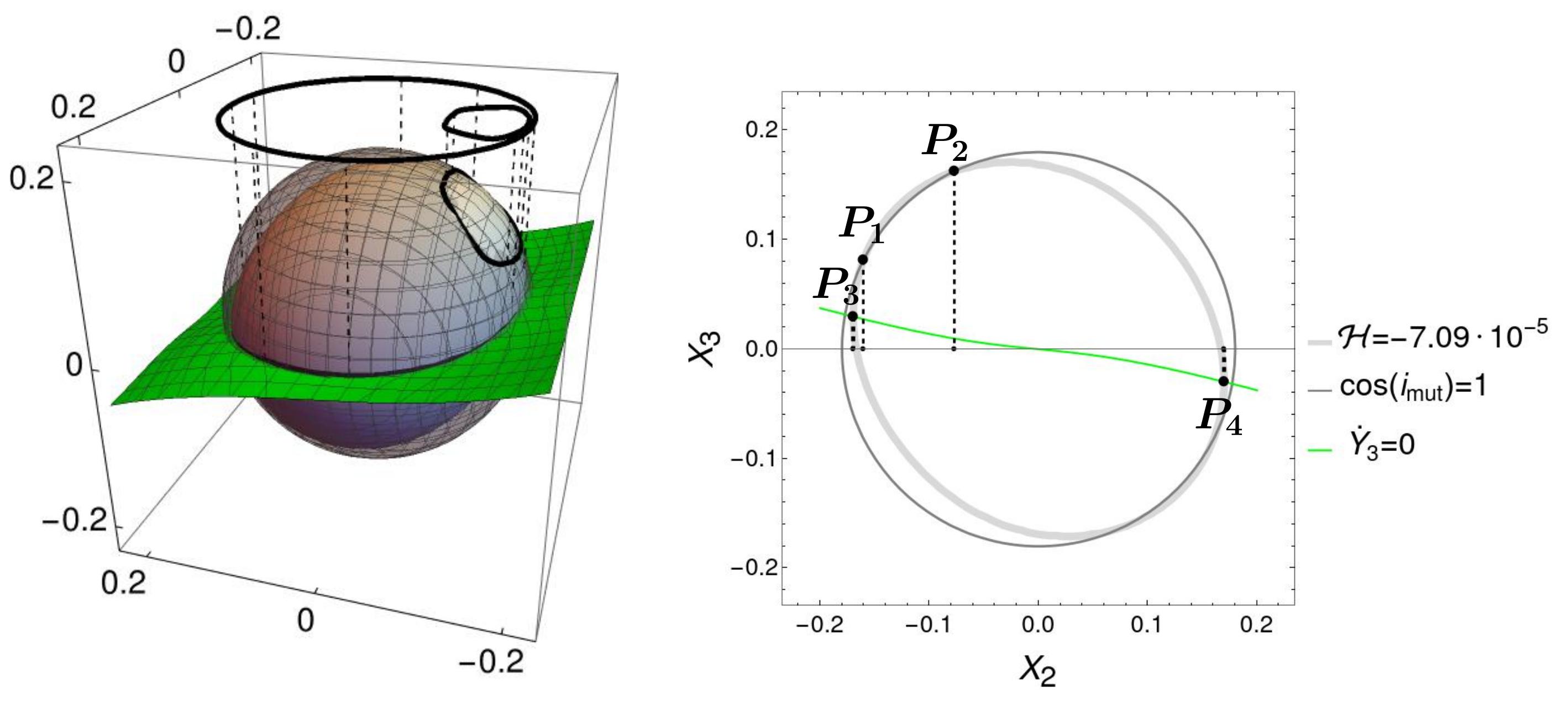}
\caption{Same as in Fig.\ref{Fig.ellisphere.Enpanel2} but for the surface $\mathcal{M}(\Escr)$ calculated at $\Escr=-7.09\cdot 10^{-5}$. The limits of possible initial conditions along $X_2$ for $X_3=Y_2=0$ are marked by the intesections $\mathcal{M}(\Escr)$ both with $\mathcal{I}_0$ and the curve $\mathcal{C}_{PL\dot{Y}_3=0}(w_2)$, for $w_2=0$ (see text).}
\label{Fig.ellisphere.Enpanel4}
\end{center}
\end{figure}

\textit{Regime 2}: for larger energies, the curve $\mathcal{C}_{PLM}(\Escr,w_2)$ still intersects the circle $\mathcal{C}_{PLI_0}$ at four points, two of which ($P_1,P_2$) are above the curve $\mathcal{C}_{PL\dot{Y}_3=0}(w_2)$, while now intersecting the curve $\mathcal{C}_{PL\dot{Y}_3=0}(w_2)$ itself at two points $(P_3,P_4)$, which are \textit{interior} to the circle $\mathcal{C}_{PLI_0}$ (Fig.\ref{Fig.ellisphere.Enpanel4}). In this case, the permissible initial conditions in the Poincar\'{e} surface of section are given by $X_2=s_2\cos(w_2)$, $Y_2=s_2\sin(w_2)$ with 
$s_2$ in one of the intervals $s_{P_3}\leq s_2\leq s_{P_1}$, or $s_{P_2}\leq s_2\leq s_{P_4}$, where $s_{P_1}$, $s_{P_2}$,  $s_{P_3}$, $s_{P_4}$ are the values of $s_2$ to which project the points of intersection $P_1$,$P_2$,$P_3$,$P_4$. In this case we have $\dot{Y}_3>0$ at the inner limits $P_1$, $P_2$, while we have $\dot{Y}_3=0$ (i.e. the orbit arrives tangently to the surface of section) at the outer limits $P_3$, $P_4$.
%----------------------------------------------------------------------------------------------------------
%----------------------------------------------------------------------------------------------------------

\begin{figure}[h!]
\begin{center}
\includegraphics[scale=.7]{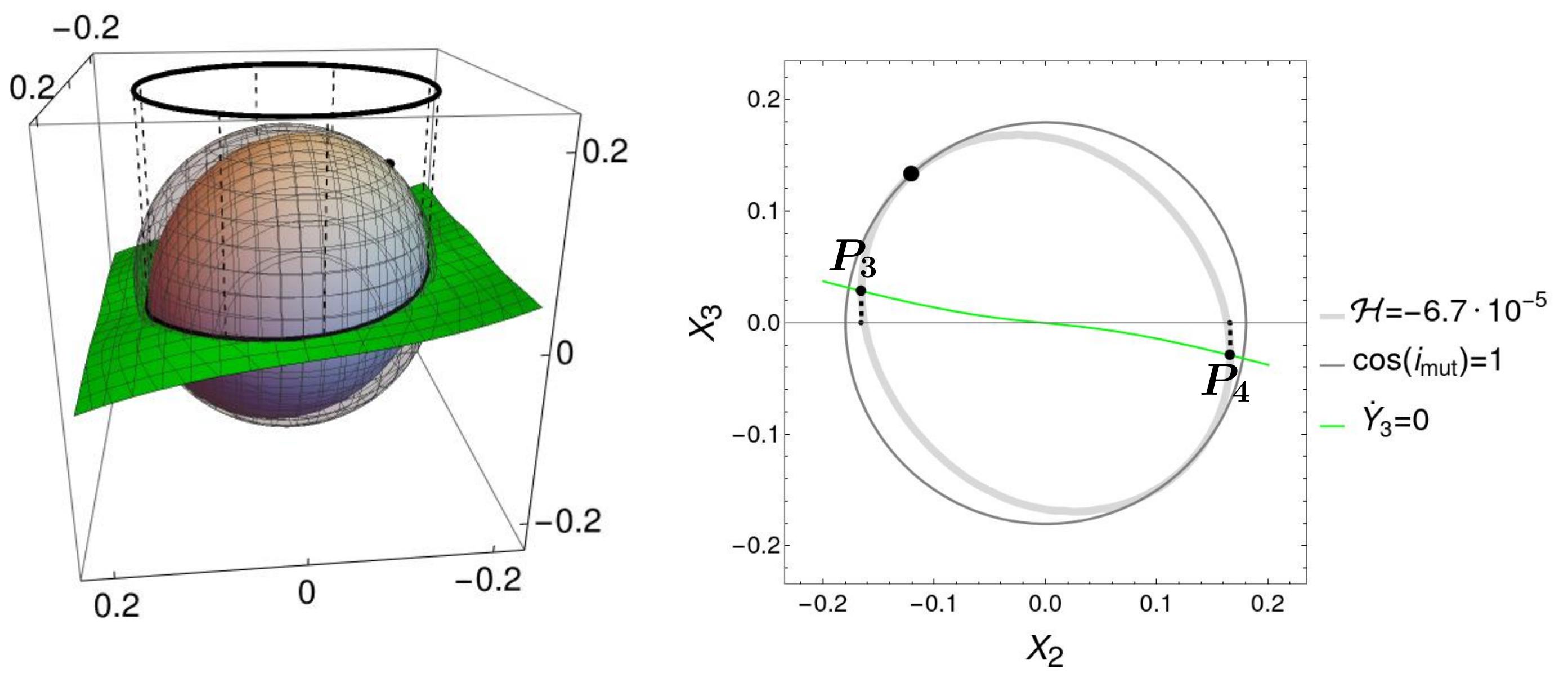}
\caption{Inner tangency between the surface $\mathcal{M}(\Escr_{2,3})$ at the energy  $\Escr_{2,3}=-6.77\times 10^{-5}$ and the sphere $\mathcal{I}_0$. The point of tangency marks a fixed point which corresponds to a planar periodic orbit of the type `mode B' (aligned apsidal corotation).  }
\label{Fig.ellisphere.Entg}
\end{center}
\end{figure}

\textit{Regime 3}: for still larger energies, the curve $\mathcal{C}_{PLM}(\Escr,w_2)$ decreases in size in such a way that it no longer intersects the circle $\mathcal{C}_{PLI_0}$. Then the only limits are posed by its intersections with the curve $\mathcal{C}_{PL\dot{Y}_3=0}(w_2)$ at the points $(P_3,P_4)$, which remain interior to the circle $\mathcal{C}_{PLI_0}$ (Fig.\ref{Fig.ellisphere.Entg}). Then, the permissible initial conditions in the Poincar\'{e} surface of section are given by $X_2=s_2\cos(w_2)$, $Y_2=s_2\sin(w_2)$ with 
$s_2$ in the interval $s_{P_3}\leq s_2\leq s_{P_4}$, and both limits correspond to $\dot{Y}_3=0$, i.e. to orbits  tangent to the surface of section.
%----------------------------------------------------------------------------------------------------------

Having fixed the value of the energy $\Escr$, and repeating the computation of the above intersection points for various lines $\mathcal{L}(w_2)$ (i.e. various choices of the argument of the perihelion $\omega_2$), we can explicitly compute the limits of the whole domain $\mathcal{D}(\Escr)$ and proceed in the computation of the phase portraits, obtaining several initial conditions within the domain $\mathcal{D}(\Escr)$. In practice, owing to the symmetries of the Hamiltonian, it is sufficient to consider only a scanning of initial conditions along the line $\mathcal{L}(w_2=0)$, i.e., $X_2=s_2$, $Y_2=0$. Note also that the regimes 1 and 2, as well as 2 and 3, are separated at the energies $\Escr_{1,2}$ and $\Escr_{2,3}$ respectively. We will see in the next section the topological differences in the phase portraits between the various regimes. We only note here that the generic regime is regime 3, which emerges beyond the energy $\Escr_{2,3}$, at which the manifold of constant energy $\mathcal{M}(\Escr_{2,3})$ has an inner tangency with the sphere $\mathcal{I}_0$. Finally, we note that both critical tangencies occuring at the energies $\Escr_{min}$ and $\Escr_{2,3}$ correspond to basic periodic orbits of the system (the apsidal corotation resonances), while the final limit $\Escr=0$ corresponds to the Kozai-Lidov fixed point of the system, at which $\i_{mut}=\i_{max}$ and $\e_2=\e_3=0$.     

%--------------------------------------------------------------
\subsubsection{Poincar\'e surface of section: precision tests}
\label{subsub:Poincare.precision} 
\begin{figure}[h!]
\begin{center}
\includegraphics[width=.99\textwidth]{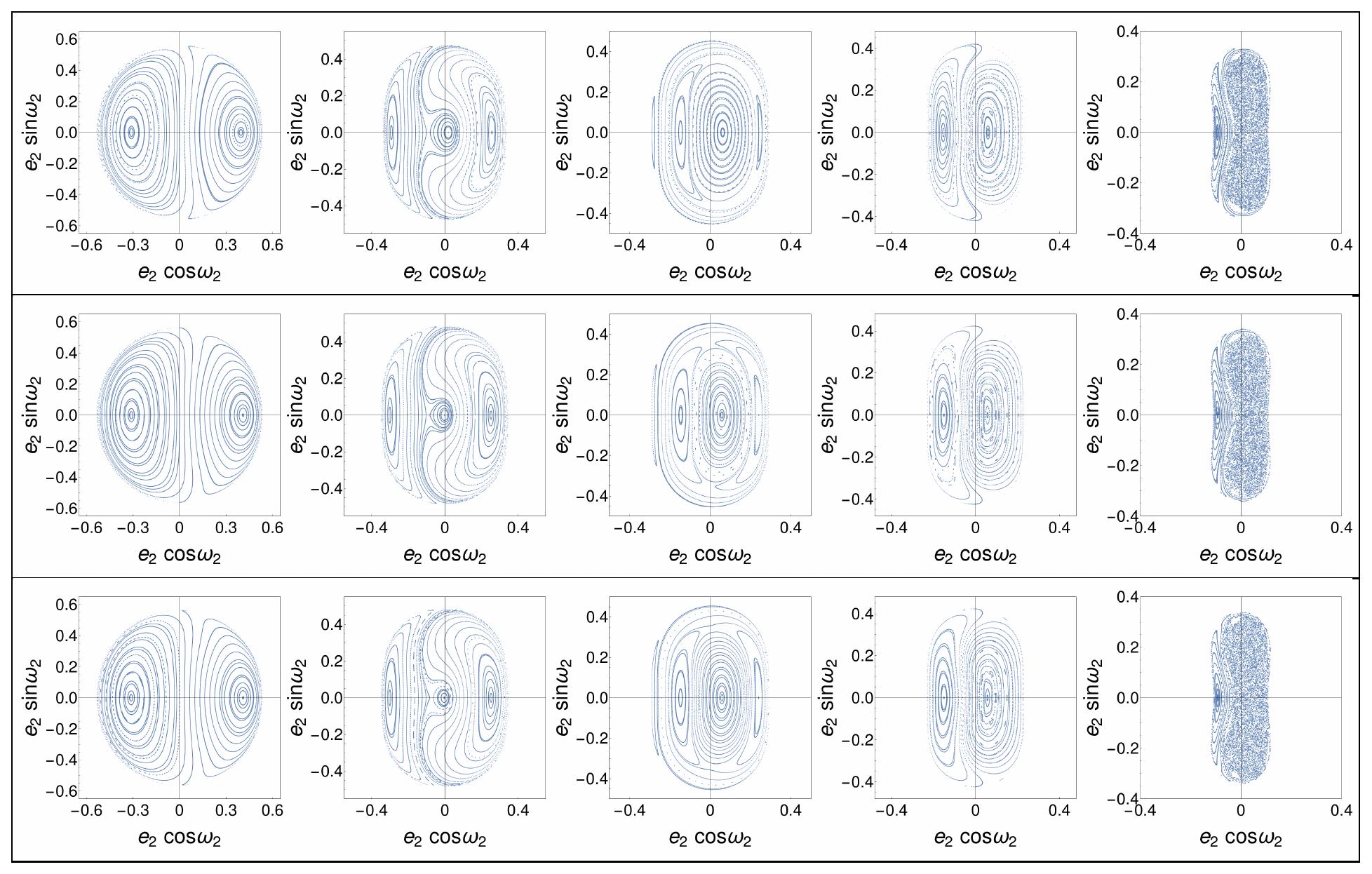}
\caption{Poincar\'{e} surfaces of section in the plane $(\e_2 \cos(\omega_2 ), \e_2 \sin(\omega_2 ))$ with $L_z$ fixed and different values of energy. The surfaces of section are computed by a numerical integration of trajectories in the Hamiltonian truncated at: \textit{Top} multipolar degree $N_{P}=5$, order $N_{bk}=10$ in the eccentricities, and energies (from left to right) $\Escr= -6.67\cdot 10^{-5},-2.53\cdot 10^{-5},-1.9\cdot 10^{-5},-1.17\cdot 10^{-5},-2.61\cdot 10^{-6}$. \textit{(Middle)} $N_{P}=6$, $N_{bk}=10$, and energies top $\Escr= -6.75\cdot 10^{-5}, -2.69\cdot 10^{-5}, -1.9\cdot 10^{-5},-1.16\cdot 10^{-5}, -2.61\cdot 10^{-6}$. \textit{(Bottom)} $N_{P}=6$, $N_{bk}=12$, and energies $\Escr= -6.75\cdot 10^{-5},-2.69\cdot 10^{-5},-1.9\cdot 10^{-5},-1.16\cdot 10^{-5},-2.59\cdot 10^{-6}$.}
\label{Fig.compara.sez1}
\end{center}
\end{figure}

%--------------------------------------------------------------
Figure~\ref{Fig.compara.sez1} shows a summary of our basic example of computation of phase portraits, in the form of Poincar\'e surfaces of section computed as explained in the previous subsection, referring to a 3D planetary system with mass, periods and $\AMD$ parameters as those estimated for the $\upsilon-$Andromedae system. We adopt the value $\mathcal{G}=4 \pi^2~AU^3/(yr^2 M_\odot)$ for Newton's gravity constant, as well as the mass parameters 
$m_0 = 1.31 M_\odot$, $m_2 = 13.98 M_J$, $m_3 = 10.25 M_J$ 
($M_\odot=$ mass of the Sun, $M_J=0.0009546 M_\odot=$ mass of Jupiter), 
and the semi-major axes $a_2 = 0.829$~AU, $a_3= 2.53$~AU (according to Table 13 of \cite{mcaetal2010}). 
For what concerns the constants $L_z$ or $\AMD$, we estimate their values adopting for the two planets' eccentricities, inclinations and arguments of periastron and of the nodes the values proposed in Table 1 of \cite{deietal2015}, namely, 
$\e_2 = 0.2445$, $\e_3 = 0.316$, $\i_2 = 11.347^\circ$, $\i_3 = 25.609^\circ$, $\omega_2 = 247.629^\circ$, $\omega_3 = 252.991^\circ$, $\Omega_2 = 248.181^\circ$, $\Omega_3 = 11.425^\circ$. As emphasized in the introduction, there are great uncertainties in the observation as regards, in particular, the estimates on the planetary eccentricity and inclination vectors. However, starting with parameters as above, we can have a representative value for the system's $\AMD$, which then serves to analyze the secular orbital dynamics under different assumptions for the planets' initial conditions. The value of the $\AMD$ is estimated from the above data, by first computing the angular momentum vector with respect to a heliocentric frame of reference whose x-axis points to the observer, then computing the rotation matrix connecting the observer's frame with Laplace's reference frame, and finally rotating the positions and velocities of both planets to the Laplace frame of reference and re-calculating their inclinations and arguments of periastron and of the nodes. This yields the values $\i_2 = 18.4748^\circ$, $\i_3 = 14.6462^\circ$, $\omega_2 = 289.049^\circ$, $\omega_3 = 235.464^\circ$, $\Omega_2 = 217.318^\circ$, $\Omega_3 = 37.3176^\circ$. Then, we obtain $Lz=0.183101~AU^2 M_\odot/yr$, which, together with $\Lambda_2=0.0873819$, $\Lambda_3=0.111923$, leads to $\AMD=L_2+L_3-L_z=0.0162044~AU^2 M_\odot/yr$.  
%--------------------------------------------------------------------------------
%--------------------------------------------------------------------------------

In Fig.\ref{Fig.compara.sez1} we already observe some of the significant changes in the phase portraits when the energy is varied from a value close to $\Escr_{min}\simeq-1.2\cdot 10^{-4}$ (slightly different in each of the models considered in the figure), to another close to $\Escr_{max}=0$. In the next section, we will return to a detailed discussion of the changes observed in the phase portraits as the energy moves from a value, in which the system is closer to the planar regime, to another, in which the system is highly inclined. Here, instead, we only discuss how these figures control the robustness of the phase portraits with respect to the model chosen, which can differ in the maximum multipole degree $N_{P}$ up to which the original Hamiltonian $\Hscr$ is expanded, before the averaging, as well as in the maximum order in book-keeping $N_{bk}$ of the Hamiltonian $\Hscr_{sec}$, which is also equal to the maximum order at which the eccentricities appear in $\Hscr_{sec}$. Checking with increasing values of $N_{P}$ and $N_{bk}$, we find that the phase portraits stabilize at $N_{P}=5$, $N_{bk}=10$. Beyond these values, the sequence of bifurcations of new fixed points and the corresponding changes in the topological features of the phase portraits become marginal, with only changes in the second significant figure observed as regards both the value of the energy $\Escr$ where a bifurcation occurs, and the position of the corresponding fixed points, separatrices, etc. 

Since we are interested only in a qualitative description of the structure and evolution of the phase portraits, for reducing computational time we choose as our basic model the one with $N_{P}=5$, $N_{bk}=10$. As an independent test, we perform a comparison between the phase portraits obtained with this model and those obtained by a completely independent Laplace-Lagrange expansion of the Hamiltonian $\Hscr$ averaged over the fast angles. In the latter case, the book-keeping process described in subsection \ref{sub:bookkeeping} must be altered for the decomposition $\Hscr=\Hscr_{planar}+\Hscr_{space}$ to naturally emerge while the Jacobi reduction is performed. Namely, we first introduce the book-keeping $\e_j\rightarrow\varepsilon \e_j$, $j=2,3$ and expand the direct term $-\mathcal{G}m_2m_3/|\mathbf{r}_2-\mathbf{r}_3|$ in powers of the orbital eccentricities up to the book-keeping order $N_{bk}=10$. The so-computed expression has the form:
\begin{align}\label{term.direct.part0}
 -{\mathcal{G}m_2m_3\over|\mathbf{r}_2-\mathbf{r}_3|}=
 \sum_{\alpha,\beta,\gamma}
 \frac{D_{\alpha,\beta,\gamma}(a_2,a_3,\e_2,\e_3,\i_2,\i_3;\varepsilon)}{(\Delta_{2,3})^{\frac{2\,s+1}{2}}} \cos(\alpha_2\lambda_2 + \alpha_3\lambda_3 + \beta_2\Omega_2 + \beta_3\Omega_3 + \gamma_2\varpi_2 + \gamma_3\varpi_3)\, ,
\end{align}
where the denominator $\Delta_{2,3}$, after the substitution $\Omega_3=\Omega_2+\pi\,$, takes the form:
\begin{equation}
\label{den.prima.riduz}
\begin{split}
\Delta_{2,3}=&a_2^2 + a_3^2 - a_2 a_3 \cos(\lambda_2 - \lambda_3) - a_2 a_3  \cos(\lambda_2 - \lambda_3)\cos(\i_2+\i_3)\\
+&a_2 a_3  \cos(\lambda_2 + \lambda_3 - 2 \Omega_2)\cos(\i_2+\i_3) - 
 a_2 a_3 \cos(\lambda_2 + \lambda_3 - 2 \Omega_2)\, ,
\end{split}\, 
\end{equation}
with $\varpi_j=\omega_j\,+\,\Omega_j$ and $\lambda_j=M_j\,+\,\varpi_j$ denoting the longitudes of the pericenter and the mean longitudes of the bodies $j=2,3$ respectively. The integer vectors $\alpha\equiv(\alpha_2,\alpha_3)$, $\beta\equiv(\beta_2,\beta_3)$, $\gamma\equiv(\gamma_2,\gamma_3)$ satisfy the D'Alembert rule $\alpha_{2}+\alpha_{3}+\beta_2+\beta_3+\gamma_2+\gamma_3=0\,$. The coefficients $D_{\alpha,\beta,\gamma}$ are polynomial in the book-keeping parameter $\varepsilon$. At this point, we introduce the book-keeping identities:
$$
\cos(i_2+i_3)=\eta\varepsilon^2(\cos(i_2+i_3)-1)+1\, ,
$$
i.e. (see Eq~\eqref{riduzione.cos.sen})
\begin{align*}
&\cos(\i_2)\cos(\i_3)=\varepsilon^2\eta\left(\cos(\i_2)\cos(\i_3)-1\right)+1\, , & 
&\sin(\i_2)\sin(\i_3)=\varepsilon^2\eta\sin(\i_2)\sin(\i_3)\, , 
\end{align*}
and expand again the expression~\eqref{term.direct.part0} with respect to the book-keeping parameter $\varepsilon$ up to the truncation order $N_{bk}=10$. This leads to an expression of the form 
\begin{align}\label{term.direct.part}
  -{\mathcal{G}m_2m_3\over|\mathbf{r}_2-\mathbf{r}_3|}=
 \frac{C(a_2,a_3,\e_2,\e_3,\i_2,\i_3;\varepsilon,\eta)}{\left(a_2^2 + a_3^2 - 2 \,a_2\,a_3\, \cos(\lambda_2 - \lambda_3)\right)^{\frac{2\,s+1}{2}}} \cos(\alpha_2\lambda_2 + \alpha_3\lambda_3 + (\beta_2+\beta_3)\Omega_2  + \gamma_2\varpi_2 + \gamma_3\varpi_3)\, ,
\end{align}
where the coefficients $C(a_2,a_3,\e_2,\e_3,\i_2,\i_3;\varepsilon,\eta)$ are polynomial in the book-keeping parameters $\varepsilon,\eta$. Finally, we perform the classical Laplace-Lagrange averaging `by scissors': 
the denominator of~\eqref{term.direct.part} is Fourier-expanded 
\begin{equation}
\label{espans.flam}
\frac{1}{\left(a_2^2 + a_3^2 - 2 \,a_2\,a_3\, \cos(\lambda_2 - \lambda_3)\right)^{\frac{2\,s+1}{2}}}=a_3^{-(2\,s+1)}\sum _{j\geq 0} b_{s+\frac{1}{2}}^{(j)}\left(\frac{a_2}{a_3}\right)\cos(j(\lambda_2-\lambda_3))\, ,
\end{equation}
where $b_{s+\frac{1}{2}}^{(j)}$ are Laplace coefficients, which can be computed numerically, as
\begin{equation}
\begin{split}
\label{Lapl.coeff}
&b_{s+\frac{1}{2}}^{(0)}\left(\alpha\right)=\frac{1}{2\pi}\int_{0}^{2\pi} \left(1+\alpha^2 - 2 \,\alpha\, \cos(\vartheta)\right)^{-(s+\frac{1}{2})}\,d\vartheta\, ,\\
&b_{s+\frac{1}{2}}^{(j)}\left(\alpha\right)=\frac{1}{\pi}\int_{0}^{2\pi} \left(1+\alpha^2 - 2 \,\alpha\, \cos(\vartheta)\right)^{-(s+\frac{1}{2})}\cos(j\vartheta)\,d\vartheta \qquad j\geq 1\, ,
\end{split}
\end{equation}
where $\alpha=a_2/a_3\,$, or via a multipolar expansion (see the Appendix~\ref{appendix:coeff Lapl}).
After performing the above expansions, the secular Hamiltonian is obtained by dropping from the original Hamiltonian all terms depending on the `fast' angles $\lambda_i$. This leads to a model in which the inclinations appear only through the combination $\cos(i_2+i_3)$. Switching, now, back to the angles $\omega_j,\Omega_j$, performing the Jacobi reduction as in subsection \ref{sub:bookkeeping} (i.e., using Eq.~\eqref{seni2i3}, collecting together all terms independent or depending on $\eta$, and restoring the numerical values of the book-keeping coefficients $\varepsilon=\eta=1\,$), we arrive at a `Laplace-Lagrange' model $\Hscr_{LL}=\Hscr_{LL,planar}+\Hscr_{LL,space}$ which has the same form as the model of Eq.~\eqref{Ham.Lapl.nostra2}. This is further processed with the introduction of Poincar\'{e} variables as in step 4 of subsection \ref{sub:bookkeeping}. 

%--------------------------------------------------------------------------------
\begin{figure}[h!]
\begin{center}
\includegraphics[width=.99\textwidth]{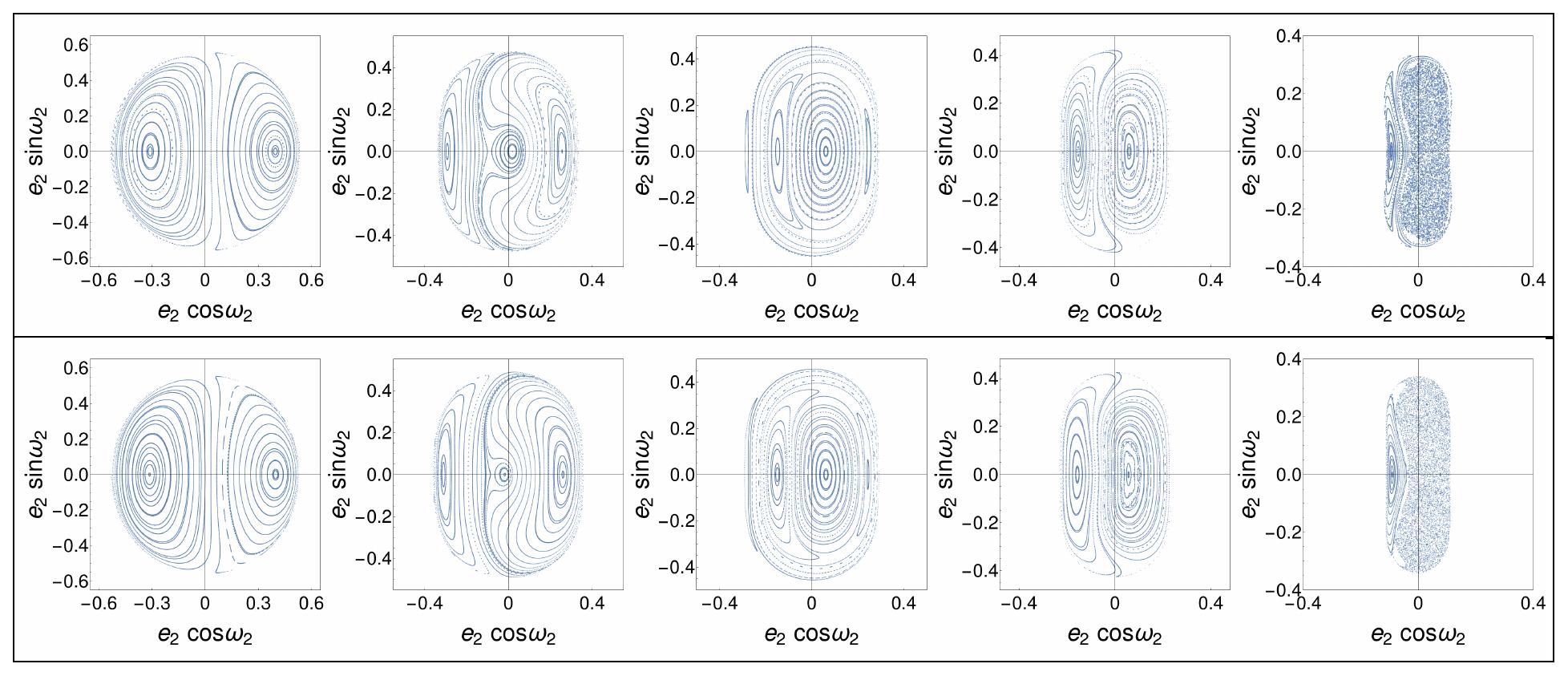}
\caption{Comparison of the Poincar\'{e} surfaces of section in the plane $(\e_2 \cos(\omega_2 ), \e_2 \sin(\omega_2 ))$ with $L_z$ fixed and different values of the energy, in two models: \textit{Top:} our basic model $N_{P}=5$, $N_{bk}=10$ (same as in the top row of Fig.\ref{Fig.compara.sez1}). \textit{Bottom:} the Laplace-Lagrange secular Hamiltonian model $\Hscr_{LL}$ (see text) truncated at order 10 in the eccentricities, with energies (from left to right) $\Escr= -6.62\cdot 10^{-5}, -2.94\cdot 10^{-5}, -1.92\cdot 10^{-5},-1.18\cdot 10^{-5}, -2.73\cdot 10^{-6}$. }
\label{Fig.compara.sez2}
\end{center}
\end{figure}
%--------------------------------------------------------------------------------
From Fig.\ref{Fig.compara.sez2} we conclude that similar remarks as those of Fig.\ref{Fig.compara.sez1} can be made as regards the comparison of the secular model $\Hscr_{sec}$ adopted in the present work, based on a multipolar expansion, and the model $\Hscr_{LL}$ obtained by the classical Laplace-Lagrange expansion in the eccentricities. We note, however, that the latter is much harder to compute, while it is obtained by using a series reversion of Kepler's equation which has a limited convergence. On the other hand, the model based on closed-form averaging of the multipole Hamiltonian expansion requires a special treatment as regards the computation of the underlying canonical transformation which eliminates (instead of `scissor cutting') the fast angles (see \cite{caveft2022}). Since in the present work we do not consider this transformation, we will hereafter deal with results derived only by use of the simple multipole and closed-form averaged model $\Hscr_{sec}$, with the truncation orders $N_{P}=5$, $N_{bk}=10$. 

%%%%%%%%%%%%%%%%%%%%%%%%%%%%%%%%%%%%%%%%%%%%%%%%%%%%%%%%%%%
\section{Dynamics}
\label{sec:dynamics}
%%%%%%%%%%%%%%%%%%%%%%%%%%%%%%%%%%%%%%%%%%%%%%%%%%%%%%%%%%%
%----------------------------
\subsection{General}
\label{subsec:general}
%----------------------------
In this section we are interested in analyzing the most important phenomena observed in the phase portraits of the Hamiltonian $\Hscr_{sec}(X_2, X_3, Y_2, Y_3; \AMD)$ computed with the basic reference model corresponding to the truncation orders $N_{P}=5$, $N_{bk}=10$. In the following subsections we will first present the general picture of the transitions taking place in the structure of the phase space as the energy increases in the range $\Escr_{min}\leq\Escr\leq 0$. Such transitions are caused, for example, by changes in the nature (e.g. stability) of the main equilibria of the system, giving birth to new families of periodic orbits which connect the families dominant in the planar-like regime (high eccentricities, low mutual inclination) with those of the highly inclined regime (low eccentricities, high mutual inclination). 

\begin{figure}[h!]
\begin{center}
\includegraphics[width=.7\textwidth]{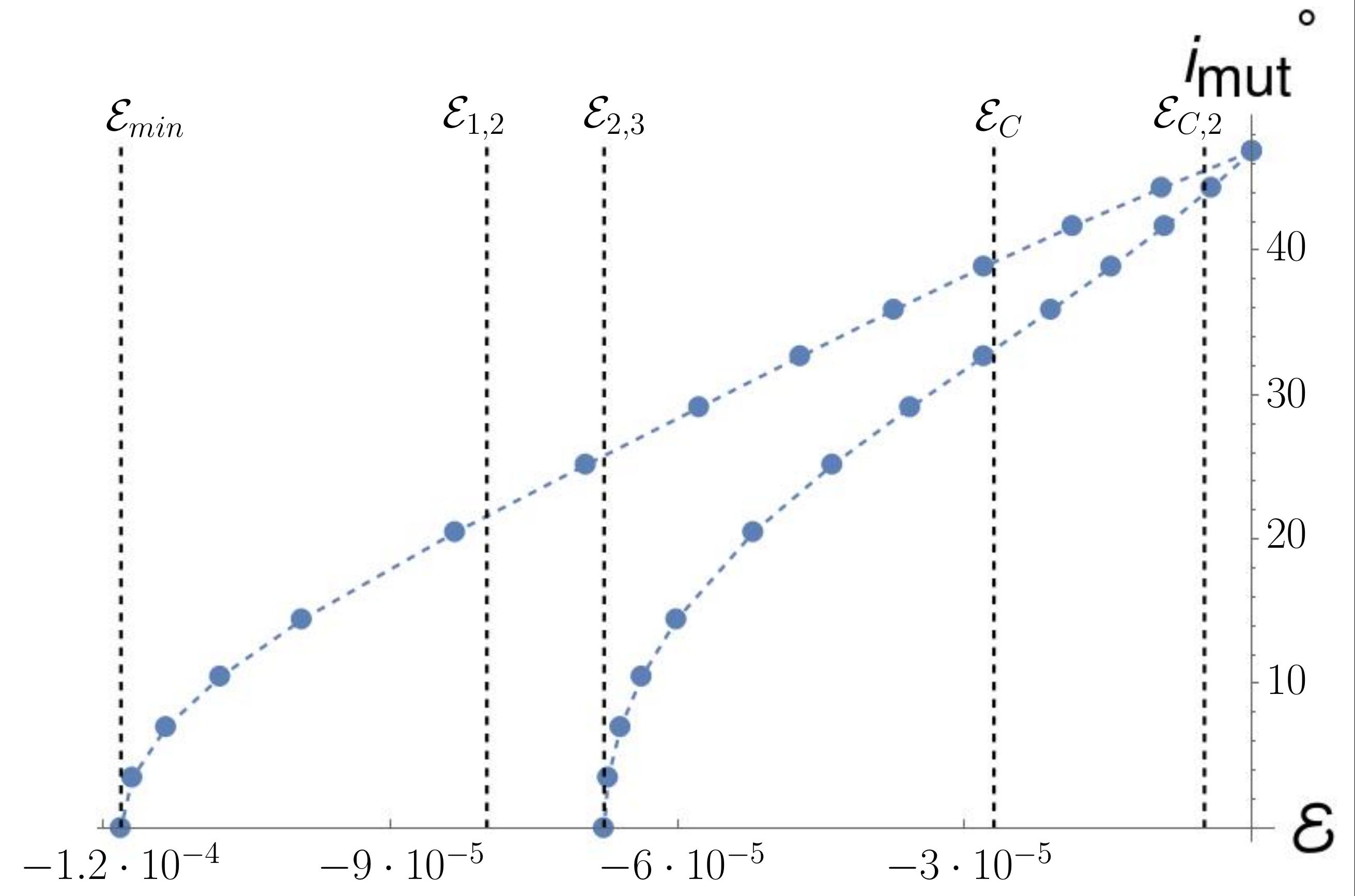}
\caption{The minimum and maximum possible values of the mutual inclination $\i_{mut}^{min}$, $\i_{mut}^{max}$, as a function of the energy $\Escr$. The vertical lines correspond to the energies $\Escr_{min}=-1.18237\cdot 10^{-4}$ (birth of the mode A - anti-aligned apsidal corotation), $\Escr_{1,2}=-8\cdot 10^{-5}$ (completion of the libration domain around the mode A), $\Escr_{2,3}=-6.77\cdot 10^{-5}$ (birth of the mode B - aligned apsidal corotation), $\Escr_C=-2.7\times 10^{-5}$ (bifurcation of the inclined Kozai-Lidov periodic orbits $C_1,C_2\,$), and $\Escr_{C,2}=-5.0\cdot 10^{-6}$ (inclined orbit $C_2$ becomes unstable, see text).}
\label{Fig.incmutene}
\end{center}
\end{figure}

In the analysis of phase portraits as in the sequel, we parametrize all transitions due to bifurcations of new periodic orbits using the fixed (in the Poincar\'e section) energy $\Escr$ as the parameter. All energies referred to below are given in units of $M_\odot AU^2/yr^2$. It is easy to see that a certain value of the energy $\Escr$ establishes a range of allowed mutual inclinations
\begin{equation}
\i_{mut}^{min}(\Escr)\leq \i_{mut}\leq \i_{mut}^{max}(\Escr)
\end{equation}
where both $\i_{mut}^{min}(\Escr)$ and $\i_{mut}^{max}(\Escr)$ are increasing functions of $\Escr$, as shown in Fig.\ref{Fig.incmutene}. 
%--------------------------------------------------------------------------------
%--------------------------------------------------------------------------------

%--------------------------------------------------------------------------------
\begin{figure}[h!]
\begin{center}
\includegraphics[width=.75\textwidth]{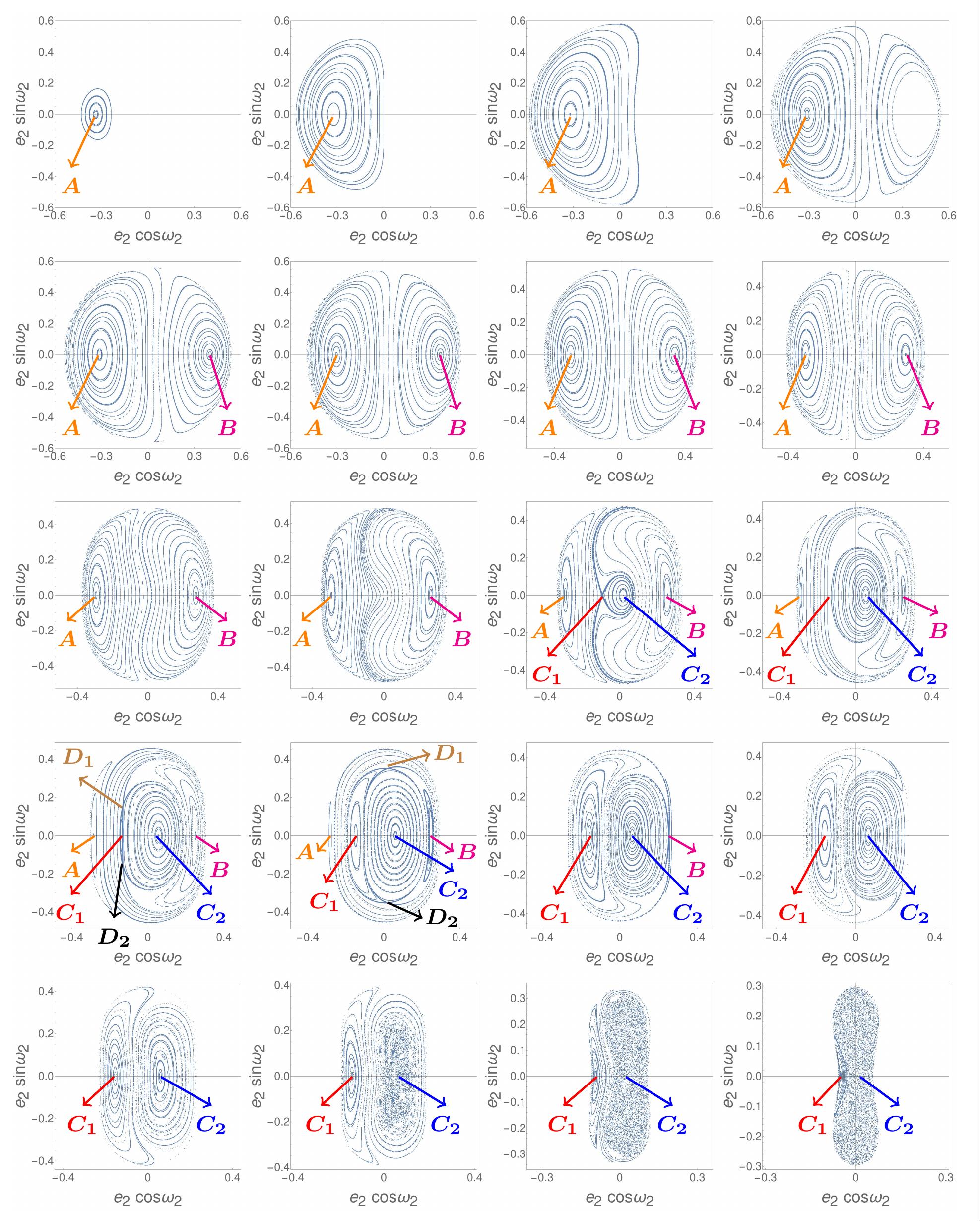}
\caption{Phase portraits (Poincar\'{e} surfaces of section in the plane $(\e_2 \cos(\omega_2 ), \e_2 \sin(\omega_2 ))$) in the basic model $N_{P}=5$, $N_{bk}=10$, and for the energies (from from top left to bottom right) $\Escr= -1.145\cdot 10^{-4}$, $-9.16\cdot 10^{-5}$, $-8.23\cdot 10^{-5}$, $-7.09\cdot 10^{-5}$, 
$-6.77\cdot 10^{-5}, -5.72\cdot 10^{-5}, -4.93\cdot 10^{-5},-3.81\cdot 10^{-5},-3.19\cdot 10^{-5},-2.76\cdot 10^{-5},-2.53\cdot 10^{-5},-2.16\cdot 10^{-5}, -2.08\cdot 10^{-5}, -1.9\cdot 10^{-5}, -1.58\cdot 10^{-5}, -1.53\cdot 10^{-5}, -1.17\cdot 10^{-5}, -7.69\cdot 10^{-6}, -2.61\cdot 10^{-6}, -7.39\cdot 10^{-7}$. The positions of the fixed points corresponding to the periodic orbits of the modes $A$, $B$, $C_1$, $C_2$, $D_1$, $D_2$ (see text) are marked by arrows.}
\label{Fig.sezioni}
\end{center}
\end{figure}
%--------------------------------------------------------------------------------
The minimum and maximum possible values of the mutual inclination $\i_{mut}^{min}$, $\i_{mut}^{max}$, as a function of the energy $\Escr$ are computed as follows: consider the ellipsoidal surface $\mathcal{I}_{\i_{mut}}$ defined by setting $Y_3=0$ in Eq.(\ref{cosimutua}) for a certain value of $\i_{mut}$. Consider the critical energy $\Escr=\Escr_{2,3}$, equal to $\Escr_{2,3}=-6.77\times 10^{-5}$ in our example. For energies $\Escr\geq\Escr_{2,3}$, we compute the values $\i_{mut}=\i_{mut}^{min}(\Escr)$ and $\i_{mut}=\i_{mut}^{max}(\Escr)$ for which the ellipsoid $\mathcal{I}_{\i_{mut}}$ comes tangent to the ellipsoidal manifold of constant energy $\mathcal{M}(\Escr)$, with $\mathcal{I}_{\i_{mut}}$ being at the interior of $\mathcal{M}(\Escr)$ (for $\i_{mut}=\i_{mut}^{max}$) or at the exterior of $\mathcal{M}(\Escr)$ (for $\i_{mut}=\i_{mut}^{min}$). For energies $\Escr_{min}\leq\Escr<\Escr_{2,3}$ (with $\Escr_{min}=-1.182\times 10^{-4}$ in our example), the condition of tangency of $\mathcal{I}_{\i_{mut}}$ at the exterior of $\mathcal{M}(\Escr)$ leads to unphysical values $\cos(\i_{mut})>1$. Thus, in this interval of energies we have $\i_{mut}^{min}=0$, while $\i_{mut}^{max}$ keeps being represented by an increasing function of $\Escr$ (Fig.\ref{Fig.incmutene}).   

The vertical lines in Fig. \ref{Fig.incmutene} indicate values of the energy where important changes take place in the structure of the phase portraits due to the birth, or change of stability character, of some of the most important families of periodic orbits of the system. The most important transitions taking place in the structure of the phase portraits are shown in Fig.\ref{Fig.sezioni}, whose details will be presented in subsequent subsections. Increasing the energy $\Escr$, these transitions appear in summary by the following sequence:\\

i) At the energy $\Escr=\Escr_{min}$, the available domain $\mathcal{D}(\Escr)$ reduces to a point, corresponding to the point of tangency of Fig.\ref{Fig.ellisphere.Enmin}. This is a fixed point of the Poincar\'e map, whose associated orbit yields two coplanar ellipses with anti-aligned pericenters precessing by the same frequency, known as the \textit{apsidal corotation orbit} (see \cite{lauetal2002},~\cite{leepea2003},~\cite{beaetal2003}). For energies $\Escr>\Escr_{min}$, the above fixed point is continued by a family of periodic orbits, called below the \textit{mode A}. These correspond physically to inclined planetary orbits whose eccentricities undergo small periodic oscillations around some non-zero constant values $\e_{2,A}$, $\e_{3,A}$ (functions of the energy), while the arguments of perihelia undergo small periodic oscillations around the fixed relation $\omega_2-\omega_3=0$ (see subsections~\ref{subsub:normalforms} and~\ref{sub:bifurcation}). For energies $\Escr_{min}<\Escr\leq\Escr_{2,3}$ (equal to $-6.77\cdot 10^{-5}$ in our numerical example), the mode A, which generalizes the anti-aligned apsidal corotation family to the non-planar case, is the unique important stable family in the surface of section (see first three panels of top row of Fig.\ref{Fig.sezioni}). The corresponding fixed point is surrounded by closed invariant curves, which represent orbits performing quasi-periodic oscillations around the configuration of anti-aligned perihelia. Up to the energy $\Escr_{1,2}$ (equal to about $-8\cdot 10^{-5}$ in our example), which marks the transition from two to four limits of permissible motion as in Figures~\ref{Fig.ellisphere.Enpanel2} and~\ref{Fig.ellisphere.Enpanel4}, only quasi-periodic orbits around the A mode exist. On the other hand for energies $\Escr_{1,2}<\Escr<\Escr_{2,3}$, we can also have trajectories with argument $\omega_3-\omega_2$ either circulating or librating around the value $\omega_3-\omega_2=\pi$ (alignment). As explained in detail in subsection \ref{sub:nearlyplanar}, the separation between the various librating or circulating regimes is not due to the presence of a dynamical separatrix, but can be explained by an integrable Hamiltonian model approximating $\Hscr_{sec}$ in the corresponding energy regime, whose phase space has the topology of a 3-sphere rather than the plane $\mathbb{R}^2$.\\

ii) For energies $\Escr_{1,2}<\Escr<\Escr_{2,3}$ there is a prohibited domain surrounding the center of the librating motions around $\omega_3-\omega_2=\pi$ (see fourth panel, top row of Fig.\ref{Fig.sezioni}). At the energy $\Escr=\Escr_{2,3}$ this domain shrinks to zero, and at the center of the librations appears a second fixed point of the Poincar\'e map, which is a periodic orbit physically corresponding to the planar aligned apsidal corotation orbit (see the first panel in the second row of Fig.~\ref{Fig.sezioni}). This also marks the inner point of tangency of the sphere $\mathcal{I}_0$ with the energy manifold $\mathcal{M}(\Escr_{2,3})$ (Fig.\ref{Fig.ellisphere.Entg}). As indicated in Fig.\ref{Fig.incmutene}, for energies $\Escr>\Escr_{2,3}$ there can no longer be any planar orbit intersecting the surface of section. However, similarly as for the mode A, the fixed point corresponding to the aligned apsidal corotation is continued as a family of off-plane periodic orbits, hereafter called the \textit{mode B}. Physically, such orbits undergo small periodic oscillations around some non-zero constant values $\e_{2,B}$, $\e_{3,B}$ (also being both functions of the energy), while the arguments of perihelia undergo small periodic oscillations around the fixed relation $\omega_2-\omega_3=\pi$ (see also subsections~\ref{subsub:normalforms},~\ref{sub:bifurcation}). The mode B also is surrounded by quasi-periodic orbits with arguments of the perihelia librating around the relation $\omega_3-\omega_2=\pi$.\\

iii) The topology of the phase space induced by the alteration between the domains of libration and circulation around the modes A and B dominates the picture obtained for the phase portraits in a large subinterval within the permissible range of values of the energy (up to about $\Escr=-3\cdot 10^{-5}$ in our example). As seen in Fig.\ref{Fig.incmutene}, this covers most cases of highly-inclined orbits, with mutual inclinations being in our example as high as $\sim 40^\circ$. However, at a critical energy $\Escr_{C}$ (equal to about $-2.7\cdot 10^{-5}   $ in our example), a saddle-node bifurcation takes place, giving rise to two new fixed points of the Poincar\'{e} map, corresponding to periodic orbits hereafter called the \textit{Kozai-Lidov} orbits $C_1$ and $C_2$. Physically, these are highly inclined orbits with planetary eccentricities $\e_2$ and $\e_3$ smaller than those of the modes A and B, tending actually to zero as $\Escr\rightarrow 0$. The detailed sequence of bifurcations related to these orbits is discussed in detail in subsection \ref{sub:Kozai} (see panels 11 to 20 of Fig.\ref{Fig.sezioni}). The most important transitions regard the orbit $C_1$, which becomes stable nearly immediately after its birth, while the orbit $C_2$ undergoes the classical Lidov-Kozai transition from stability to instability, accompanied by the appearance of chaotic motions around it. Such phenomena appear in a small range of energies near the limit $\Escr=0$, where the whole phase space shrinks again to a unique point at the origin of the surface of section, corresponding to two circular orbits having the maximum possible mutual inclination ($\i_{mut}^{max}(\Escr=0)\simeq 46^\circ\,$ in our example). In fact, Fig.\ref{Fig.incmutene} indicates that in this Lidov-Kozai regime, the range in possible inclinations for all orbits becomes narrow, being limited to values around the critical $\Escr_{C,2}$ where the periodic orbit $C_2$ underhgoes the Lidov-Kozai instability.\\

The above was a synoptic account of the main phenomena occuring in the phase space as the energy (and, hence, mutual inclination) of the orbits increases and the system gradually shifts from a regime where the dynamics is qualitatively similar to the one of the planar problem, to the Lidov-Kozai regime dominated by nearly circular and highly inclined orbits. In the next subsections we examine these phenomena in more detail, following the same sequence as the one of their appearance by increasing the value of the energy $\Escr$.

%-------------------------------------------------
\subsection{Planar-like regime}
\label{sub:nearlyplanar}
%------------------------------------------------
In the present section we will discuss in detail the structure of the phase portraits in the range of energies $\Escr_{2,3}\leq\Escr<\Escr_C$, were the dominant periodic orbits are the Modes A and B, which generalize the apsidal corotations (anti-aligned and aligned, respectively) of the planar case. It was already mentioned that the phase portait in this case contains two domains where the argument $\omega_2-\omega_3$ librates (around the values $0$ and $\pi$ respectively), separated by a domain where $\omega_2-\omega_3$ circulates, as in panels 5 to 10 of Fig.\ref{Fig.sezioni}. Owing to its similarity with the phase portrait of the planar problem, this will be called the \textit{planar-like} regime. It was emphasized, however, that the maximum mutual inclination can be quite high in this regime (see Fig.\ref{Fig.incmutene}), thus the analogy with the planar case stems from the dynamics, and not necessarily from the degree of coplanarity of the orbits in this regime. Excluding the energetically prohibited domains (as specified in subsection \ref{subsub:Poincare.section}), this can be extended to cover the cases where the librational domains are only partly covered by quasi-periodic orbits, as in panels 1-4 of Fig.\ref{Fig.sezioni}.  

%--------------------
\subsubsection{Integrable approximation of the Hamiltonian}
\label{subsub:integrable}
%--------------------
The main qualitative features of the planar-like regime, as well as precise computations regarding its periodic and surrounding quasi-periodic orbits, can be obtained in the context of an integrable approximation for the Hamiltonian $\Hscr_{sec}$, stemming from the splitting $\Hscr_{sec}=\Hscr_{planar}+\Hscr_{space}$ as in Eq.~\eqref{hamplsp}. Starting from
\begin{equation}
\Hscr_{sec}(a_2, a_3, \e_2, \e_3, w_2, w_3; L_z)\,=\Hscr_{planar}(w_2-w_3,W_2,W_3)
+\Hscr_{space}(w_2,w_3,W_2,W_3; L_z)\, .
\end{equation}
and splitting $\Hscr_{space}$ into those terms which depend only on the difference $w_2-w_3=\omega_3-\omega_2$, denoted by $\Hscr_{0,{space}}$, and those which do not, denoted by $\Hscr_{1,{space}}$, we arrive at the following decomposition of the Hamiltonian
\begin{equation}\label{hamdecompo}
\Hscr_{sec}=\underbrace{\Hscr_{planar}( w_2-w_3,W_2,W_3) + \Hscr_{0,space}(w_2-w_3,W_2,W_3; L_z)}_{\textbf{\textit{ Integrable part}}:=\Hscr_{int}}+ \Hscr_{1,space}(w_2,w_3,W_2,W_3; L_z)\, .
\end{equation}
The first two terms in the above expression give rise to a 2 degrees of freedom integrable Hamiltonian
\begin{equation}\label{hamint}
\Hscr_{int}( w_2-w_3,W_2,W_3;\AMD)= \Hscr_{planar}( w_2-w_3,W_2,W_3)+ \Hscr_{0,space}( w_2-w_3,W_2,W_3;\AMD)
\end{equation}
whose second integral is $W_2+W_3$. 

\begin{figure}[h!]
\begin{center}
\includegraphics[width=.75\textwidth]{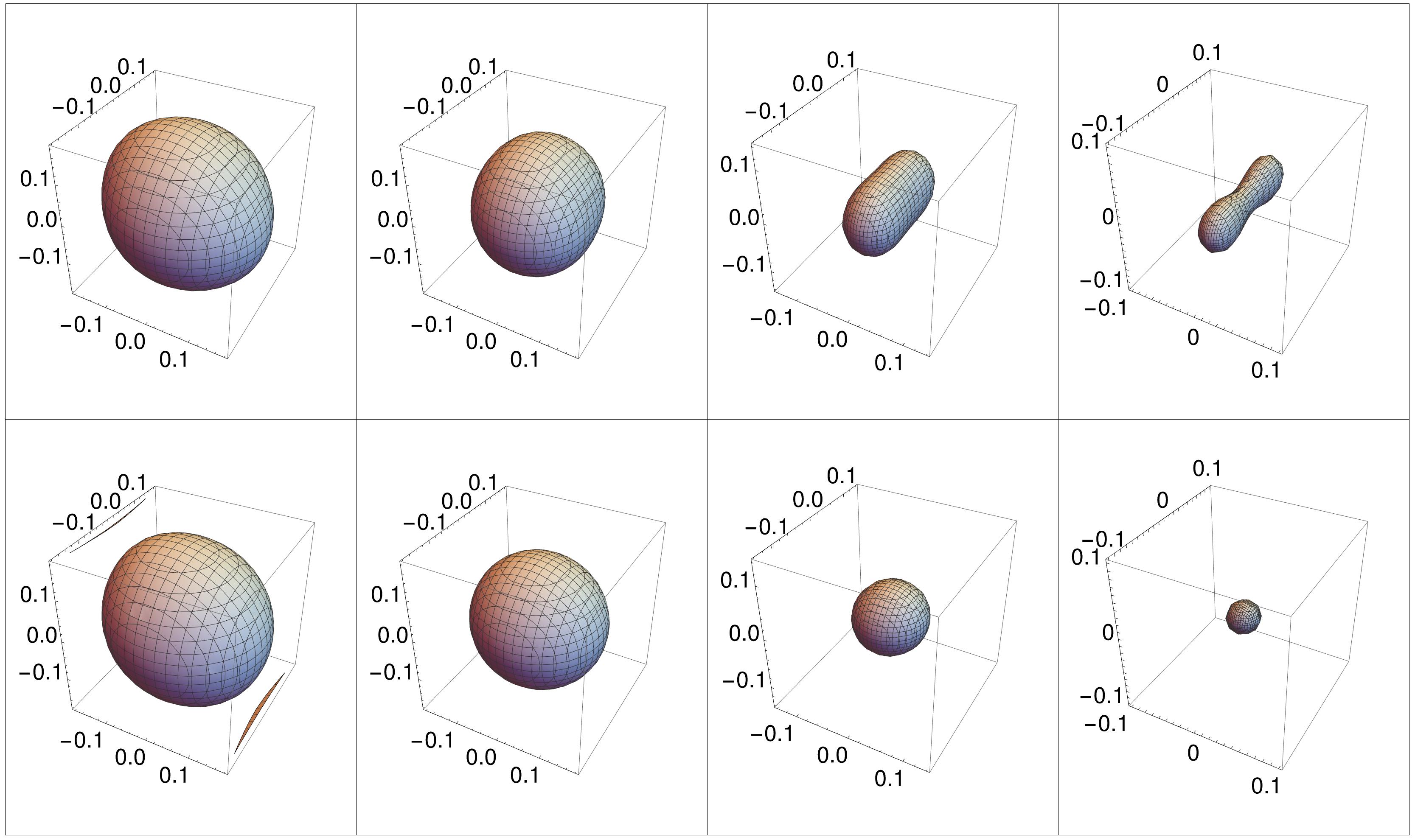}
\caption{The top row shows the manifolds of constant energy $\mathcal{M}(\Escr)$ compared to the manifolds of constant energy $\mathcal{M}_{int}(\Escr)$ of the integrable model $\Hscr_{int}$ (bottom row), for the energies  (from left to right) $\Escr_1=-8\cdot 10^{-5}$ ($\Escr_1<\Escr_{2,3}$), $\Escr_2=-4\cdot 10^{-5}$ ($\Escr_{2,3}<\Escr_2<\Escr_{C,2}$), $\Escr_3=-5\cdot 10^{-6}$ ($\Escr_3=\Escr_{C,2}$), $\Escr_4=-5\cdot 10^{-7}$ ($\Escr_4>\Escr_{C,2}$).}
\label{Fig.enecont}
\end{center}
\end{figure}

A fact hidden in the process of Jacobi reduction is that a decomposition of the Hamiltonian as in Eq.(\ref{hamdecompo}) yields a relative importance of the terms $\Hscr_{int}$ and $\Hscr_{1,space}$ \textit{varying 
with the energy level} $\Escr$ at which the orbits are computed. This is due to the fact that, after throwing apart constants, all the terms in $\Hscr_{space}$ stem from substitutions of the inclinations $\i_2$ and $\i_3$ depending only on the small quantity $1-\cos(\i_2+\i_3)$ and being of order second or higher in the eccentricities, according to Eq.~\eqref{riduzione.cos.sen} and~\eqref{seni2i3}. Thus, both $\Hscr_{0,space}$ and $\Hscr_{1,space}$ contain terms weighted by factors $(1-\cos(\i_2+\i_3))^s(\e_j^2+\ldots)$ (with $s\geq 1$, $j=2,3$). On the other hand, the terms $\Hscr_{planar}$ contain no factors $(1-\cos(\i_2+\i_3))^s$ and are of degree quadratic or higher in the eccentricities. Thus, due to the bound between increasing energy $\Escr$ and increasing mutual inclination (Fig.\ref{Fig.incmutene}), the relative importance of the terms $\Hscr_{1,space}$ with respect to the terms $\Hscr_{int}$ in the Hamiltonian rises as the energy (and hence the level of mutual inclination) increases. This is demonstrated graphically in Fig.\ref{Fig.enecont}, which shows a comparison between the form of the manifolds    
of constant energy $\mathcal{M}(\Escr)$ and $\mathcal{M}_{int}(\Escr)$ (Eqs.(\ref{manifene}) and (\ref{manifeneint})) computed at four different energy levels chosen as $\Escr_{min}<\Escr_1<\Escr_{2,3}$, $\Escr_{2,3}<\Escr_2<\Escr_{C,2}$, $\Escr_3=\Escr_{C,2}$, $\Escr_{C,2}<\Escr_4<0$. While the manifolds of $\Hscr_{int}$ remain always ellipsoidal-like, we note the progressive change of the form of the manifolds of constant energy in the complete model from an ellipsoidal to a peanut-shaped form, as the energy increases. This is caused by the growing importance of some terms in $\Hscr_{1,space}$, in particular the terms $\cos(2\omega_2)$ quadratic in $\e_2$, which are the same terms causing the transition to the Lidov-Kozai regime (see subsection \ref{sub:Kozai}). In fact, near the energy $\Escr_{C,2}$, where the orbit $C_2$ turns from stable to unstable, the manifold $\mathcal{M}(\Escr)$ under the complete model becomes nearly cylindrical, marking the change of its section with the plane $(X_2,Y_2)$ from elliptic-like to hyperbolic-like, as implied by the Lidov-Kozai mechanism.
%--------------------------------------------------------------------------------
%--------------------------------------------------------------------------------

Given the above, we will now focus on a description of the phase portraits in the energy regime (roughly identified as $\Escr<\Escr_C$) where the dynamics induced by $\Hscr_{sec}$ can be well approximated by the dynamics of $\Hscr_{int}$. In this regime, the following canonical transformation proves useful in semi-analytical (normal form) calculations related to the periodic orbits A and B and their surrounding quasi-periodic orbits:
\begin{align}
\label{trasf.can.int.plane}
& \psi=w_2-w_3\, , & &\Gamma=\frac{W_2-W_3}{2}\, , \notag\\
& \varphi=w_2+w_3\, , & &J=\frac{W_2+W_3}{2}~~.
\end{align}
The Hamiltonian in the new variables reads (apart from a constant)
\begin{equation}
\label{Ham.new.var}
\Hscr_{sec}(\psi, \varphi,\Gamma, J)=\Hscr_{int}( \psi, \Gamma; J)+\eta\,\Hscr_{1,space}(   \psi, \varphi, \Gamma, J)\, .
\end{equation}
Figure~\ref{Fig.hopfpc} shows the phase portrait (in the representative planes $(\e_2 \cos(\omega_2), \e_2 \sin(\omega_2)$ and $(\omega_2, \e_2)\,$) corresponding to the Hamiltonian $\Hscr_{int}$ at the energy $\Escr=-6.6\cdot 10^{-5}$. The phase portrait, computed as a Poincar\'e surface of section $\mathcal{P}(\Escr=-6.6\cdot 10^{-5}, \, \AMD=0.0162044)$ (see Eq.~\eqref{pcsec}), yields invariant curves equivalent to those obtained by the continuous flow after treating $\Hscr_{int}$ as a one degree of freedom Hamiltonian in the variables $(\psi,\Gamma)$, with $J$ serving as parameter.  

 We observe that with the integrable model $\Hscr_{int}$ we obtain a phase portrait with features qualitatively very similar to those of the phase portraits in the `planar-like' regime under the complete Hamiltonian (e.g. the panels 5-10 in Fig.\ref{Fig.sezioni}). In particular, the modes A and B of the integrable model are found as fixed points of $\Hscr_{int}$, given by the solutions of the equations  
\begin{equation}
\label{eq.moto.apsidal}
\begin{cases}
\displaystyle{\dot{\psi}=\frac{\partial\Hscr_{int}}{\partial\Gamma}=0}
\\[2ex]
\displaystyle{\dot{\Gamma}=-\frac{\partial\Hscr_{int}}{\partial\psi}=0}
\end{cases} \, .
\end{equation} 
Since $J$ is an integral of motion of $\Hscr_{int}$, setting $J=K=$ constant implies that any solution $( \psi_{\ast},\,\Gamma_{\ast})$ satisfying~\eqref{eq.moto.apsidal} is a periodic orbit with period given by $T_{\varphi}=2\pi/\omega_{\varphi}\,$, where $\displaystyle{\dot{\varphi}=\omega_{\varphi}=\frac{\partial\Hscr_{int}}{\partial J}\vert_{(\psi=\psi_{\ast},\,\Gamma=\Gamma_{\ast}\, , J=K})}\,$. In particular, the modes $A\,$ and $B\,$ are given, respectively, by $(\psi^{(A)}=0,\,\Gamma^{(A)}\, ,K^{(A)})$  and $( \psi^{(B)}=\pi,\,\Gamma^{(B)},\, K^{(B)})\,$, where $(\psi^{(A)},\,\Gamma^{(A)})\,$, $(\psi^{(B)},\,\Gamma^{(B)})\,$ are solutions of~\eqref{eq.moto.apsidal}. In fact, since the Hamiltonian $\Hscr_{int}$ depends only on the harmonics $\cos(k\psi)$, with $k\in\naturali\,$, the partial derivative in the second of Eqs.~\eqref{eq.moto.apsidal} yields only $\sin(k\psi)\,$ terms, hence, it admits the solution $\psi=\psi^{(A)}=0$ and $\psi=\psi^{(B)}=\pi$. Given now that $\Omega_3=\Omega_2+\pi$, the condition $\omega_3=\omega_2\,$, i.e. $\psi=0=\psi^{(A)}\,$, implies $\varpi_3=\varpi_2+\pi$ (perihelia anti-aligned), while the condition $\omega_3=\omega_2+\pi\,$, i.e. $\psi=\pi=\psi^{(B)}\,$, implies (modulo $2\pi$) $\varpi_3=\varpi_2$ (perihelia aligned). On the other hand, substituting one of the angles, $\psi^{(A)}$ or $\psi^{(B)}$, in the first of the Eqs.~\eqref{eq.moto.apsidal}, we obtain an algebraic equation of the form $\dot{\psi}(\psi=\psi^{(A)},\Gamma,J)=0$ or $\dot{\psi}(\psi=\psi^{(B)},\Gamma,J)=0$. This, together with the constant energy condition $\Hscr_{int}(\psi=\psi^{(A)},\Gamma,J)=\Escr$, or $\Hscr_{int}(\psi=\psi^{(B)},\Gamma,J)=\Escr$, can be solved to yield the pairs of values $\Gamma^{(A)},J=K^{(A)}$, or $\Gamma^{(B)},J=K^{(B)}$. Finally, the frequency of the apsidal precession for anyone of the periodic orbits A, B is given by $\nu_{apsidal}=-\omega_\varphi/2=-\dot{\varphi}/2=-(1/2)(\partial\Hscr_{int}(\psi,J,\Gamma)/\partial J)$, with $(\psi,J,\Gamma)$ substituted with one of the solutions A or B. 

%--------------------------------------------------------------------------
\subsubsection{The phase space of $\Hscr_{int}\,$: Hopf variables}
\label{subsub:hopf}
%--------------------------------------------------------------------------
An alternative method to compute the equilibria A and B stems from the use of a particular set of variables, called the \textit{Hopf variables} \cite{cusbat1997}, which, besides the computation of the equilibria, provides a global mapping of the phase space of the integrable Hamiltonian $\Hscr_{int}$ to the 3-sphere, thus allowing for a clear identification of all possible orbital dynamical regimes. We introduce the variables $(\sigma_1,\, \sigma_2,\, \sigma_3\,)$ defined by: 
\begin{equation}\label{def.Weyl}
\sigma_1=X_2 X_3+Y_2 Y_3\, ,\qquad
\sigma_2=Y_2 X_3-Y_3 X_2\, ,\qquad
\sigma_3=\frac{1}{2}\left(X_2^2+Y_2^2-X_3^2-Y_3^2\right)\, ,
\end{equation}
satisfying the Poisson algebra $\poisson {\sigma_i} {\sigma_j} =-2\,\epsilon_{i j k}\sigma_k\, $, where $\epsilon_{i j k}$ is the Levi-Civita symbol and $i,j,k=1,2,3$. Furthermore, we introduce the variable 
\begin{equation}\label{def.sigma0}
\sigma_0=\frac{1}{2}\left(X_2^2+Y_2^2+X_3^2+Y_3^2\right)
\end{equation}
which is a Casimir invariant of the previous algebra, since all Poisson brackets $\{\sigma_i,\sigma_0\}$, $i=1,2,3\,$, vanish. From the definition~\eqref{def.Weyl} it follows that 
\begin{equation}\label{def.Weyl.Delau}
\sigma_1=2\sqrt{J+\Gamma}\sqrt{J-\Gamma}\cos(\psi)\, ,\qquad
\sigma_2=-2\sqrt{J+\Gamma}\sqrt{J-\Gamma}\sin(\psi)\, ,\qquad
\sigma_3=W_2-W_3=2\Gamma\, .
\end{equation}
We also have the relation $\sigma_0=W_2+W_3=2J$, as well as
\begin{equation}\label{cond.sphere}
\sigma_1^2+\sigma_2^2+\sigma_3^2=\sigma_0^2=4J^2\, .
\end{equation}

Then, given the values of $(\sigma_1,\sigma_2,\sigma_3)$, the values of $\Gamma$, $J$ and $\psi$ can be computed unequivocally using the relations (\ref{def.Weyl.Delau}) and (\ref{cond.sphere}). Furthermore, since $J=\sigma_0/2$, and since the only trigonometric terms in the Hamiltonian $\Hscr_{int}$ are terms $\cos(k(w_3-w_2))=\cos(k\psi)$, $k=\mathbb{N}^{\ast}$, it follows that $\Hscr_{int}=\Hscr_{int}(\sigma_1,\sigma_3;\sigma_0)$, i.e. the Hamiltonian $\Hscr_{int}$ does not depend on $\sigma_2$. This implies that, fixing a value of $\sigma_0$ (i.e. of the integral $J$), the continuous in time phase flow obtained by solving the equations
\begin{equation}\label{sigmaflow}
\dot{\sigma}_1=\{\sigma_1,\sigma_3\}{\partial\Hscr_{int}\over\partial\sigma_3},~~~ 
\dot{\sigma}_2=
\{\sigma_2,\sigma_1\}{\partial\Hscr_{int}\over\partial\sigma_1}+
\{\sigma_2,\sigma_3\}{\partial\Hscr_{int}\over\partial\sigma_3},~~~ 
\dot{\sigma}_3=-\{\sigma_1,\sigma_3\}{\partial\Hscr_{int}\over\partial\sigma_1},~~~ 
\end{equation}
yields a flow equivalent to the one obtained under the Hamiltonian $\Hscr_{int}(\psi,\Gamma;J)$, i.e., treating $J$ as a parameter. Due to the constrain on the sphere (Eq.(\ref{cond.sphere})), the curves of the flow (\ref{sigmaflow}) are given by the intesection of the constant energy surface $\Hscr_{int}(\sigma_1,\sigma_3;\sigma_0)=\Escr$ with the sphere, i.e., they are closed curves which can be mapped to invariant curves in the plane $(\Gamma,\psi)$. These are geometrically equivalent to the invariant curves of the Poincar\'e surface of section of $\Hscr_{int}$ treated as a 2DOF system, mapping $(X_2,Y_2)$ as $X_2=-\sqrt{2(\Gamma+J)}\cos(\psi-\pi)$, $Y_2=\sqrt{2(\Gamma+J)}\sin(\psi-\pi)$. It should be stressed, however, that the above Hamiltonian reduction only yields a geometric equivalence of the two curves, since in the 1DOF reduced system the flow is continuous, while in the 2DOF full system the curves are traced stroboscopically, and they correspond to the intersection of the 2D invariant tori involving both the angles $(\psi,\varphi)$ with the selected surface of section.    

%--------------------------------------------------------------------------------
\begin{figure}[h!]
\begin{center}
\includegraphics[width=0.7\textwidth]{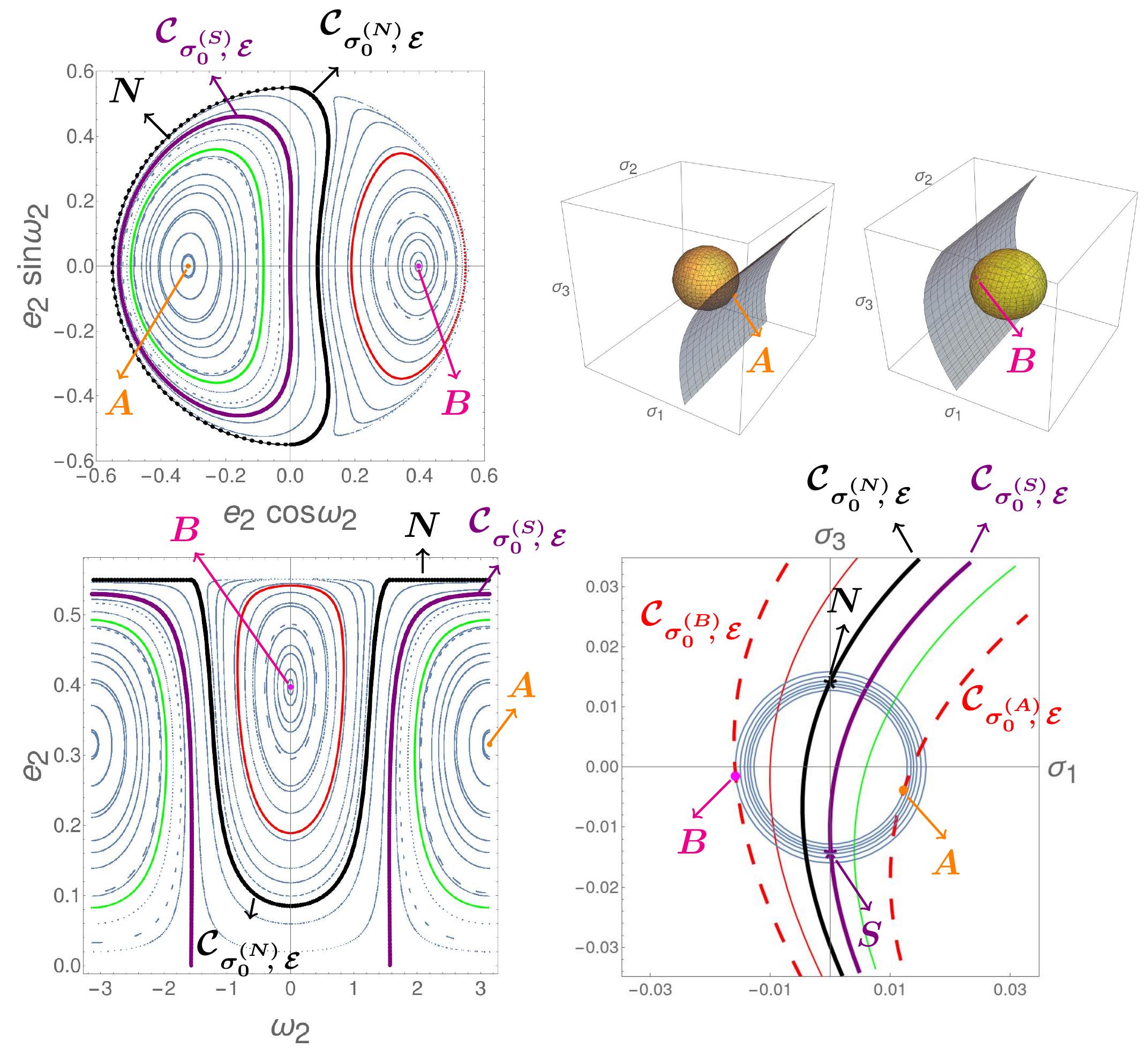}
\caption{Left column: The phase portrait of the integrable Hamiltonian $\Hscr_{int}$ at the energy $\Escr=-6.6\cdot 10^{-5}$, projected in the variables $(\e_2\cos\omega_2,\e_2\sin\omega_2)$ (top), or simply $(\omega_2,\e_2)$ (bottom). Right column, top: At the values $\sigma_0=\sigma_0^{(A)}$ and $\sigma_0=\sigma_0^{(B)}$, the corresponding spheres $\mathcal{S}_{\sigma_0^{(A)}}$, $\mathcal{S}_{\sigma_0^{(B)}}$ become tangent to the energy surfaces $\mathcal{C}_{\sigma_0^{(A)},\,\Escr}$, $\mathcal{C}_{\sigma_0^{(B)},\,\Escr}$. The points of tangency yield the position of the fixed points $A$ and $B$ in the surface of section (see text). Right column, bottom: the intersection of the spheres $\mathcal{S}_{\sigma_0}$ and of the energy surfaces $\mathcal{C}_{\sigma_0,\,\Escr}$ with the plane $(\sigma_1,\sigma_3)$ for $\sigma_2=0$, for various values of $\sigma_0$. The intersection of one sphere with one energy surface yields a curve on the sphere which is projected to a curve in the above plane. For a particular value of $\sigma_0=\sigma_0^{(S)}$, the curve (thick purple) passes through the south pole $S$ of the corresponding sphere $\mathcal{S}_{\sigma_0}$. This corresponds to a trajectory forming a closed curve in the Poincar\'e section, which surrounds mode A and passes through the origin. This curve delimits the domain of orbits whose angle $\psi=\omega_3-\omega_2$ librates around the value $\psi=0$. At a different value of $\sigma_0=\sigma_0^{(N)}$ the curve of constant energy (thick black in the right column, bottom) passes through the corresponding sphere's north pole $N$. This corresponds to a curve in the surface of section which surrounds the previous curve as well as the origin. In particular, the whole black curve, except for the point $N\,$, yields the part of the corresponding curve in the Poincaré section (left, top figure) contained in the positive semi-plane, while the point $N$ itself inflates to the part of the corresponding closed curve in contained in the negative semi-plane. The domain in the surface of section between the thick purple and the thick black curves corresponds to orbits whose argument $\psi=\omega_3-\omega_2$ circulates. All trajectories beyond the outer delimiting curve exhibit librations of the argument $\omega_3-\omega_2$ around the value $\pi$, characteristic of the B-mode.}
\label{Fig.hopfpc}
\end{center}
\end{figure}
%--------------------------------------------------------------------------------
The above information can now be used to the purpose of analyzing the structure of the phase portraits using the mapping of each invariant curve in the sphere (\ref{cond.sphere}) to the corresponding curve in the usual Poincar\'e surface of section. To this end, let
\begin{equation}
\label{sphere}
\mathcal{S}_{\sigma_0}=\lbrace(\sigma_1, \sigma_2, \sigma_3)\in\reali^3\, : \, \sigma_1^2+\sigma_2^2+\sigma_3^2=\sigma_0^2 \rbrace
\end{equation}
denote the sphere corresponding to the value $\sigma_0=2J$ of the integral $J$, and 
\begin{equation}
\label{energy.curve}
\mathcal{C}_{\sigma_0,\, \Escr} =\lbrace(\sigma_1, \sigma_2, \sigma_3)\in\reali^3\, : \, \Hscr_{int}(\sigma_0, \sigma_1, \sigma_3)=\Escr\rbrace\, ,
\end{equation} 
denote the energy surface in the space $(\sigma_1,\sigma_2,\sigma_3)\in\mathbb{R}^3$ corresponding to a fixed energy value $\Escr$. We can have a physical trajectory for all values of $\sigma_0$ (i.e., of $J$) for which the surfaces $\mathcal{S}_{\sigma_0}$ and $\mathcal{C}_{\sigma_0,\, \Escr}$ intersect, limited by two values of $\sigma_0\,$, that are $\sigma_0^{(A)}$ and $\sigma_0^{(B)}$ (corresponding to $J^{(A)}=\sigma_0^{(A)}/2$ and $J^{(B)}=\sigma_0^{(B)}/2\,$), where the two surfaces become tangent (Fig.\ref{Fig.hopfpc}). By the non-dependence of $\Hscr_{int}$ on $\sigma_2$, the constant energy surface $\mathcal{C}_{\sigma_0,\, \Escr}$ is normal to any plane $(\sigma_1,\sigma_3)$ with $\sigma_2=const$. Hence, at a tangency point of $\mathcal{S}_{\sigma_0}$ with $\mathcal{C}_{\sigma_0,\, \Escr}\,$ we necessarily have that $\sigma_2=0$, as well as the tangency condition 
$$
\mathrm{rank}\begin{pmatrix}
\displaystyle{2\sigma_1} & \displaystyle{2\sigma_2} & \displaystyle{2\sigma_3} \\
\displaystyle{\frac{\partial \Hscr_{int}}{\partial \sigma_1}} & \displaystyle{0} & \displaystyle{\frac{\partial \Hscr_{int}}{\partial \sigma_3}}
\end{pmatrix}=1~~.
$$
The latter condition implies that
$$
\dot{\sigma}_2=\sigma_3\frac{\partial \Hscr_{int}}{\partial \sigma_1}-\sigma_1\frac{\partial \Hscr_{int}}{\partial \sigma_3}=0
$$
that is, the point of tangency is a fixed point of the flow. Up to terms of second order in the variables $\sigma_i$ (i.e. of fourth order in the eccentricities), we find 
$$
\Hscr_{int}=A\sigma_1^2+B\sigma_3^2+C\sigma_1\sigma_3+D(\sigma_0)\sigma_1+E(\sigma_0)\sigma_3+F(\sigma_0)+...
$$ 
where $A,B,C$ are constants, while the functions $D(\sigma_0)$ and $E(\sigma_0)$ are linear in $\sigma_0$ and $F(\sigma_0)$ contains terms linear and quadratic in $\sigma_0$. The quadratic form $A\sigma_1^2+B\sigma_3^2+C\sigma_1\sigma_3$ yields hyperbolas, being $A,B,C$ such that $C^2>AB\,$. Thus, for any permissible value $\sigma_0$ (or, equivalently, of the integral $J\,$), the surface $\mathcal{C}_{\sigma_0,\, \Escr}$ intersects the plane $\sigma_2=0$ along hyperbola-like curves (Fig.\ref{Fig.hopfpc}). The two points of tangency occur at the values $\sigma_0^{(A)}=2J^{(A)}$ and $\sigma_0^{(B)}=2J^{(B)}$. We find that $\sigma_0^{(A)}<\sigma_0^{(B)}$, while, checking the sign of $\cos(w_3-w_2)$ for the corresponding fixed points, we identify the left tangency (see Fig.\ref{Fig.hopfpc}) as the B-mode and the right tangency as the A-mode at the given level of energy. 

With the help of the bottom-right panel of Fig.\ref{Fig.hopfpc} it is possible, now, to interpret the form of the phase portraits as in the left column of the same figure. To this end, we specify the correspondence between the various curves of the phase flow on the sphere, obtained by the intersections between the surfaces $\mathcal{S}_{\sigma_0}$ and $\mathcal{C}_{\sigma_0,\Escr}$ as $\sigma_0$ is altered in the interval $\sigma_0^{(A)}\leq\sigma_0\leq\sigma_0^{(B)}$, and the mapping of these curves to the surface of section $(\e_2\cos\omega_2,\e_2\sin\omega_2)$. Since $\Hscr_{int}$ does not depend on $\sigma_2$, all the curves produced by intersections of the surfaces $\mathcal{S}_{\sigma_0}$ and $\Cscr_{\sigma_0,\,\Escr}$ contain points which lie in the meridian circle produced by the intersection of $\mathcal{S}_{\sigma_0}$ with the plane $\sigma_2=0$. In particular, the points of tangency $A$ and $B$ belong to this meridian. Besides these points, there are two critical curves which separate domains of libration of the angle $\omega_2-\omega_3$ around the value 0 (mode A), or $\pi$ (mode B), from domains where the angle $\omega_2-\omega_3$ circulates.   

By varying, now, the value of $\sigma_0$ in the interval $\sigma_0^{(A)}\leq\sigma_0\leq\sigma_0^{(B)}$ we progressively obtain curves on the sphere which pass from a librating domain around the fixed point $A$ to a circulating domain, and then to a librating domain around the fixed point $B$. The first such transition occurs at a value $\sigma_0^{(S)}$ where the curve corresponding to the intersection between $\mathcal{S}_{\sigma_0^{(S)}}$ and $\mathcal{C}_{\sigma_0^{(S)},\,\Escr}$ passes from the south pole $S$ of the sphere $\mathcal{S}_{\sigma_0^{(S)}}$. The coordinates of the south pole are $\sigma_1^{(S)}=\sigma_2^{(S)}=0$, $\sigma_3^{(S)}=-\sigma_0^{(S)}$, implying $J=-\Gamma$ or $W_2+W_3=-(W_2-W_3)$, hence $W_2=0$. This means a curve in the Poincar\'e section (thick purple) which crosses the origin $\e_2=0$. As clear from Fig.\ref{Fig.hopfpc}, we stress the well known fact that this curve means no real separatrix in the surface of section, generated by any kind of unstable periodic orbit, but it merely reflects the singularity induced by projecting an (all continuous) transition taking place in the phase space of the integrable Hamiltonian $\Hscr_{int}$, which is the 3-sphere, to the usual Poincar\'e section applicable to the full problem, i.e., the plane $(\e_2\cos\omega_2,\e_2\sin\omega_2)$. 

Passing, now, the value $\sigma_0=\sigma_0^{(S)}$, we have curves of the sphere which are projected to invariant curves still surrounding the fixed point $A$, but for which the argument $\omega_2-\omega_3$ (or $\omega_2-\pi$, in the surface of section) circulates. A second limit of the circulation domain occurs at a value $\sigma_0^{(N)}$ where the curve corresponding to the intersection between $\mathcal{S}_{\sigma_0^{(N)}}$ and $\mathcal{C}_{\sigma_0{(N)}\, , \Escr}$ passes from the north pole $N$ of the sphere $\mathcal{S}_{\sigma_0^{(N)}}$. We readily find that the whole curve in the sphere $\mathcal{S}_{\sigma_0^{(N)}}$, except for the pole $N$ itself, yields the open black curve in the surface of section corresponding to the positive semi-plane $\e_2\cos(\omega_2)\geq 0$ (or $-\pi/2\leq \omega_2\leq \pi/2$), while the north pole itself inflates to the dotted semicircle obtained in the negative semi-plane $\e_2\cos(\omega_2)< 0$ (or $\pi/2< \omega_2<3\pi/2$). Here again the singularity is not real but only due to the choice of the variables representing the surface of section (see e.g.~\cite{pau1983},~\cite{henlib2004},~\cite{micetal2006}). In fact, $N$ has coordinates $\left(\sigma_0^{(N)},\, \sigma_1^{(N)}=0,\,\sigma_2^{(N)}=0,\,\sigma_3^{(N)}=\sigma_0^{(N)}\right)\,$, implying $W_3=0$, i.e. $\e_3=0$. However, the equality $\sigma_3^{(N)}=\sigma_0^{(N)}$ implies also $X_2^2+Y_2^2=2\sigma_0$, i.e. a circle on the section $Y_3=0$. Together with the condition $\dot{Y}_3\geq 0$ of the Poincar\'e section, this implies the semi-circle $X_2\geq 0$, i.e. the dotted part of the black curve in the top-left panel of Fig.\ref{Fig.hopfpc}. It can be shown that the two parts or the curve join each other smoothly at two limiting values $X_2=0,Y_2=\pm Y_{2,max}=\pm \sqrt{2\sigma_0^{(N)}}$. In fact, the semi-circle $Y_2=\pm \sqrt{2\sigma_0^{(N)}-X_2^2}$, $X_2>0\,$, corresponding to the inflation of the north pole, yields $\lim_{X_2\rightarrow 0^+}(d Y_2/ d X_2)=0$, while the open curve $Y_2=Y_2(X_2)$, corresponding to all other points of the intersection of  $\mathcal{S}_{\sigma_0^{(N)}}$ and $\mathcal{C}_{\sigma_0{(N)}, \, \Escr}$ except for the pole, yields the limit 
\begin{equation}
\label{derivata.curva.critica}
\lim_{X_2\rightarrow 0^-}\frac{d Y_2}{d X_2}=
\lim_{X_2\rightarrow 0^-}
\left(-\frac{\frac{\partial \Hscr_{int}}{\partial X_2}}
{\frac{\partial \Hscr_{int}}{\partial Y_2}}\right)~~.
\end{equation}
From the form of $H_{int}\,$, recalling Eq.~\eqref{conseguenza.Lib-Henr}, we readily find $\partial\Hscr_{int}/\partial X_2=0$, $\partial\Hscr_{int}/\partial Y_2\neq 0$ 
at the north pole limit $X_2=X_3=Y_3=0$, $Y_2=\pm Y_{2,max}$. 

Finally, for $\sigma_0$ in the interval $\sigma_0^{(N)}<\sigma_0\leq\sigma_0^{(B)}$ we find invariant curves in the sphere $\mathcal{S}_{\sigma_0}$ mapped to closed invariant curves around the tangency corresponding to the B-mode fixed point, which yield also closed curves in the Poincar\'e section for which the argument $\omega_2-\omega_3$ (or $\omega_2-\pi$ in the section) librates around the value $\pi$, i.e., $\omega_2$ librates around the value $\omega_2=0$ of the B-mode. 

%-----------------------------------------------------------------------------------------
\subsubsection{Semi-analytical (normal form) determination of the periodic orbits A and B}
\label{subsub:normalforms}
%------------------------------------------------------------------------------------------
In the previous subsection we have seen how the existence of the A and B modes, which generalize the apsidal corotation periodic orbits of the model $\Hscr_{planar}$ to the spatial case, can be established within the framework of the integrable model $\Hscr_{int}=\Hscr_{planar}+\Hscr_{0,space}$. In the present subsection we discuss how to recover semi-analytically the periodic orbits A and B under the full Hamiltonian $\Hscr_{sec}=\Hscr_{int}+\Hscr_{1,space}$. Besides its relevance in establishing the existence of these orbits in the full model (up to an exponentially small remainder), a computation of the orbits A and B using normal forms allows to obtain a semi-analitycal representation of the long term time series of the orbital elements for these planetary trajectories. Since the modes A and B are among the most probable expected endstates of the formation process for exoplanetary systems, such a representation can be of use also in the interpretation of the observational data regarding the planetary orbital configurations in such systems.   

Let $(\psi_*,\Gamma_*, J_\ast)$ be a fixed point of the integrable Hamiltonian $\Hscr_{int}(\psi,\Gamma;J)$, corresponding to one of the modes A or B. Introduce the translation
\begin{equation}\label{traaps}
\psi=\psi_{\ast}+\delta\psi,~~\Gamma=\Gamma_{\ast}+\delta\Gamma,~~J=J_{\ast}+\delta J.
\end{equation}
The transformation $(\psi,\varphi,\Gamma,J)\rightarrow (\delta\psi,\varphi,\delta\Gamma,\delta J)$ is canonical. The Hamiltonian $\Hscr_{sec}$ in the new variables reads:
\begin{equation}\label{hamdelta}
\Hscr_{sec}(\delta\psi,\varphi,\delta\Gamma,\delta J)=\Hscr_{int}(\delta\psi,\delta\Gamma,\delta J)
+ \Hscr_{1,space}(\delta\psi,\varphi,\delta\Gamma,\delta J)\, .
\end{equation}
We then have the following\\
\\
\noindent
{\bf Proposition:} There is a near-to-identity canonical transformation 
\begin{equation}\label{phiper}
(\delta\psi,\varphi,\delta\Gamma,\delta J)=\Phi^{(r)}(\delta\t{\psi},\t{\varphi},\delta\t{\Gamma},\delta\t{J})~~,
\end{equation}
real-analytic in a suitably defined domain, and obtained by a composition of $r$ Lie canonical transformations, with $r$ a sufficiently high integer, such that the Hamiltonian in the variables $(\delta\t{\psi},\t{\varphi},\delta\t{\Gamma},\delta\t{J})$ takes the form:
\begin{equation}\label{hamnophi}
\begin{split}
\Hscr^{(r)}(\delta\t{\psi},\t{\varphi},\delta\t{\Gamma},\delta\t{J})=
Z^{(r)}(\delta\t{\psi},\delta\t{\Gamma},\delta\t{J})
+R^{(r)}(\delta\t{\psi},\t{\varphi},\delta\t{\Gamma},\delta\t{J})~
\end{split}
\end{equation}
with $||R^{(r)}||<<||Z^{(r)}||$ for a suitably defined norm in the domain of the transformation. The superscript $(r)$ in the above expressions means the expression derived after substituting the original variables $(\delta\psi,\varphi,\delta\Gamma,\delta J)$ with the new canonical variables obtained by the transformation inverse to the $r-$step transformation (\ref{phiper}): 
\begin{equation}\label{phiper}
(\delta\t{\psi},\t{\varphi},\delta\t{\Gamma},\delta\t{J})=
\left(\Phi^{(r)}\right)^{-1}(\delta\psi,\varphi,\delta\Gamma,\delta J)~~.
\end{equation}
The functions $Z^{(r)}$ and $R^{(r)}$ are hereafter called the `normal form' and `remainder' respectively.\\ 
\\
The procedure by which to obtain the transformation $\Phi^{(r)}$ as well as its inverse is summarized in Appendix B, while a detailed demonstration of the above proposition is given elsewhere (Mastroianni and Efthymiopoulos, in preparation). In summary, this is a classical Birkhoff-like normalization procedure by which the angle $\varphi$ is eliminated from the Hamiltonian via a sequence of Lie canonical transformations (see \cite{eft2012}). 

The physical meaning of the above normalization procedure is the following: in the planar-like regime, we have $||\Hscr_{1,space}||<<\Hscr_{int}$. Then, through the normalization we find a new set of variables in which, except for a very small remainder,  the dynamics locally (around the equilibrium $(\psi_*,\Gamma_*,J_*)$) is given by the normal form $Z^{(r)}(\t{\psi},\t{\Gamma},\t{J})$, with
\begin{equation}\label{tildevar}
\t{\psi}=\psi_*+\delta\t{\psi},~~\t{\Gamma}=\Gamma_*+\delta\t{\Gamma},~~\t{J}=J_*+\delta\t{J}~~. 
\end{equation}
The phase flow induced by the integrable Hamiltonian $Z^{(r)}$ has the same structure as the one of the Hamiltonian $\Hscr_{int}$ analyzed in the previous subsection, differing by it just in the fact that $Z^{(r)}$ contains terms arising from the normalization of $\Hscr_{1,space}$ up to a very small remainder. In particular, Hamilton's flow under the normal form  in the transformed variables admits a periodic orbit given by 
\begin{eqnarray}\label{nfper}
\t{\psi}_*&=&\psi_*+\delta\t{\psi}_*~~, \nonumber\\
\t{\Gamma}_*&=&\Gamma_*+\delta\t{\Gamma}_*~~, \\
\t{J}_*&=&J_*+\delta\t{J}_*~~, \nonumber\\
\t{\varphi}(t)&=&\t{\varphi}(0)+\nu_*t~~,\nonumber
\end{eqnarray}
where $(\t{\psi}_*$, $\t{\Gamma}_*,\t{J}_*)$ are computed by the system of algebraic equations
 \begin{equation}\label{compnfper}
\begin{cases}
\displaystyle{\dot{\t{\psi}}=\frac{\partial Z^{(r)}}{\partial\t{\Gamma}}=0}
\\[2ex]
\displaystyle{\dot{\t{\Gamma}}=-\frac{\partial Z^{(r)}}{\partial \t{\psi}}=0}\\[1.3ex]
Z^{(r)}(\t{\psi},\,\t{\Gamma},\,\t{J}\,)=\Escr
\end{cases}
\end{equation} 
and the frequency $\nu_*$ is given by
\begin{equation}\label{nustar}
\nu_*=\left({\partial Z^{(r)}
\over\partial\t{J}}\right)_{\t{\psi}=\t{\psi}_*,\t{\Gamma}=\t{\Gamma}_*,\t{J}=\t{J}_*}~~.
\end{equation}

%----------------------------------------------------------------------------------------------------------
\begin{figure}[h!]
\centering
\includegraphics[width=0.7\textwidth]{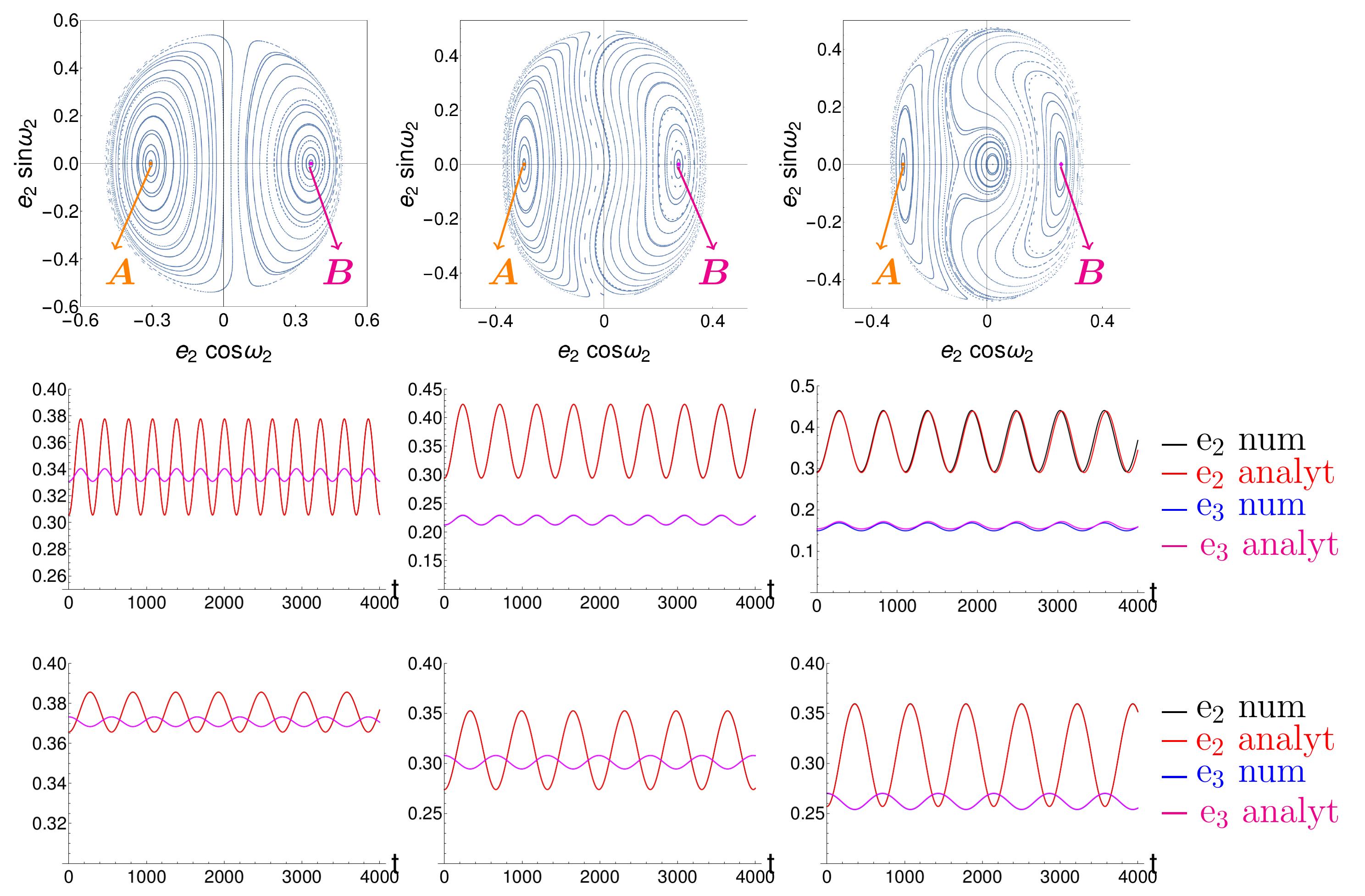}
\caption{First row: Poincar\'e surfaces of section for the values of the energy (from left to right), $\Escr=-5.72\cdot 10^{-5},\, -3.19\cdot 10^{-5},\, -2.53\cdot 10^{-5}\,$. Second and third rows: time series of the evolution of the eccentricities for both planets along the mode $A$ (\textit{anti-apsidal}, second row) and $B$ (\textit{apsidal}, third row). The curves in red and in magenta show the time series for the eccentricities $\e_2$ and $\e_3$ respectively as computed semi-analytically. The black and blue curves, instead, show the time series $\e_2$ and $\e_3$ as computed through a numerical evaluation of the periodic orbits.} 
\label{Fig.normal.apsi}
\end{figure} 
%----------------------------------------------------------------------------------------------------------
Figure \ref{Fig.normal.apsi} shows an example of the comparison between the semi-analytical computation of the periodic orbits A, B on the basis of the normal form flow of $Z^{(r)}$, and a full numerical computation of the same orbits. The semi-analytical computation of the periodic orbits proceeds by the following steps:

i) we use the tangency method of subsection \ref{subsub:hopf} to compute the fixed points of the integrable Hamiltonian $\Hscr_{int}$, first in the Hopf variables and then in the variables $(\psi$, $\Gamma$, $J)\,$, obtaining $\psi_*$, $\Gamma_*$, $J_*$. 

ii) Using an appropriate expansion of the Hamiltonian, as well as the method of composition of Lie series (see Appendix B), we then obtain the transformation $\Phi^{(r)}$ and its inverse $\left(\Phi^{(r)}\right)^{-1}$, as well as the normal form $Z^{(r)}$ representation of the full Hamiltonian. 

iii) Implementing a Newton method, we then compute the root $(\t{\psi}_*$, $\t{\Gamma}_*,\t{J}_*)$ of the system of algebraic equations~\eqref{compnfper}, as well as the frequency~\eqref{nustar}. This yields the time evolution of all 
four quantities $(\delta\t{\psi}=\t{\psi}_*-\psi_*,\delta\t{\Gamma}=\t{\Gamma}_*-\Gamma_*,\delta\t{J}=\t{J}_*-J_*)$ (fixed) and $\t{\varphi}(t)$ as in Eq.(\ref{nfper}). 

iv) We finally obtain the semi-analytical approximation to the time flow of all four variables $\psi(t)=\psi_*+\delta\psi(t)$, $\varphi(t)$, $\Gamma(t)=\Gamma_*+\delta\Gamma(t)$, $J(t)=J_*+\delta J(t)$, through the normalizing transformation
\begin{equation}\label{nfperflow}
(\delta\psi(t),\varphi(t),\delta\Gamma(t),\delta J(t))
=\left(\Phi^{(r)}(\delta\t{\psi},\t{\varphi},\delta\t{\Gamma},\delta\t{J})\right)_{ \delta\t{\psi}=\delta\t{\psi}_*,\t{\varphi}=\t{\varphi}(t), \delta\t{\Gamma}=\delta\t{\Gamma}_*,\delta\t{J}=\delta\t{J}_*}~~. 
\end{equation}
This can be further transformed into a time series of the evolution of the orbital elements along the periodic orbit through the equations
\begin{align}\label{nfperele}
&\omega_2=\frac{-\psi-\varphi}{2}\, , & &W_2=J+\Gamma\, , \notag\\
& \omega_3=\frac{\psi-\varphi}{2}\, , & &W_3=J-\Gamma\,,
\end{align}
which allow to recover the evolution of the arguments of perihelia and eccentricities 
$$
\e_j=\sqrt{1-(1-W_j/\Lambda_j)^2},~~j=2,3 
$$
as well as (equivalently to Eq.~\eqref{cosi23e23})
\begin{align}\label{nfpertheta}
\Theta_2&=\frac{(\AMD-W_2-W_3)(2 \Lambda_3+W_2-W_3-\AMD)}{2 (\Lambda_2+\Lambda_3-\AMD)}\\
\Theta_3&=\frac{(\AMD-W_2-W_3)(2 \Lambda_2-W_2+W_3-\AMD)}{2 (\Lambda_2+\Lambda_3-\AMD)}\, \notag
\end{align}
which allow to compute the time series for the inclinations 
$$
\i_j=\cos^{-1}\left(1-\frac{\Theta_j}{\Lambda_j-W_j}\right) 
$$
of the planetary periodic orbits. 

As shown in Fig.\ref{Fig.normal.apsi}, the semi-analytical representation of the periodic orbits fits with precision the numerical ones for both the modes A and B. As regards the physical interpretation, we note that the main effect of the perturbation $\Hscr_{1,space}$ is to induce a periodic oscillation in the eccentricities (and, hence, also the inclinations) of both planets, which are no longer constant, contrary to what holds for the classical apsidal corotation orbits in the planar case, or for the orbits of the integrable approximation $\Hscr_{int}$. From Fig.\ref{Fig.normal.apsi} it is evident that the amplitude of the oscillation of the eccentricities of both planets increases with the energy $\Escr$, and, hence, with the mutual inclination. Most, notably, however, we point out the ability of the semi-analytical theory to well represent the orbits of both modes A and B in the regime \textit{after} the onset of the saddle-node bifurcation giving rise to the periodic orbits $C_1$ and $C_2$, as in the third panel of Fig.\ref{Fig.normal.apsi} (see subsection \ref{subsec:general}). This implies that, while at the corresponding level of energies the structure of the phase portraits is altered considerably with respect to the one of $\Hscr_{int}$, the representation of the motions by a local normal form computed in the neighborhood of each of the modes A or B yields the correct picture of the dynamics, both qualitatively and quantitatively. 

%-----------------------------------------------------------------
\subsection{Transition regime between planar-like and Lidov-Kozai: 
sequences of bifurcations}
\label{sub:bifurcation}
%-----------------------------------------------------------------

The planar-like regime discussed in the previous subsection characterizes the structure of the phase portraits of the full system up to the energy $\Escr_C$, where the orbits $C_1,C_2$ are generated by a saddle node bifurcation in the taking place in the neighborhood of the mode B, as in the transition seen in the third row of Fig.\ref{Fig.sezioni}. We call \textit{transition regime} the one holding at energies in the interval $\Escr_{C}\leq\Escr_{C,2}$.

%--------------------------------------------------------------------------------
\begin{figure}[h!]
\begin{center}
\includegraphics[width=0.5\textwidth]{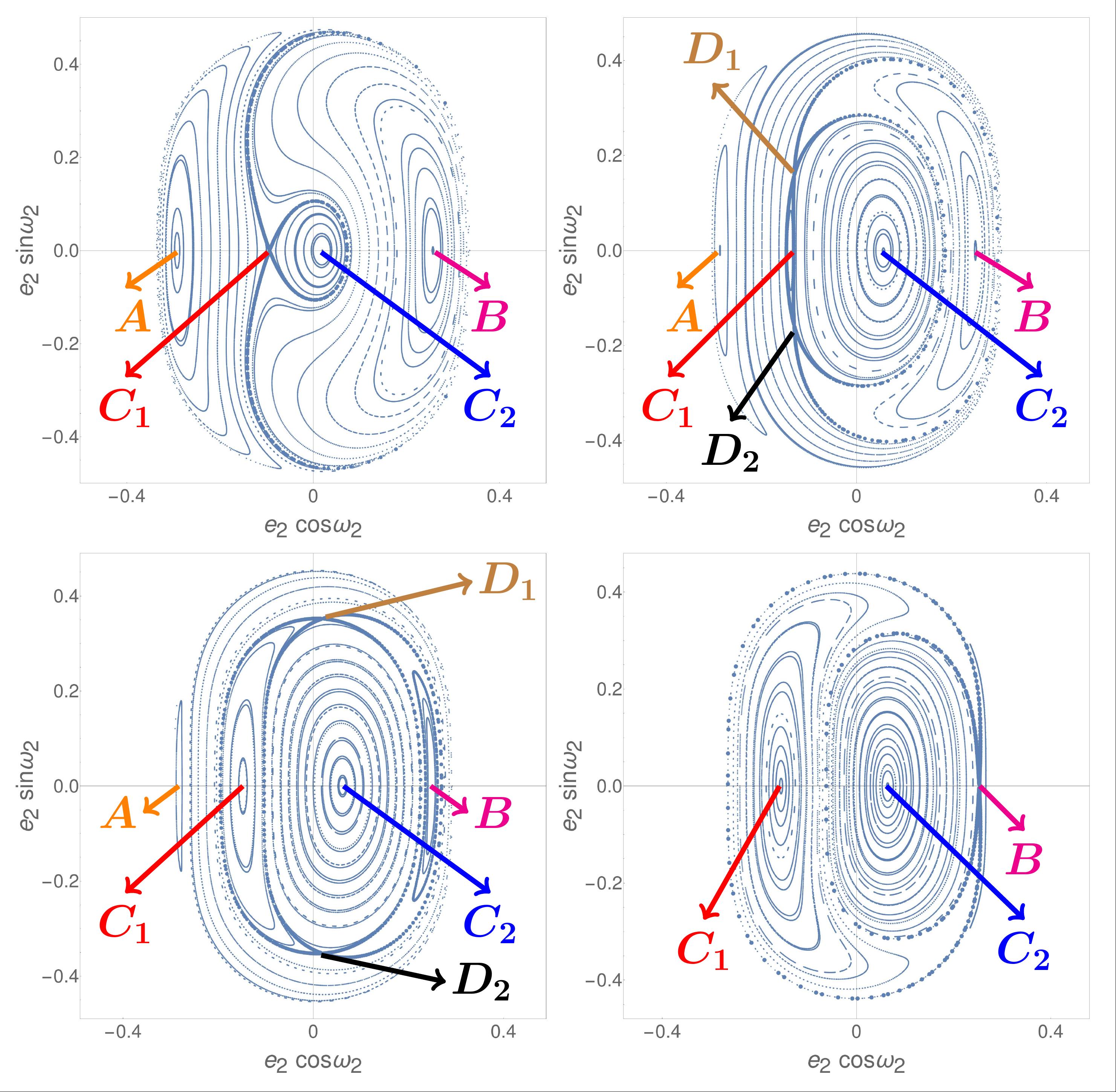}
\caption{Poincaré surfaces of section for some values of the energy illustrating the sequence of bifurcations taking place in the `transition regime' (see text): $\Escr=-2.53\cdot 10^{-5}$  (top left), $\Escr=-2.08 \cdot 10^{-5}$ (top right), $\Escr=-1.9\cdot 10^{-5}$ (bottom left), $\Escr=-1.58\cdot 10^{-5}$ (bottom right), corresponding, respectively, to panels $11\,$, $13\,$, $14$ and $15$ of Figure~\ref{Fig.sezioni}.}
\label{Fig.sezionizoom}
\end{center}
\end{figure}
%--------------------------------------------------------------------------------
%--------------------------------------------------------------------------------
\begin{figure}[h!]
\begin{center}
\includegraphics[width=1.\textwidth]{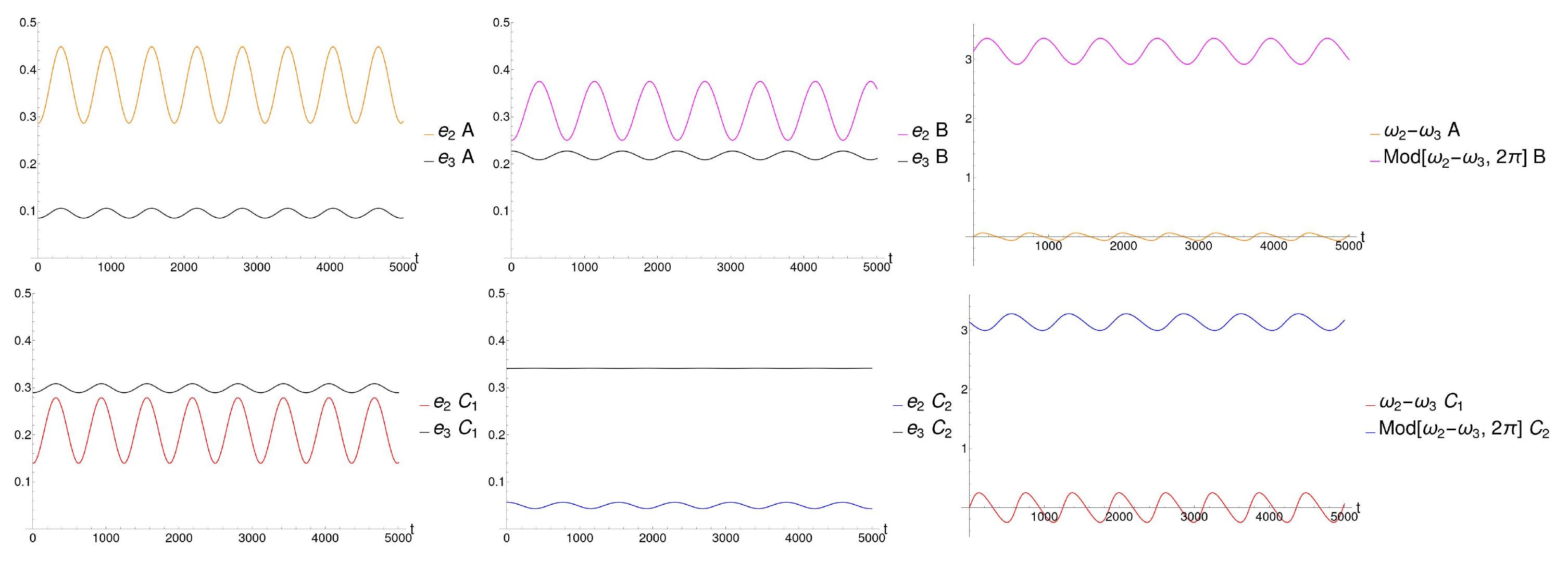}
\caption{Top row: the evolution of the orbital eccentricities $\e_2,\,\e_3$ for the periodic orbits corresponding to the modes A (top left panel: top curve $\e_2$, bottom curve $\e_3$), and B (top middle panel: top curve $\e_2$, bottom curve $\e_3$). The top right panel shows the librations of the argument $\omega_2-\omega_3$ for the B-mode (top curve) or the A-mode (bottom curve). Bottom row: the evolution of the orbital eccentricities $\e_2,\,\e_3$ for the periodic orbits corresponding to the fixed points $C_1$ (bottom left panel: top curve $\e_3$, bottom curve $\e_2$), and $C_2$ (bottom middle panel: top curve $\e_2$, bottom curve $\e_3$).The bottom right panel shows the librations of the argument $\omega_2-\omega_3$ for the orbit $C_2$ (top curve) or $C_1$ (bottom curve). The above time series are computed at the value of the energy $\Escr=-2.08\cdot 10^{-5}\,$, corresponding to the $13$-th panel of Figure~\ref{Fig.sezioni}.}
\label{Fig.abc12}
\end{center}
\end{figure}
%--------------------------------------------------------------------------------
Figure \ref{Fig.sezionizoom} presents in detail the sequence of bifurcations taking place across the transition regime, which eventually lead to turning unstable the periodic orbit of mode B. At the beginning of the transition, for energies slightly larger than $\Escr_C$, the orbit $C_1$ turns from unstable to stable by a pitchfork bifurcation, which gives rise to an unstable periodic orbit yielding two fixed points, $D_1,D_2$, in the Poincar\'e surface of section. As the energy increases, these fixed points move initially away from the fixed point of the orbit $C_1$, while later (for still larger energy) they approach the fixed point of the mode B. Finally, at a second critical value of the energy the fixed points $D_1,D_2$ collide with the B-mode. This terminates the D-family of periodic orbits, by an inverse pitchfork bifurcation which renders the B-mode unstable.    

Figure \ref{Fig.abc12} illustrates the evolution of the eccentricity vectors for all four periodic orbits $A,B,C_1,C_2$. 
In all four cases the eccentricities of both planets oscillate periodically, while the argument $\omega_2-\omega_3$ undergoes small librations around one of the values $0$ or $\pi$. 

%-------------------------------------------------------------------------------- 
\subsection{Lidov-Kozai regime}
\label{sub:Kozai}
%-------------------------------------------------------------------------------- 
We finally discuss the transition seen in the last row of Figure~\ref{Fig.sezioni}, in which the periodic orbit $C_2$ turns from stable to unstable via the \textit{Lidov-Kozai mechanism}. This is accompanied by a large volume of trajectories around $C_2$ becoming chaotic. In the case of a test inner particle ($m_2=0$) at circular orbit $\e_2=0$, the mutual inclination is a preserved quantity, equal to the inclination of the test particle $\i_{mut}=\i_2$ (since $m_2=0$ the Laplace plane coincides with the constant orbital plane of the outer particle). Furthermore, the stability character of the so-called Lidov-Kozai state (~\cite{lid1962},~\cite{koz1962},~\cite{lidzig1976},
~\cite{itooht2019}) depends only on the value of the inclination. In particular, in the quadrupolar approximation, the transition from stability to instability occurs at a critical inclination equal to $\cos^{-1}\sqrt{\frac{3}{5}}\sim 39^{\circ}.2\,$ (see~\cite{nao2016} for a review). 

Here, instead, we will use a criterion analogous to the one of the classical Lidov-Kozai mechanism in order to obtain an estimate of the critical energy $\Escr_{C,2}$ at which the orbit $C_2$ turns from stable to unstable in the framework of the quadrupolar approximation to the secular Hamiltonian $\Hscr_{sec}$ in the full three body problem, i.e., for $m_2\neq 0$. The quadrupolar Hamiltonian, apart from constants, reads
\begin{equation}
\label{Ham.quadrupolar.c}
\Hscr_{quad}=-\frac{3\,\cgrav\,m_2\,m_3}{8\,a_3}\left(\frac{a_2}{a_3}\right)^2\frac{1}{\left(1-\e_3^2\right)^{3/2}}\Fscr_{quad}\, ,
\end{equation}
where
\begin{equation}
\label{Ham.quadrupolar}
\Fscr_{quad}=-\frac{1}{3}-\frac{\e_2^2}{2}+\frac{3}{2}\,\e_2^2\,\theta^2+\theta^2+\frac{5}{2}\,\e_2^2\left(1-\theta^2\right)\cos(2\,\omega_2)\, ,
\end{equation}
with $\theta=\cos(\i_2+\i_3)\,$. Following the process of Jacobi reduction, without any book-keeping control, ammounts to expanding the cosine of the mutual inclinations as $\cos(\i_2+\i_3)=\cos(\i_2)\cos(\i_3)-\sin(\i_2)\sin(\i_3)\,$, we can use expression~\eqref{seni2i3}, that automatically cancel the dependence of the Hamiltonian on the mutual inclination. The Hamiltonian takes now the form (apart from constants):
\begin{equation}\label{hamquad}
\begin{split}
\Hscr_{quad}=&-\frac{3 \,a_2 \mathcal{G} \left(3 \,\e_2^2+2\right) \,m_3^3}{64 \,a_3^2 \left(1-\,\e_2^2\right) \sqrt{1-\,\e_3^2} \,m_2}-\frac{\,a_2^2 \mathcal{G} \left(3 \,\e_2^2+2\right) \,m_2 \,m_3}{32 \,a_3^3 \left(1-\,\e_3^2\right)^{3/2}}+\frac{3 \,a_2^3 \mathcal{G} \left(3 \,\e_2^4-\,\e_2^2-2\right) \,m_2^3}{64 \,a_3^4 \left(1-\,\e_3^2\right)^{5/2} \,m_3}\\
&+L_z^2\left(\frac{3 \,a_2 \left(3 \,\e_2^2+2\right) \,m_3}{32 \,a_3^3 \left(1-\,\e_2^2\right) \left(1-\,\e_3^2\right)^{3/2} \,m_0 \,m_2}+\frac{3 \,a_2^2 \left(3 \,\e_2^2+2\right) \,m_2}{32 \,a_3^4 \left(1-\,\e_3^2\right)^{5/2} \,m_0 \,m_3}\right)\\
&-\frac{3 \,a_2 \left(3 \,\e_2^2+2\right)\,L_z^4}{64 \,a_3^4 \mathcal{G} \left(1-\,\e_2^2\right) \left(1-\,\e_3^2\right)^{5/2} \,m_0^2 \,m_2 \,m_3}\\
&+\cos(2\,\omega_2)\Bigg[\frac{15 \,a_2 \mathcal{G} \,\e_2^2 \,m_3^3}{64 \,a_3^2 \left(1-\,\e_2^2\right) \sqrt{1-\,\e_3^2} \,m_2}-\frac{15 \,a_2^2 \mathcal{G} \,\e_2^2 \,m_2 \,m_3}{32 \,a_3^3 \left(1-\,\e_3^2\right)^{3/2}}+\frac{15 \,a_2^3 \mathcal{G} \,\e_2^2 \left(1-\e_2^2\right) \,m_2^3}{64 \,a_3^4 \left(1-\,\e_3^2\right)^{5/2} \,m_3}\\
&\phantom{+\cos(2\,\omega_2)\Bigg[}+L_z^2\left(-\frac{15 \,a_2 \,\e_2^2 \,m_3}{32 \,a_3^3 \left(1-\,\e_2^2\right) \left(1-\,\e_3^2\right)^{3/2} \,m_0 \,m_2}-\frac{15 \,a_2^2 \,\e_2^2 \,m_2}{32 \,a_3^4 \left(1-\,\e_3^2\right)^{5/2} \,m_0 \,m_3}\right)\\
&\phantom{+\cos(2\,\omega_2)\Bigg[}+\frac{15 \,a_2 \,\e_2^2\,L_z^4}{64 \,a_3^4 \mathcal{G} \left(1-\,\e_2^2\right) \left(1-\,\e_3^2\right)^{5/2} \,m_0^2 \,m_2 \,m_3}\Bigg]\, .
\end{split}
\end{equation}
It can be observed that the secular Hamiltonian at the quadrupolar level does not depend on the argument of the pericenter of the outer planet $\omega_3\,$, therefore the system is integrable (a fact known as the “happy coincidence”, see~\cite{lidzig1976} or~\cite{nao2016} for a review). In particular, the non-dependence of the Hamiltonian on $\omega_3$ implies that the eccentricity of the outer planet $\e_3$ is a conserved quantity. 

Using $\e_3$ as a parameter, the Hamiltonian $\Hscr_{quad}$ can now be regarded as a one degree of freedom system. This can obtain a polynomial form by passing to the Poincaré variables $(X_2, Y_2)$ described by~\eqref{Poincare1}. After such a substitution, the quadratic part of the above Hamiltonian is given by
\begin{equation}
\label{Ham.quad.quadrupolar}
\Hscr_{2,quad}(X_2, Y_2; G_3, L_z)=\frac{\mathfrak{a}(G_3, L_z)}{2}\, Y_2^2-\frac{\mathfrak{b}(G_3, L_z)}{2}\, X_2^2 \, ,
\end{equation}
where the coefficients $\mathfrak{b}$ and $\mathfrak{a}$ are functions of $G_3=L_3\sqrt{1-\e_3^2}=$const and of $L_z\,$; in particular
\begin{equation}
\label{ab}
\begin{split}
&\mathfrak{b}= \frac{3\, \mathcal{G}^2\, L_2^3 \,\left(3 \,G_3^2 - L_2^2 + L_z^2\right) \,m_0 \,m_3^7}{8 \,G_3^5\, L_3^3 \,m_2^3}\\
&\mathfrak{a}=-\frac{3\, \mathcal{G}^2\, L_2 \left(5\, G_3^4 - 4\, G_3^2\, L_2^2 + 3\, L_2^4 - 10 \,G_3^2\, L_z^2 - 
    8\, L_2^2\, L_z^2 + 5\, L_z^4\right) m_0\, m_3^7}{16\, G_3^5\, L_3^3\, m_2^3}\, .
\end{split}
\end{equation}
Now, in order to find the critical value of the energy $\Escr_{C,2}$, for fixed $L_z$, at which the periodic orbit $\e_3=const$, $\e_2=0$ becomes unstable,  it is sufficient to compute the eigenvalues of the Jacobian matrix of the Hamiltonian vector field:
\begin{equation}
\label{matrix.M}
M=\begin{pmatrix}
\displaystyle{\frac{\partial \dot{X_2}}{\partial X_2}} & \displaystyle{\frac{\partial \dot{X_2}}{\partial Y_2}}\\[2.5ex]
\displaystyle{\frac{\partial \dot{Y_2}}{\partial X_2}}& \displaystyle{\frac{\partial \dot{Y_2}}{\partial Y_2}}
\end{pmatrix}=
\begin{pmatrix}
\displaystyle{\frac{ \partial^2 \Hscr_{quad}}{\partial X_2\,\partial Y_2}} & \displaystyle{\frac{\partial^2 \Hscr_{quad}}{\partial^2 Y_2}}\\[2.5ex]
\displaystyle{-\frac{\partial^2 \Hscr_{quad}}{\partial^2 X_2}}& \displaystyle{-\frac{\partial^2 \Hscr_{quad}}{\partial Y_2\,\partial X_2}}
\end{pmatrix}=
\begin{pmatrix}
\displaystyle{0} & \displaystyle{\mathfrak{a}}\\
\displaystyle{\mathfrak{b}}& \displaystyle{0}
\end{pmatrix}\, .
\end{equation}
The transition occurs at a critical value of $\e_3=\e_{3,C_2}$ at which the eigenvalues of $M$ pass from imaginary to real. The critical energy is then given by $\Escr_{C,2}=\Hscr_{quad}(\e_2=0,\e_{3,C_2};L_z)$. The critical value $\e_{3,C_2}$ can be computed by the following

\begin{Proposition}
\label{prop.Kozai}
Consider the secular Hamiltonian developed up to a quadrupolar expansion. Define the quantities
\begin{equation}
\begin{split}
\label{critical.points}
&A=\frac{1}{5}\left(4L_2^2+5 L_3^2(1-\e_3^2)-L_2\sqrt{L_2^2+60 L_3^2(1-\e_3^2)}\right)\, ,\\
&B=\frac{1}{5}\left(4L_2^2+5 L_3^2(1-\e_3^2)+L_2\sqrt{L_2^2+60 L_3^2(1-\e_3^2)}\right)\, ,\\
&C=L_2^2-3 L_3^2 (1-\e_3^2)\, ,
\end{split}
\end{equation}
with $L_2=m_2\sqrt{\mathcal{G} m_0 a_2}>0\,$, $L_3=m_3\sqrt{\mathcal{G} m_0 a_3}>0\,$.
 Then, the following cases hold:\\
 \begin{description}
\item[Case i)]: $0< L_2<\sqrt{3}L_3$ $\wedge$ $0\leq \e_3\leq \sqrt{1-\frac{L_2^2}{3 L_3^2}}\,$, the periodic orbit $C_2$ is Floquet-stable if $0<L_z^2<A\,$ or $L_z^2 > B\,$ and Floquet-unstable if $A<L_z^2<B\,$.
\item[Case ii)]:
 $0< L_2<\sqrt{3}L_3$ $\wedge$ $ \sqrt{1-\frac{L_2^2}{3 L_3^2}}< \e_3 < \sqrt{1-\frac{L_2^2}{4 L_3^2}}\,$ \textbf{or} $\sqrt{3} L_3 \leq L_2< 2 L_3$ $\wedge$ $0\leq \e_3 < \sqrt{1-\frac{L_2^2}{4 L_3^2}}\,$, the periodic orbit $C_2$ is Floquet-stable if $C<L_z^2<A\,$ or $L_z^2>B\,$  and Floquet-unstable if $0<L_z^2<C\,$ or $A<L_z^2<B\,$ .
\item[Case iii)]:
 $0< L_2\leq 2 L_3$ $\wedge$ $  \e_3 = \sqrt{1-\frac{L_2^2}{4 L_3^2}}\,$ , the periodic orbit $C_2$ is Floquet-stable if $L_z^2>B\,$ and Floquet-unstable if $0<L_z^2<B\,$ .\\
\item[Case iv)]:
$0< L_2\leq 2 L_3$ $\wedge$ $ \e_3 > \sqrt{1-\frac{L_2^2}{4 L_3^2}}\,$ \textbf{or} if $ L_2> 2 L_3$ , the periodic orbit $C_2$ is Floquet-stable if $A<L_z^2<C\,$ or $L_z^2>B\,$ and Floquet-unstable if $0<L_z^2<A\,$ or $C<L_z^2<B\,$.
 \end{description}
\end{Proposition}
%\end{prop.Kozai}
Moreover, having the critical points of $L_z^2\,$, it is easy to compute the critical values for the mutual inclination, being  (by Eq.~\eqref{cosimutua})
\begin{align}
\label{i.max}
&\cos(\i_2+\i_3)=\frac{L_z^2-\Lambda_2^2-\Lambda_3^2+\Lambda_2^2\,\e_2^2+\Lambda_3^2\,\e_3^2}{2\Lambda_2\Lambda_3\sqrt{1-\e_2^2}\sqrt{1-\e_3^2}}\, , &
&\max \i_{mut}=\arccos\left(\frac{L_z^2-L_2^2-L_3^2}{2 \,L_2\, L_3}\right)\, .
\end{align}
The proof of the previous Proposition is reported in the Appendix~\ref{subappendix:proof Kozai}. We readily find that in the limit $m_2\rightarrow 0$, $e_3\rightarrow 0$, the Kozai angles $\i=39.2^\circ $ and $\i=140.77^\circ$ are recovered (see~\cite{nao2016}). In fact, for $\e_3=0\,$, the critical values $A\,$, $B$ and $C$ become
\begin{align}
\label{val.crit.circolare}
&A=\frac{1}{5}\left(4L_2^2+5 L_3^2-L_2\sqrt{L_2^2+60 L_3^2}\right)\, , &
&B=\frac{1}{5}\left(4L_2^2+5 L_3^2+L_2\sqrt{L_2^2+60 L_3^2}\right)\, , & &C=L_2^2-3\,L_3^2\, .
\end{align}
Then, at the limit $m_2\rightarrow 0\,$ we readily obtain that 
\begin{align*}
&\left(\cos i_{mut}\right)_{Lz^2=A}=\frac{A-L_2^2-L_3^2}{2 L_2 L_3}=-\frac{\sqrt{L_2^2+60\,L_3^2}+L_2}{10\,L_3}\;\; \xrightarrow{L_2\rightarrow 0}\;\; -\sqrt{\frac{3}{5}}\\
&\left(\cos i_{mut}\right)_{Lz^2=B}=\frac{B-L_2^2-L_3^2}{2 L_2 L_3}=\phantom{-}\frac{\sqrt{L_2^2+60\,L_3^2}-L_2}{10\,L_3} \;\; \xrightarrow{L_2\rightarrow 0}\;\; \phantom{-}\sqrt{\frac{3}{5}}\, .
\end{align*} 

%--------------------------------------------------------------------------------
\begin{figure}[h!]
\begin{center}
\includegraphics[width=0.3\textwidth]{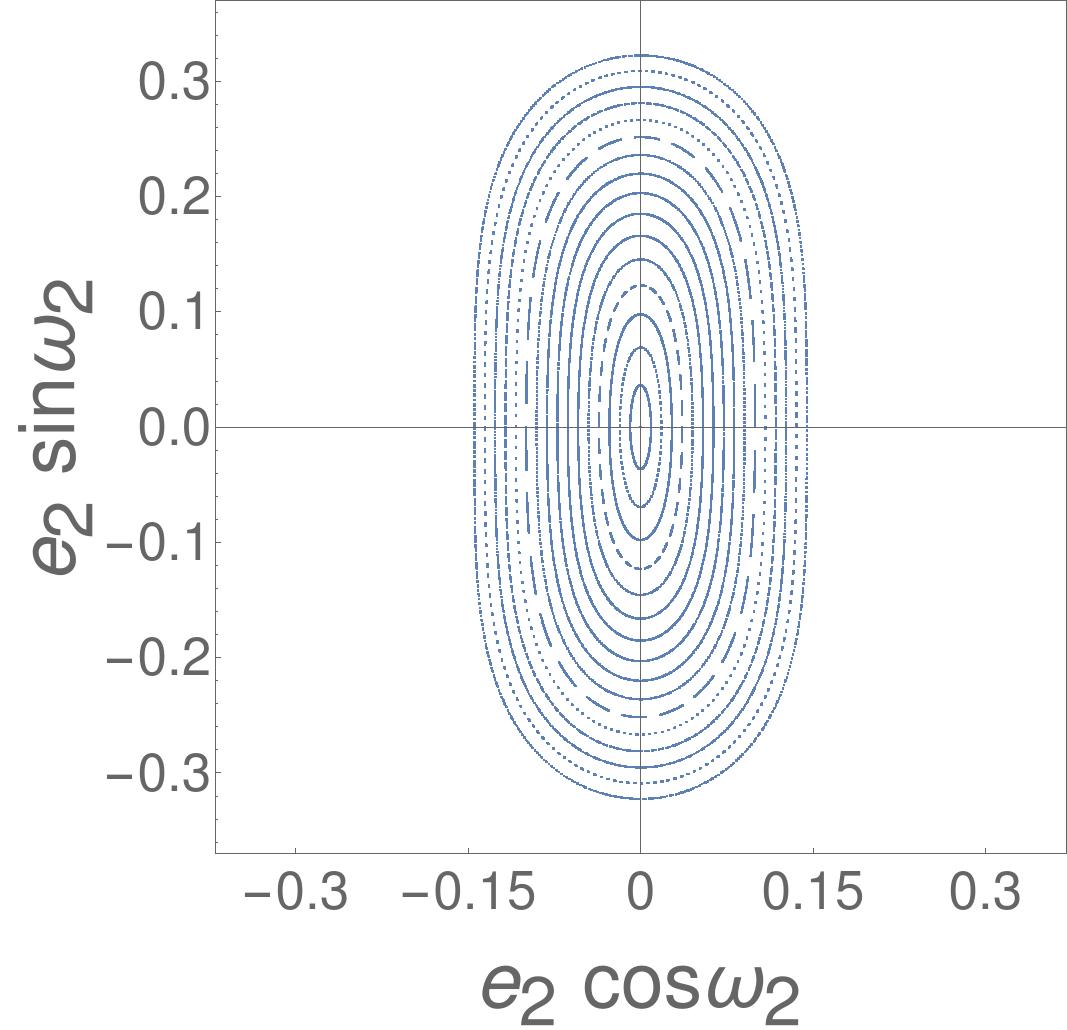}
\includegraphics[width=0.3\textwidth]{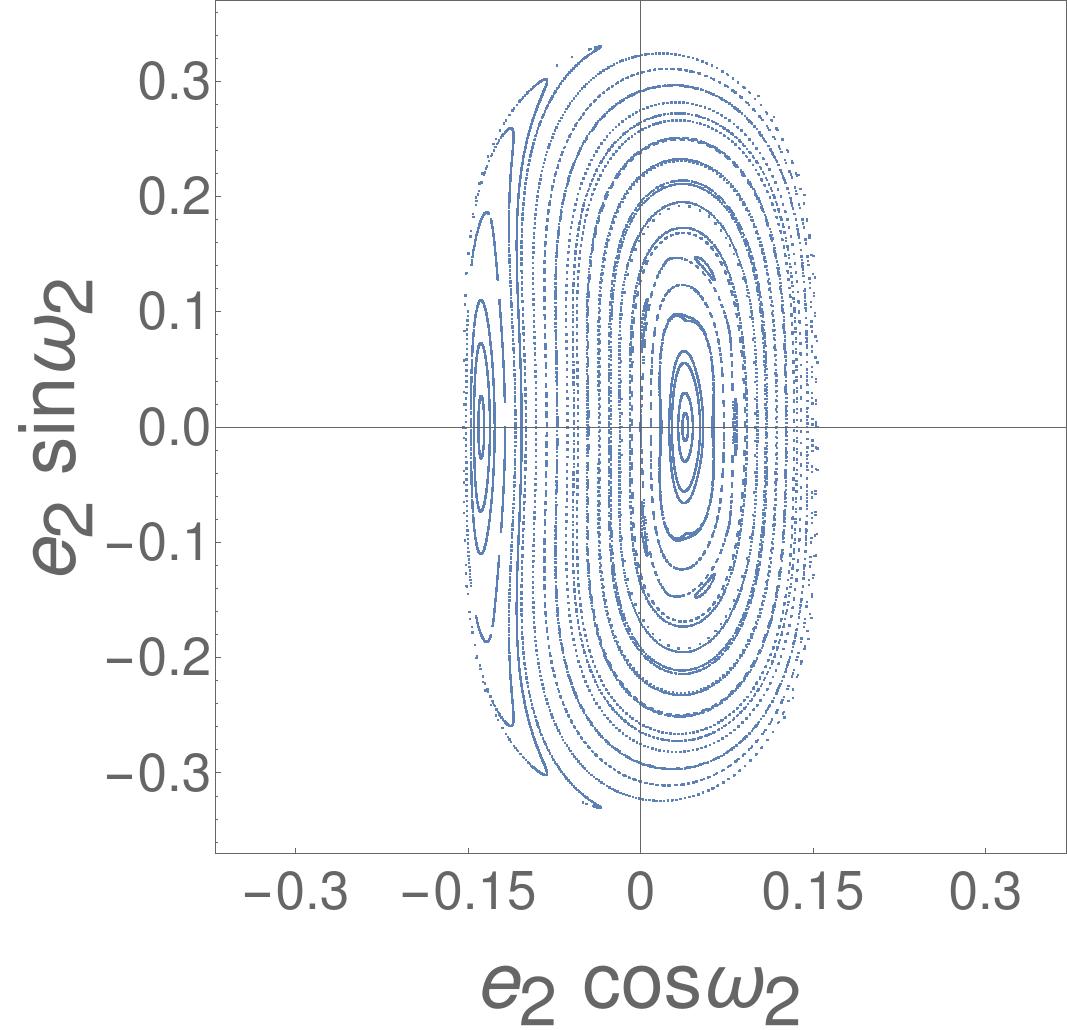}\\
\includegraphics[width=0.3\textwidth]{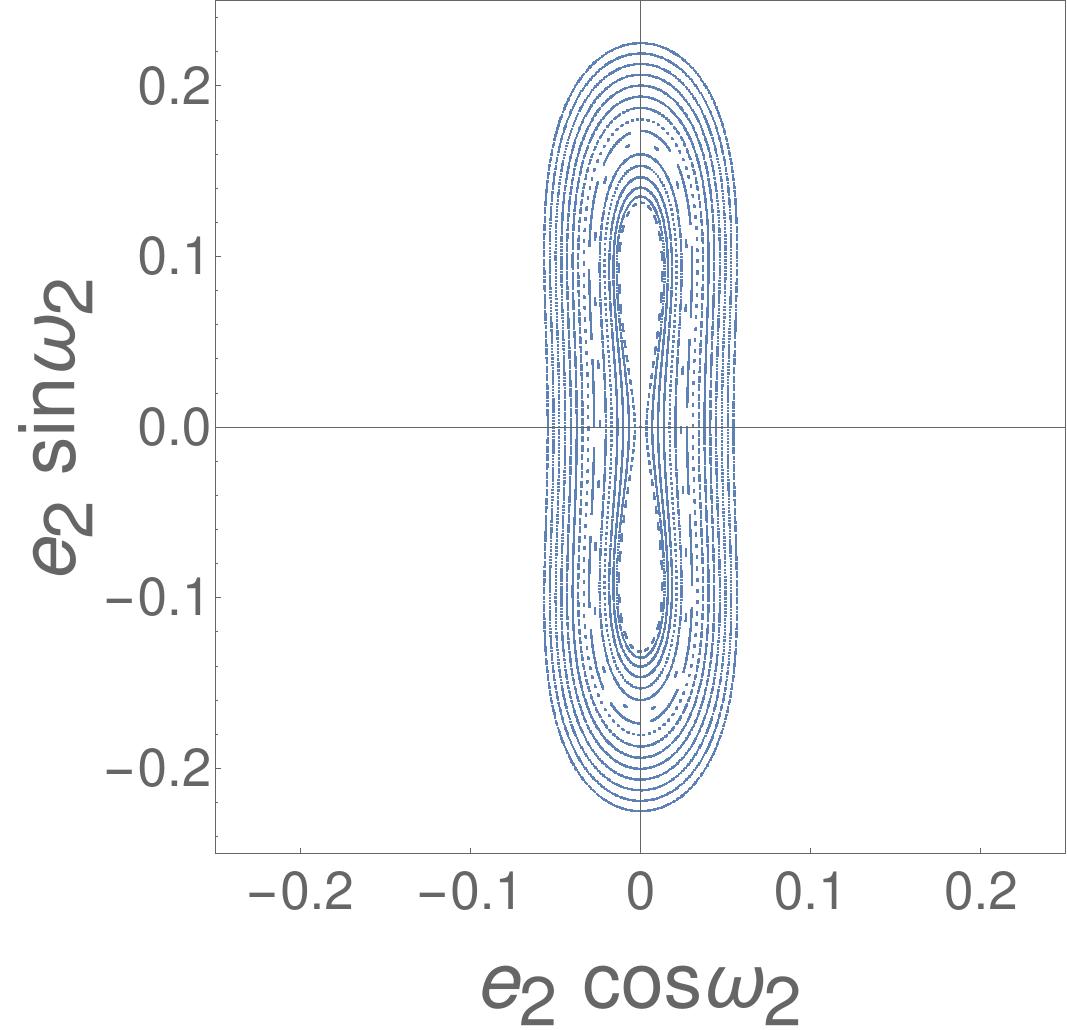}
\includegraphics[width=0.3\textwidth]{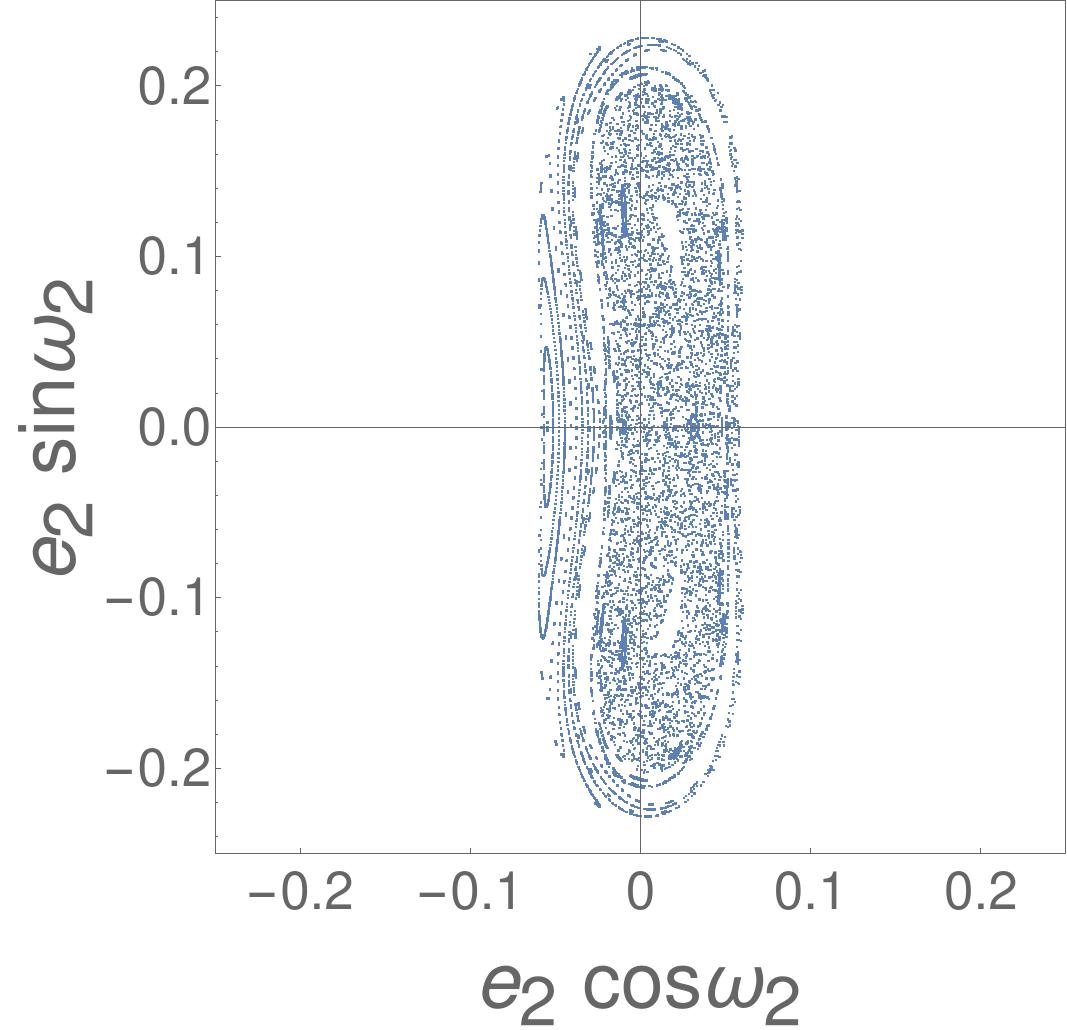}
\caption{Comparison between the phase portraits (Poincar\'e surfaces of section) obtained with the quadrupolar approximation $\Hscr_{quad}$ (left column) and the octupolar approximation $\Hscr_{oct}$ (right column), at the energy levels $\Escr=-4.9\cdot 10^{-6}$ (top), and $\Escr=-6.9\cdot 10^{-7}$ (bottom). The quadrupolar approximation yields the phase portrait of an integrable system, which contains a figure-8 separatrix for energies beyond $\Escr_{C,2}$. The octupolar approximation is necessary to obtain both periodic orbits $C_1$ and $C_2$, as well as the chaotic zone around $C_2$ caused by the overlapping of resonances in the neighborhood of this orbit.} 
\label{Fig.kozaioct}
\end{center}
\end{figure}
%--------------------------------------------------------------------------------
Note that, while the above proposition strictly establishes the limit of $\Escr_{C,2}$ only in the quadrupolar approximation, in practice we find that the estimate is quite precise when the full Hamiltonian with higher order multipoles is considered. For example, applying the above criterion in the numerical example of Fig.\ref{Fig.sezioni} yields the estimate $\Escr_{C,2}\approx -2.2\cdot 10^{-6}$, which is in agreement with the value of the energy where the transition is observed in the numerical phase portraits. On the other hand, since the quadrupolar approximation yields an integrable Hamiltonian, for any value of the energy $\Escr>\Escr_{C,2}$ $\Hscr_{quad}$ yields a figure-8 phase portraits with invariant curves surrounding the `frozen' (stable) orbits on both sides of the unstable orbit $\e_2=0$. However, this picture changes by adding just the octupolar terms to the model, given by
\begin{equation}
\Hscr_{oct}=\Hscr_{quad}+\t\Hscr_{oct}\, ,
\end{equation}
where
\begin{equation}
\label{Ham.octupolar.c}
\t\Hscr_{oct}=\frac{75\,\cgrav\,m_2\,m_3}{64\,a_3}\left(\frac{a_2}{a_3}\right)^3\frac{1}{\left(1-\e_3^2\right)^{5/2}}\,\e_2\,\e_3\,\Fscr_{oct}
\end{equation} 
with
\begin{equation}
\begin{split}
\label{Ham.octupolar}
\Fscr_{oct}\,=&\,\frac{1}{40}\left(3\,\e_2^2+4\right)\left(1+11\theta-5\theta^2-15\,\theta^3\right)\cos(\omega_2-\omega_3)\\
+&\frac{1}{40}\left(3\,\e_2^2+4\right)\left(1-11\theta-5\theta^2+15\,\theta^3\right)\cos(\omega_2+\omega_3)\\
+& \frac{7}{8}\,\e_2^2(\theta^2-1)(1+\theta)\cos(3\omega_2-\omega_3)+\frac{7}{8}\,\e_2^2(\theta^2-1)(1-\theta)\cos(3\omega_2+\omega_3)\, ,
\end{split}
\end{equation} 
with $\theta=\cos(\i_2+\i_3)\,$. Note that the above formula is equivalent, after some algebraic operations, to the formulas given in~\cite{miggoz2011} and~\cite{nao2016}, see also~\cite{litnao2011},~\cite{naoetal2013a},~\cite{foretal2000}. As depicted in Fig.\ref{Fig.kozaioct}, the adition of more harmonics besides $\cos(2\omega_2)$ via the octupole approximation implies the creation of a large domain of resonance overlap, mostly between the islands around the frozen orbits and the island of the $C_1$ orbit, which appears already in the octupole approximation. This implies, in turn, the disappearance of most quasi-periodic trajectories around the frozen orbits, and the appearance, instead, of a large domain of chaotic orbits, in accordance to what is observed in the last two panels of Fig.\ref{Fig.sezioni} for the full model. 

Finally, we note that the use of the above proposition for the determination of critical values for the transition to instability has to be done in conjunction with the test that $\cos(\max i_{mut})$ does not obtain unphysical values $|\cos(i_{mut})|>1$. Such an unphysical value can occur in \textbf{case ii)}, for the critical value $L_z^2=C\,$ and in \textbf{case iv)}, for the critical value $L_z^2=A\,$. For example, in the second case of ii), setting $\sqrt{3}L_3\leq L_2< 2\, L_3$ and $\e_3=0\,$, we find $C=L_2^2-3\,L_3^2$ which, together with $L_z^2=C\,$, leads to $\cos(\max i_{mut})=-2\,L_3/L_2<-1$ iff $L_2< 2\,L_3$. Similarly, in the second case of iv), where $L_2>2\,L_3$, setting $\e_3=0\,$, by the relation $L_z^2=A=\frac{1}{5}\left(4L_2^2+5 L_3^2-L_2\sqrt{L_2^2+60 L_3^2}\right)\,$, we obtain $\cos(\max i_{mut})=-(\sqrt{L_2+60\,L_3^2}+L_2)/(10\,L_3)$, which is physical only iff $L_2\leq 2\,L_3\,$.

%-------------------------------------------------------------------------------
\subsection{Where does the $\upsilon-$Andromedae system lie in phase space?}
%-------------------------------------------------------------------------------

%--------------------------------------------------------------------------------
\begin{figure}[h!]
\begin{center}
\includegraphics[width=0.4\textwidth]{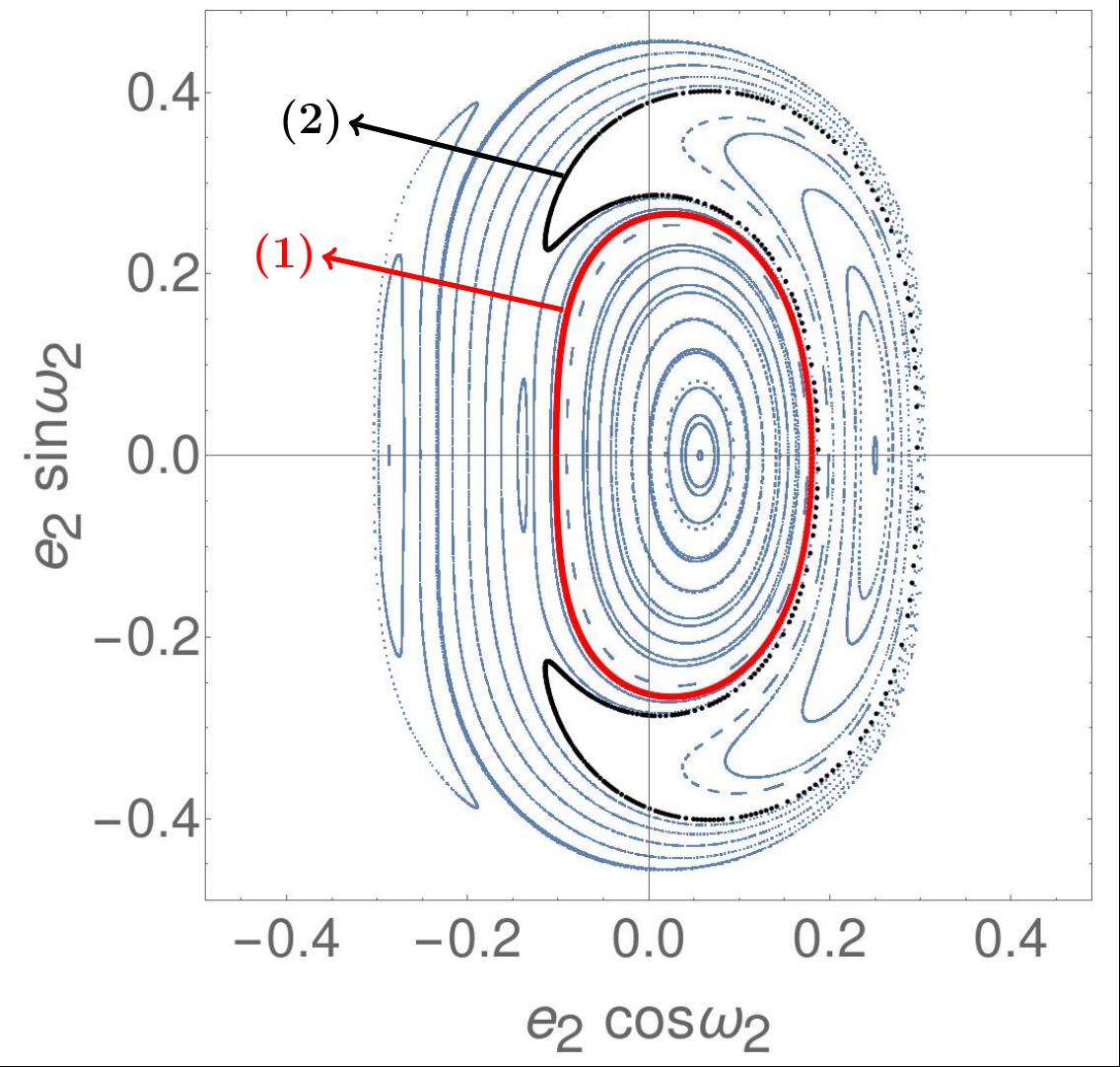}
\caption{Poincar\'e surface of section for the energy $\Escr=-2.081\cdot 10^{-5}$. The red curve (1) shows the orbit obtained by adopting our basic assumption as regards the initial conditions of the $\upsilon-$Andromedae system (see subsection \ref{subsub:Poincare.precision}), i.e. (after reduction to the Laplace plane) $\e_2=0.2445$, $\e_3=0.316$, $\omega_2=289.049^\circ$, $\omega_3=235.464^\circ$. The black curve (2) shows the trajectory obtained by changing by only about 10\% the eccentricity $\e_2$ and argument of the periastron $\omega_2$, while maintaining the energy and argument of periastron of the outer planet invariant ($\e_2=0.269$, $\omega_2=296.1^\circ$, $\omega_3=235.464^\circ$, $\e_3=0.299$).} 
\label{Fig.ups}
\end{center}
\end{figure}
%--------------------------------------------------------------------------------
We finally comment on the implications of the above analysis of the phase space structure on the interpretation of the data given by astronomical observations. Figure \ref{Fig.ups} shows the phase portrait (surface of section) for the value of the energy $\Escr=-2.081\cdot 10^{-5}$, corresponding to the one found after reducing to the Laplace plane the data for the observed $\upsilon$-Andromedae system (see subsection \ref{subsub:Poincare.precision}). The invariant curve (1) (marked in red) corresponds to the actual initial conditions of the system. According to the figure, this is a quasi-periodic orbit surrounding the family $C_2$. We note, however, that the orbit is very close to the separatrix between the $C_2$ librational domain and the domain of stability surrounding the fixed point of the mode B orbit. As a consequence, a change of the initial conditions $(\e_2,\omega_2)$ by less than 10\% results in a trajectory ((2), black curve) which undergoes quasi-periodic oscillations around the (aligned) apsidal corotation orbit of the system. The proximity of the real trajectory to one of the separatrices generated by the families $C_1,\,C_2$ is noteworthy, as it is in contrast with the basic scenario of evolution of planetary orbits, according to which the dissipative phase of planetary migration should lead to an endstate close to one of the stable fixed points of the system (see, for example, \cite{beaetal2012}).

%%%%%%%%%%%%%%%%%%%%%%%%%%%%%%%%%%%%%%%%%%%%%%%%%%%%%%%%%%%
\section{Parametric study}
\label{sec:parametric}
%%%%%%%%%%%%%%%%%%%%%%%%%%%%%%%%%%%%%%%%%%%%%%%%%%%%%%%%%%%

%--------------------------------------------------------------------------------
\begin{figure}
\begin{center}
\includegraphics[width=0.85\textwidth]{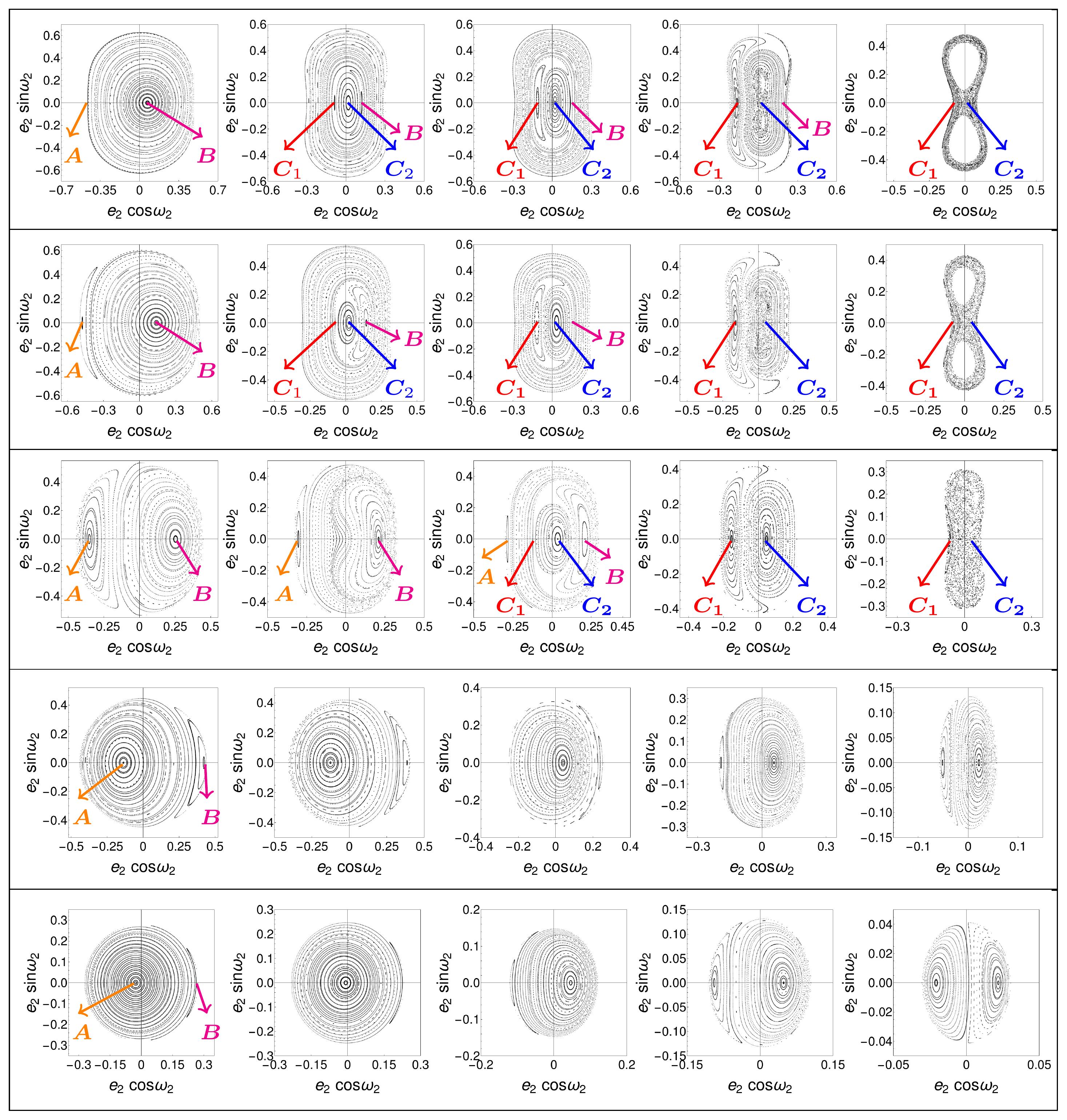}
\caption{Poincar\'{e} surfaces of section in the plane $(\e_2 \cos(\omega_2 ), \e_2 \sin(\omega_2 ))$ with $L_z$ fixed such that max$\,i_{mut}$ is $45^\circ\,$ and
different values of energy. We consider the models $a_2/a_3=1/3$ and (from top to bottom) $m_2/m_3=1/10,\,1/3,\,1,\,3,\,10\,$. The values of the energy (from top to bottom) are (from left to right), top $\Escr=-2.71\cdot 10^{-6}, -1.01\cdot 10^{-6}, -9.24\cdot 10^{-7},-5.73\cdot 10^{-7}, -8.23\cdot 10^{-8} $, $\Escr=-9.03\cdot 10^{-7},-3.37\cdot 10^{-7},-2.87\cdot 10^{-7}, -1.76\cdot 10^{-7},  -2.35\cdot 10^{-8} $, $\Escr=-2.68\cdot 10^{-7},-1.39\cdot 10^{-7},-1.14\cdot 10^{-7},-6.37\cdot 10^{-8}, -6.98\cdot 10^{-9}  $, $\Escr=-1.12\cdot 10^{-6},-9.26\cdot 10^{-7},-3.37\cdot 10^{-7}, -2.08\cdot 10^{-7},-1.67\cdot 10^{-8}  $, bottom $\Escr= -2.75\cdot 10^{-6}, -1.91\cdot 10^{-6},-4.72\cdot 10^{-7},-3.54\cdot 10^{-7}, -2.91\cdot 10^{-8}\,$. The surfaces of section have been computed by a numerical integration of trajectories in a Hamiltonian averaged in closed form with a multipolar expansion truncated at degree $6$  and expanded up to order $12$ in the eccentricities.} 
\label{Fig.sezioni.ger:13}
\end{center}
\end{figure}
%--------------------------------------------------------------------------------
The purpose of the present section is to provide an overview of how the form of the phase portraits (Poincar\'e sections) is varied with the energy $\Escr$, and hence with an increase in the level of mutual inclination, in various types of planetary systems differing from our so far main example as regards the choice of mass ratio $m_2/m_3$ as well as of semi-major axis ratio $a_2/a_3$ between the two planets. A thorough parametric study of the latter question is beyond our present scope. Instead, here we focus on only one central aspect of this study, namely the question of how generic is the description, as in section \ref{sec:dynamics}, of the transition from the planar-like to the Lidov Kozai regime, when either the mass ratio $m_2/m_3$ or the semi-major axis ratio $a_2/a_3$ are altered, including the so-called \textit{hierarchical} limits, which correspond to mass ratio limits $m_2/m_3\rightarrow 0$ (restricted three-body problem with the inner planet being a test particle), or $m_2/m_3\rightarrow\infty$ (the outer planet is a test particle), and semi-major axis ratio $a_2/a_3<<1$ (outer planet way further from the star than inner planet). To produce a suite of numerical experiments covering most cases of practical interest, we consider in the sequel five different values of the mass ratio $m_2/m_3$ representative of all possible mass hierarchies (or lack thereof) namely $m_2/m_3=1/10,\,1/3,\,1,\,3,\,10$, produced by the corresponding combination of the masses $m_0=1M_\odot\,$ and $1\,M_{J}\,$, $3\,M_{J}$ or $10\,M_{J}$ for $m_2$ and $m_3\,$, while we consider the semimajor axes ratio $a_2/a_3=1/7$ ($a_2=1\,AU$, $a_3=7\,AU$) as a case representing distance hierarchy, or $a_2/a_3=1/3$ ($a_2=1\,AU$, $a_3=3\,AU$) as a case of no distance hierarchy. In each one of the previous cases, a comparison with the analysis of section \ref{sec:dynamics} requires keeping fixed the value of the $\AMD$ (or $L_z$) while computing phase portraits with a varying value of the energy $\Escr$. To find a relevant value for $L_z$ in each experiment, we consider a uniform value of the maximum possible mutual inclination $\i_{max}$ (Eq.\ref{imutuamax}) in all the experiments, set as $\i_{mut}=45^\circ$. This typically proves to be slightly above the limit of the Lidov-Kozai instability (which is equal to $39.2^\circ\,$ in the quadrupolar approximation of the restricted three body problem with $m_2=0$ and turns to be about $\sim 42.5^\circ\,$ in our main numerical example of section \ref{sec:dynamics}). By Eq.(\ref{imutuamax}), this implies setting
$$
L_z=\sqrt{\Lambda_2^2+\Lambda_3^2+\sqrt{2}\Lambda_2\Lambda_3}
$$
with $\Lambda_2,\Lambda_3$ specified by the choice of masses and semi-major axes in each experiment. Finally, in order to get some safety from any accuracy issues, we make computations with a Hamiltonian expanded at slightly larger orders with respect to those of the main numerical example of the previous section, i.e. multipole order 6, and order 12 in the eccentricities. To arrive at the final secular model, all Hamiltonians are processed, and their phase portraits computed, in the same way as in section \ref{sec:hampc}.   

Figure ~\ref{Fig.sezioni.ger:13} shows the phase portraits obtained for increasing energy $\Escr$ in the case $a_2/a_3=1/3$, and in various mass hierarchy scenarios, namely, for the values $m_2/m_3$ (from top to bottom row) $1/10$, $1/3$, 
$1/1$, $3/1$, $10/1$.  The values of the energies whose corresponding phase portraits are shown are selected in each panel of Fig.\ref{Fig.sezioni.ger:13} so as to visualize the most important changes observed in the phase-space structure as the energy increases. 

Our main numerical example (Fig.\ref{Fig.sezioni}) exhibits a similar structural change in the phase portraits, with increasing energy, as in the fully non-hierarchical case $m_2/m_3=1$ illustrated in the third row of Fig.\ref{Fig.sezioni.ger:13}. However, in Fig.\ref{Fig.sezioni.ger:13} it is clear that the transition from the planar-like regime, where the modes A and B dominate the phase space, through the saddle-node bifurcation giving rise to the orbits $C_1$ and $C_2$ (transition regime), and, finally, the Lidov-Kozai regime (orbit $C_2$ becomes unstable and surrounded by a chaotic figure-8 separatrix-like layer) is generic in all cases $m_2/m_3\leq 1$ (top three rows of Fig.\ref{Fig.sezioni.ger:13}) and follows by the same sequence of bifurcations. As the ratio $m_2/m_3$ tends to small values, the main differences observed with respect to the non-hierarchical case are: i) in the planar like regime (i.e. top row, first panel), the domain of quasi-periodic orbits around the aligned apsidal corotation orbit (B-mode) occupies most of the available phase space, and ii) at energies beyond the one of the Lidov-Kozai instability, there is a significant domain of regular orbits obtained around two stable periodic orbits of non-zero eccentricity $\e_2$ with fixed points along the axes $\omega_2=\pm\pi/2$ in the surface of section (the so-called `frozen orbits'). Both properties (i) and (ii) can be interpreted by analogy with the restricted three-body problem $(m_2=0)$, in which, for low inclinations, the only fixed point of the quadratic (Laplace-Lagrange) secular Hamiltonian is associated with an eccentric orbit of eccentricity $\e_2$ equal to the forced value induced by the outer perturber and pericenter aligned with the one of the outer perturber (see \cite{fer2007}). On the other hand, in the cases $m_2/m_3>1$ we do not observe at all the transition to the Lidov-Kozai regime. Again, this can be interpreted in analogy with the RTBP, in which the outer Lidov-Kozai instability ($m_3\rightarrow 0$) occurs at an inclination ($\approx 63^\circ$; see \cite{zei1910}) much higher than the maximum mutual inclination $\i_{max}=45^\circ$ considered in our examples. The case of a very high mutual inclination has been considered in the literature (see for example~\cite{hannao2020} and references therein), but it appears of rather theoretical interest compared to available observations on the orbital inclinations in exoplanetary systems.            

%--------------------------------------------------------------------------------
\begin{figure}
\begin{center}
\includegraphics[width=0.85\textwidth]{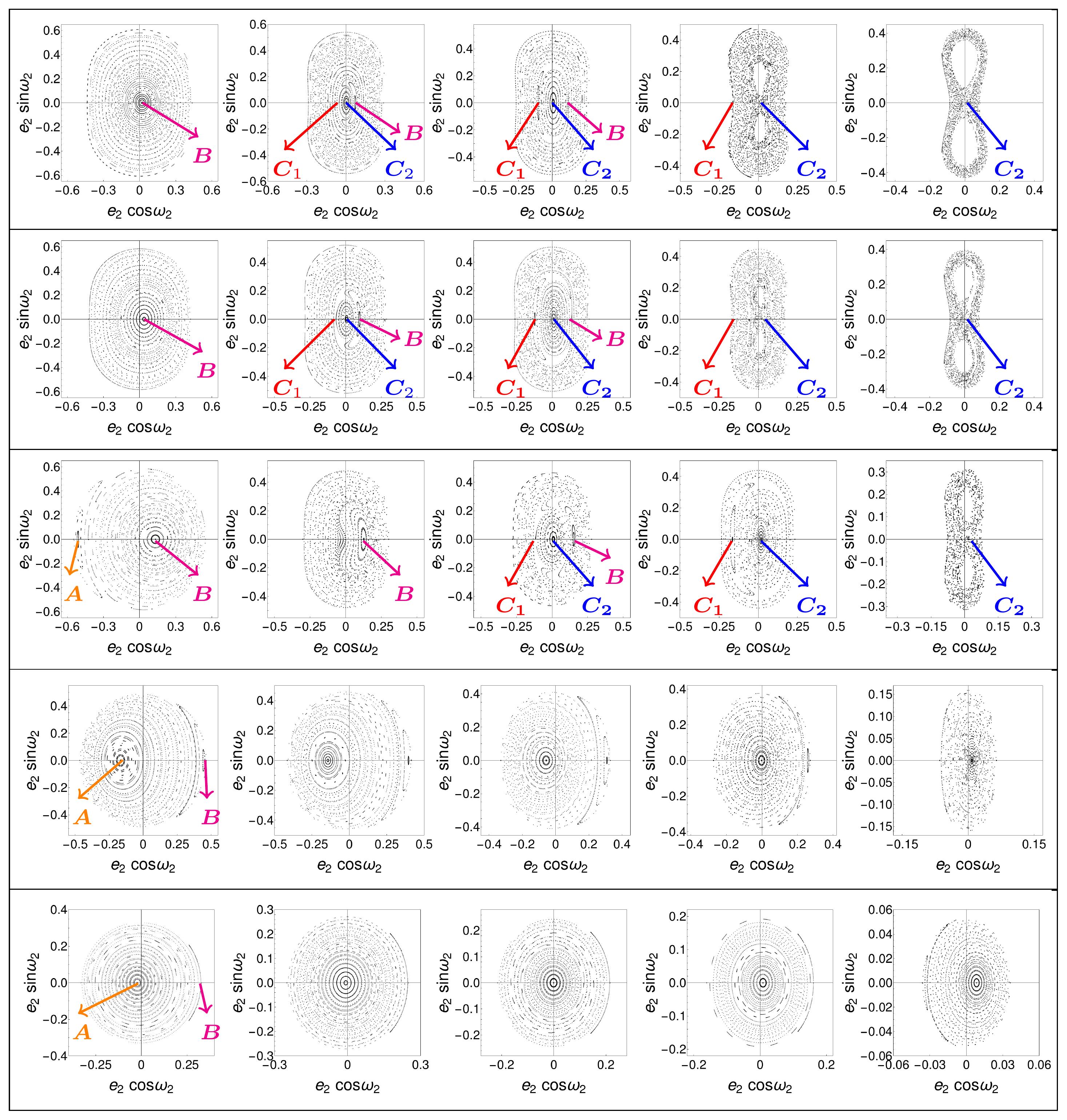}
\caption{Poincar\'{e} surfaces of section in the plane $(\e_2 \cos(\omega_2 ), \e_2 \sin(\omega_2 ))$ with $L_z$ fixed such that max$\,i_{mut}$ is $45^\circ\,$ and different values of energy. We consider the models $a_2/a_3=1/7$ and (from top to bottom) $m_2/m_3=1/10,\,1/3,\,1,\,3,\,10\,$. The values of the energy (from top to bottom) are (from left to right), top $\Escr=-1.54\cdot 10^{-7},-6.71\cdot 10^{-8},-5.78\cdot 10^{-8},-2.46\cdot 10^{-8}, -6.15\cdot 10^{-9} $, $\Escr=-4.5\cdot 10^{-8},-2.05\cdot 10^{-8}, -1.7\cdot 10^{-8},-7.16\cdot 10^{-9}, -1.79\cdot 10^{-9} $, $\Escr=-2.75\cdot 10^{-8},-8.25\cdot 10^{-9},-7.01\cdot 10^{-9},-4.97\cdot 10^{-9},  -5.51\cdot 10^{-10} $, $\Escr=-7.5\cdot 10^{-8},-5.94\cdot 10^{-8},-3.69\cdot 10^{-8}, -2.5\cdot 10^{-8},  -1.38\cdot 10^{-9} $, bottom $\Escr=-2.22\cdot 10^{-7},-1.21\cdot 10^{-7},-8.96\cdot 10^{-8},-4.91\cdot 10^{-8}, -2.42\cdot 10^{-9}\,$. The surfaces of section have been computed by a numerical integration of trajectories in a Hamiltonian averaged in closed form with a multipolar expansion truncated at degree $6$  and expanded up to order $12$ in the eccentricities.} 
\label{Fig.sezioni.ger:17}
\end{center}
\end{figure}
%--------------------------------------------------------------------------------
Fig.\ref{Fig.sezioni.ger:17}, now, shows how the phase portraits evolve with the energy when, in addition to altering the mass ratio $m_2/m_3$, the limit of distance hierarchy is also approached ($a_2/a_3=1/7$). As an overall observation, we note again that the full sequence of bifurcations leading from the planar-like to the Lidov-Kozai regime is realized only in the cases $m_2/m_3\leq 1$. However, the main difference, with respect to the case $a_2/a_3=1/3$, is that in the hierarchical in distance case the phase portraits contain many more regular orbits, with a cosiderably large domain of stability around the frozen orbits surviving the system's perturbations even at the mass ratio $m_2/m_3=1$ (third row of Fig.\ref{Fig.sezioni.ger:17}. This can be interpreted by the fact that the distance hierarchy brings the system closer to the dynamics of the integrable Hamiltonian $\Hscr_{quad}$ (Eq.\eqref{hamquad}), reducing the relative importance of perturbations including and beyond the octupolar one. 

%%%%%%%%%%%%%%%%%%%%%%%%%%%%%%%%%%%%%%%%%%%%%%%%%%%%%%%%%%%%%%%%%%%%%%%%%%%
\section{Concluding remarks}
\label{sec:conclusions}
%%%%%%%%%%%%%%%%%%%%%%%%%%%%%%%%%%%%%%%%%%%%%%%%%%%%%%%%%%%%%%%%%%%%%%%%%%%
As a main concluding remark on what was exposed in sections \ref{sec:hampc} to \ref{sec:parametric} above, we wish to point out again the twofold purpose of the present paper. On one hand, we aim to establish a convenient \textit{formalism} for the study of the 3D planetary three body problem, allowing for a straightforward classification, visualization, and semi-analytical study of the most important phenomena appearing in the phase space of secular motions as the mutual inclination between the planetary orbits increases from zero to a value ($\sim 45^\circ$) beyond the onset of the Lidov-Kozai instability. On the other, we aim to provide ourselves a study of some of these phenomena in some detail. In particular, we focus on the connection between a chain of major transitions observed in the phase space structure as the mutual inclination increases, and the associated bifurcations (and/or the change of stability character) of the most important periodic orbits of the problem. In physical terms, the hereby proposed formalism serves to analyze the gradual transition of the system between two well distinct regimes. One of them is a nearly-integrable regime, called above `planar-like', dominated by periodic orbits (A and B modes) analogous to the apsidal corotations of the planar case, as well as their surrounding quasi-periodic orbits. The second is the regime called above `Lidov-Kozai'. This is characterized by the highly inclined periodic orbits emerging in the octupolar approximation (orbits $C_1$ and $C_2$), one of which ($C_2$) undergoes the well known Lidov-Kozai transition from stable to unstable beyond a critical value of the mutual inclination. 

As also emphasized in the introduction, our presently proposed formalism is strongly motivated by the need to provide a unifying conceptual framework in which to understand and put into context several results, distinct or partly overlapping with those presently reported, obtained in past literature by use of quite diverse methods and representations of the problem under exam. The following is a summary of our main conclusions, serving also as a navigation guide to the main methodological steps proposed in the preceeding sections of the paper.\\
\\
\noindent
i) Section \ref{sec:hampc} exposed the methodology for obtaining a suitable secular Hamiltonian model, as well as our choice of variables and representation for phase portraits, aiming to present in a unified way both the planar-like regime, where apsidal corotation orbits dominate, and the Lidov-Kozai regime, where nearly circular highly inclined orbits dominate. The use of a particular book-keeping method was explained in subsection \ref{sub:bookkeeping}, leading, after Jacobi reduction, to a Hamiltonian decomposition of the form $\Hscr_{sec}=\Hscr_{planar}+\Hscr_{space}$, where the term $H_{planar}$ is integrable and the term $\Hscr_{space}$ depends on the system's $\AMD$ (or, equivalently, the value of the system's angular momentum $L_z$). We argue on the merits stemming from a choice of simple definition for the Poincar\'e surface section as in \ref{subsub:Poincare.section}, allowing for a straightforward mapping of phase portraits to the evolution of the corresponding orbits in terms of Keplerian elements. This is accompanied by a detailed analysis of the permissible domains of initial conditions on the surface of section as the energy $\Escr$ is altered. It is demonstrated that these limits are defined by the intersections between any two of three important manifolds, namely the manifold of constant energy $\mathcal{M}(\Escr)$, the manifold of zero mutual inclination ${\cal{I}}_{0}$, and the condition of tangency of an orbit with the surface of section ($\dot{Y}_3=0$ in the Poincar\'{e} variables (\ref{Poincare1})). In particular, we explain how the tangencies between $\mathcal{M}(\Escr)$ and ${\cal{I}}_0$ at two particular values of the energy $\Escr_A\,(=\Escr_{min})$, $\Escr_B\,(=\Escr_{2,3})$ mark the birth of the planar anti-aligned and aligned apsidal corotation orbits respectively. Finally, we test the precision of various models defined either by a closed-form scissors-averaged model obtained from a multipole expansion of the Hamiltonian truncated at degree $N_P$, or by the usual (Jacobi-reduced) Laplace-Lagrange expansion. In both the previous cases, we introduce an additional `book-keeping' order of truncation $N_{bk}$, with a book-keeping collecting at the same time powers of the eccentricities, of the mutual inclination and of the system's $\AMD$. As discussed in subsection \ref{subsub:Poincare.precision}, we find that, in general, a multipole expansion with $N_P=5$ or $6$ is sufficient to represent even non-hierarchical systems with semi-major axis ratio between the two planets $a_2/a_3=1/3$ or smaller. The multipole model has the salient features of being easy to scissors-average in closed form, while its lowest order truncations (quadrupolar and octupolar) yield formulas sufficiently small to offer analytical insight into the results. Comparison with the Laplace-Lagrange model (harder to obtain) shows the practical equivalence of the phase portraits at $N_P=5$ and with order of expansion as high as $N_{bk}=10$. On the other hand, obtaining the canonical transformation corresponding to the averaging, or going to second order in the masses, seems quite harder to achieve with the closed-form approach.\\
\\
\noindent
ii) Section \ref{sec:dynamics} gives a qualitative overview of the main mechanism behind the transition from the `planar-like' to the `Lidov-Kozai' regime, i.e. the birth of the families $C_1,C_2$ via a saddle-node bifurcation around the aligned AC (mode B), accompanied by the transition of the family $C_1$ from unstable to stable via a pitchfork bifurcation, and, finally, the de-stabilization of the B-mode via an inverse pitchfork transformation. These phenomena are described qualitatively via the corresponding phase portraits, i.e., surfaces of section obtained numerically in a basic example by increasing the value of the energy $\Escr$. We demonstrate how this latter increase is connected to the overall rise of the level of mutual inclination of the orbits (Fig.\ref{Fig.incmutene}). Finally, we numerically observe the Lidov-Kozai instability in the same phase portraits which turns the family $C_2$ from stable to unstable at a critical energy $\Escr_{C,2}$. \\
\\
\noindent
iii) Based on the form of the phase portraits, we identify three basic regimes in which a two-planet system with inclined orbits can be found: 

- the `planar-like' regime is characterized by a phase-portrait similar to the one of an integrable model $\Hscr_{int}$ subsection \ref{sub:nearlyplanar}, which is the 3D analogue of the integrable model $\Hscr_{planar}$. As pointed out in past literature (see references in subsection~\ref{subsub:hopf}) this is a model whose natural phase space is the two-sphere, represented by the Hopf variables of Eq.(\ref{def.Weyl.Delau}). We first implement a tangency method (~\cite{cusbat1997},~\cite{paletal2006},~\cite{marpuc2016}) to locate the periodic orbits of the integrable secular model. Then, by a normal form approach, we demonstrate how the perturbations to the integrable model can be normalized and absorbed in the definition of the periodic orbits corresponding to the apsidal corotations in the non-planar case. The main conclusion is that both the A (anti-aligned) and B (aligned) modes of the integrable model turn into orbits whose eccentricities are not exactly constant, but undergo periodic oscillations with a secular period easy to compute semi-analytically. At the same time, the pericenters of the planets in either mode do not remain locked exactly to the alignment or anti-alignment condition. Instead, the difference $\omega_2-\omega_3$ also undergoes oscillations around the central value (0 for anti-alignment, $\pi$ for alignment). The amplitude of these oscillations grows with the energy, and hence the level of mutual inclination,  where each of these periodic orbits is computed.

- the `transition regime', starting with the birth of the families $C_1$, $C_2$ and ending with the termination of the  intermediate ($D_{1,2}$) family at a collision with the B-mode rendering the latter unstable

- the`Lidov-Kozai' regime, in which the structure of the phase space is similar to the one provided by the octupolar model. As well known in literature, the turning of the orbit $C_2$ from stable to unstable can be predicted analytically in the framework of the (integrable) quadrupolar approximation. We adapt this prediction to a proposition (subsection \ref{sub:Kozai}) allowing to compute approximately the value of the transition energy $\Escr_{C,2}$ and of the corresponding critical inclination for a system with fixed $\AMD$. Actually, a careful investigation of the same proposition shows that it can be applied also independently of the choice of parameter, i.e., using the outer planet's eccentricity (which is constant in the context of the quadrupolar model) as well as altering the value of the systems $\AMD$ (or angular momentum). At any rate, the result obtained by such an analysis is only approximate, at least for non-hierarchical systems, since the octupolar terms introduce a significant perturbation to the quadrupolar model. It is noteworthy that the effect of the octupolar terms can be locally absorbed by a perturbation theory eliminating all harmonics except for $\cos(2\omega_2)$ (attempts to provide such a theory in the framework of the restricted three body problem are nearly as old as the subject itself, see \cite{itooht2019} for a review). We are aware of no work attempting to provide the thresholds for the onset of the Lidov-Kozai instability in the framwework of the general three-body problem at truncations $N_P\geq 3$ by analytical means (normal forms), thus, this is proposed as a subject for further study. 

iv) We give an overview of how the picture of the phase-space structure and dynamics depicted as above applies to systems in which we have either a mass hierarchy, or a distance hierarchy. As regards models with mass hierarchy, we give numerical evidence that a necessary condition to obtain the full chain of transitions from the `planar-like' to the `Lidov-Kozai' regime is that $m_2$ be smaller than $m_3$. This is justified by the fact that the critical mutual inclination for obtaining the Lidov-Kozai instability is $\approx 39^\circ$ in the limit $m_2<<m_3$, but it raises, instead, to $\approx 63^\circ$ in the limit $m_3<<m_2$. Such a high value of the mutual inclination is rather improbable in exoplanetary systems. On the other hand, a quick comparison of Figs.\ref{Fig.incmutene} and \ref{Fig.sezioni} shows that the so-called `planar-like' regime corresponds to a dynamical characterization (or, simply, the similarity of phase portraits with the planar case) rather than a really small value of the mutual inclination. For example, in our basic model the planar-like regime persists up to the energy $\Escr_C$ of birth of the $C_1,C_2$ families, which corresponds to trajectories whose mutual inclination can be as high as $35^\circ$. On the other hand, as can be deduced from Fig.\ref{Fig.sezioni.ger:17}, in models with distance hierarchy we arrive rather quickly to forms of the phase portraits reminiscent of those obtained in the Lidov-Kozai regime, i.e., with the $C_1$ and $C_2$ orbits rather, than the mode A and B orbits, dominating the dynamics. Insight to this phenomenon is provided by the fact that the octupolar model gives a far better approximation to the full Hamiltonian in distance-hierarchical cases. In that model, however, the harmonics $\cos(\omega_2-\omega_3)$, which is the basis for the appearance of apsidal corotation orbits, has smaller amplitude than the harmonics $\cos(2\omega_2)$ (already present at quadrupolar level). This implies that the dominant periodic orbits are $C_1$ and $C_2$, as well as the frozen orbits arising after the Lidov-Kozai instability. Note that the orbits $C_1$ and $C_2$ have aligned and anti-aligned apsides, while, by the position of the corrsponding fixed points in the phase portraits, we readily deduce that they have smaller eccentricities than the one of the A mode (which remains stable at all values of the energy). Thus, once again, the difference between these periodic orbits is dynamical rather than morphological, i.e., they play distinct roles in shaping the phase space locally around them. \\
\\
{\bf Acknowledgements:} C.E. acknowledges the support of MIUR-PRIN 20178CJA2B ‘New frontiers of Celestial Mechanics: theory and applications’.

%%%%%%%%%%%%%%%%%%%%%%%%%%%%%%%%%%%%%%%%%%%%%%%%%%%%%%%%%%%%%%%%%%%%%%%%%%
\bibliographystyle{abbrvnat}
%\nocite{*}
\bibliography{bibliography}
%%%%%%%%%%%%%%%%%%%%%%%%%%%%%%%%%%%%%%%%%%%%%%%%%%%%%%%%%%%%%%%%%%%%%%%%%%

\clearpage

\noindent
{\Large\bf Appendix}

\appendix

%--------------------------------------------------------------------
\section{Laplace coefficients}
%--------------------------------------------------------------------
\label{appendix:coeff Lapl}
In this section we show how to compute the Laplace coefficients (calculated in Subsection~\ref{subsub:Poincare.precision} numerically through equations~\eqref{Lapl.coeff}) via a multipolar expansion. We can state the following
\newtheorem{Lapl.coeff}{Lemma}
\setcounter{Lapl.coeff}{0}
\label{Lapl.coeff.lemma}
\begin{Lapl.coeff}
The Laplace coefficients, allowing to compute the direct part of the Hamiltonian~\eqref{ham.3BP}, i.e. (see~\eqref{espans.flam})
$$ \frac{1}{\left(a_2^2 + a_3^2 - 2 \,a_2\,a_3\, \cos(\lambda_2 - \lambda_3)\right)^{\frac{2\,s+1}{2}}}=a_3^{-(2\,s+1)}\sum _{j\geq 0} b_{s+\frac{1}{2}}^{(j)}\left(\alpha\right)\cos(j(\lambda_2-\lambda_3))\, ,$$
can be computed as
\begin{equation}
\label{Lapl.coeff.analitic}
\begin{split}
b_{s+\frac{1}{2}}^{(0)}\left(\alpha\right)&= 1+\sum_{k=1}^{+\infty}(-1)^{k}\, \frac{\frac{l}{2}(\frac{l}{2}+1)\ldots(\frac{l}{2}+k-1)}{k!}\,\alpha^{2k}\\
&\phantom{=}+\sum_{h\in\Ascr_2}\sum_{k=h}^{\infty}(-1)^{k-h} \frac{(s+\frac{1}{2})(s+\frac{3}{2})\ldots(s-\frac{1}{2}+k)}{\left(\frac{h}{2}\right)!\left(\frac{h}{2}\right)!(k-h)!}\alpha^{2k-h}\, ,\\
b_{s+\frac{1}{2}}^{(j)}\left(\alpha\right)&=\sum_{h\in\Ascr_j}\sum_{k=h}^{\infty}2(-1)^{k-h} \frac{(s+\frac{1}{2})(s+\frac{3}{2})\ldots(s-\frac{1}{2}+k)}{\left(\frac{h-j}{2}\right)!\left(\frac{h+j}{2}\right)!(k-h)!}\alpha^{2k-h}\, ,\quad j\geq 1\, ,
\end{split}
\end{equation}
where $\alpha=a_2/a_3\,$ and $\Ascr_j=\big\lbrace h\in\naturali\,:\,h=\begin{cases}
2i\phantom{+1\:\,}\quad\textit{if j is even}\\
2i+1\quad\textit{if j is odd}
\end{cases} ,\,i\in\naturali\,,\: i\geq \lfloor \frac{j}{2}\rfloor \bigg\rbrace\, $, $j\geq 1\,$.
\end{Lapl.coeff}
\begin{proof}
We are interested in the expansion of the following quantity
\begin{equation}
\label{start}
\frac{1}{\left(a_2^2 + a_3^2 - 2 \,a_2\,a_3\, \cos(\lambda_2 - \lambda_3)\right)^{\frac{2\,s+1}{2}}}=a_3^{-(2s+1)}\frac{1}{\left(1+\alpha^2-2\alpha \cos(\lambda_2 - \lambda_3)\right)^{\frac{2\,s+1}{2}}}\, ,
\end{equation}
with $\alpha=a_2/a_3\,$; for simplicity, we can define $l=2s+1$ and $\sigma=\lambda_2-\lambda_3\,$ and we can Taylor expand the previous quantity\footnote{It is sufficient to think $2\alpha\cos(\sigma)-\alpha^2:=\varepsilon$ and to Taylor expand, in $\varepsilon=0\,$, the function $(1-\varepsilon)^{-l/2}\,$.}, having 
\begin{equation}
\frac{1}{\left(1+\alpha^2-2\alpha \cos(\sigma)\right)^{\frac{l}{2}}}=1+\sum_{k=1}^{+\infty}\frac{\frac{l}{2}(\frac{l}{2}+1)\ldots(\frac{l}{2}+k-1)}{k!}\,\left(2\alpha \cos(\sigma)-\alpha^2\right)^k\, .
\end{equation}
Moreover, we can expand the quantity $\left(2\alpha \cos(\sigma)-\alpha^2\right)^k\,$, obtaining:
\begin{equation}
\begin{split}
\frac{1}{\left(1+\alpha^2-2\alpha \cos(\sigma)\right)^{\frac{l}{2}}}=&1+\sum_{k=1}^{+\infty}\frac{\frac{l}{2}(\frac{l}{2}+1)\ldots(\frac{l}{2}+k-1)}{k!}\,\left(2\alpha \cos(\sigma)-\alpha^2\right)^k\, \\
=& 1+\sum_{k=1}^{+\infty}\frac{\frac{l}{2}(\frac{l}{2}+1)\ldots(\frac{l}{2}+k-1)}{k!}\,\left[\sum_{h=0}^{k}\frac{k!}{h!(k-h)!}(-1)^{k-h}\alpha^{2(k-h)}(2\alpha\cos(\sigma))^h\right]\,\\
=& 1+\sum_{k=1}^{+\infty}\sum_{h=0}^{k}(-1)^{k-h}\, \frac{\frac{l}{2}(\frac{l}{2}+1)\ldots(\frac{l}{2}+k-1)}{h!(k-h)!}\alpha^{2k-h}\,2^h\,\cos(\sigma)^h\,.
\end{split}
\end{equation}
After having explicitely written the sum corresponding to $h=0\,$, it is possible to reverse the order of the sums (over $k$ and $h$), arriving to
\begin{equation}
\label{intermedia}
\begin{aligned}
\frac{1}{\left(1+\alpha^2-2\alpha \cos(\sigma)\right)^{\frac{l}{2}}}&=1+\sum_{k=1}^{+\infty}(-1)^{k}\, \frac{\frac{l}{2}(\frac{l}{2}+1)\ldots(\frac{l}{2}+k-1)}{k!}\alpha^{2k} \\
&\phantom{=}+\sum_{h=1}^{+\infty}\sum_{k=h}^{+\infty}(-1)^{k-h}\,2^h\, \frac{\frac{l}{2}(\frac{l}{2}+1)\ldots(\frac{l}{2}+k-1)}{h!(k-h)!}\alpha^{2k-h}\,\cos(\sigma)^h \\
&:= \sum_{h=0}^{+\infty}D_{h}\cos(\sigma)^h\, ,
\end{aligned}
\end{equation}
where we can define $D_h\,$ (with $h\geq 0$) as:
\begin{equation}
\label{Dh} D_h=
\begin{cases}
\displaystyle{1+\sum_{k=1}^{+\infty}(-1)^{k}\, \frac{\frac{l}{2}(\frac{l}{2}+1)\ldots(\frac{l}{2}+k-1)}{k!}\,\alpha^{2k}}\quad\quad\:\:\:\textit{for}\quad h=0\\
\displaystyle{\sum_{k=h}^{+\infty}(-1)^{k-h}\,2^h\, \frac{\frac{l}{2}(\frac{l}{2}+1)\ldots(\frac{l}{2}+k-1)}{h!(k-h)!}\,\alpha^{2k-h}}\quad\textit{for}\quad h\geq 1
\end{cases}.
\end{equation}
Therefore, in order to compute the Laplace coefficients, we need to expand also the $\cos(\sigma)^h\,$ in~\eqref{intermedia}; indeed, from one hand (Eq.~\eqref{intermedia}) we have 
$$ \frac{1}{\left(a_2^2 + a_3^2 - 2 \,a_2\,a_3\, \cos(\sigma)\right)^\frac{l}{2}}=a_3^{-l}\sum _{h\geq 0} D_h\cos(\sigma)^h\, ,
$$
and on the other one (Eq.~\eqref{espans.flam})
$$ \frac{1}{\left(a_2^2 + a_3^2 - 2 \,a_2\,a_3\, \cos(\sigma)\right)^\frac{l}{2}}=a_3^{-l}\sum _{j\geq 0} b^{(j)}_\frac{l}{2}(\alpha)\cos(j\sigma)\, .
$$
Then, using again the binomial formula, we have
\begin{equation}
\label{esp.coseno}
\cos(\sigma)^h=\left(\frac{e^{i\sigma}+e^{-i\sigma}}{2}\right)^h=\frac{1}{2^h}\sum_{k=0}^{h} \binom{h}{k}e^{ik\sigma} e^{-i(h-k)\sigma}=\frac{1}{2^h}\sum_{k=0}^{h} \binom{h}{k}\cos((2k-h)\sigma),
\end{equation}
where we applied the parity of the cosine. Then, inserting the previous expression~\eqref{esp.coseno} in Eq.~\eqref{intermedia}, we obtain
\begin{equation}
\label{final0}
\begin{aligned}
\frac{1}{\left(1+\alpha^2-2\alpha \cos(\sigma)\right)^{\frac{l}{2}}}=&D_0+\sum_{h=1}^{+\infty}\sum_{k=0}^{h} \frac{D_{h}}{2^h}\frac{h!}{k!(h-k)!}\cos((h-2k)\sigma)\\
=&D_0+\sum_{h=1}^{+\infty}\,\sum_{j\in\Bscr_h } \frac{D_{h}}{2^h}\frac{h!}{\left(\frac{h-j}{2}\right)!\left(\frac{h+j}{2}\right)!}\cos(j\sigma)\, ,
\end{aligned}
\end{equation}
where we define $j=h-2k$ and the set
$$
\Bscr_h=\left\lbrace j\in\interi \, : \, j=
\begin{cases}
2n \quad \textit{if $h$ is even}\,, \, n\in\interi\,,\,-\frac{h}{2}\leq n \leq \frac{h}{2}\\
2n+1 \quad \textit{if $h$ is odd}\,, \, n\in\interi\,,\,\lfloor-\frac{h}{2}\rfloor\leq n \leq \lfloor\frac{h}{2}\rfloor
\end{cases} \right\rbrace\, , \qquad h\geq 1\, .
$$
On the other hand, the previous set of indexes can be thought also as
$$
\Bscr_h=
\begin{cases}
\Bscr_{h_-} \cup \Bscr_{h_+} \cup \lbrace j=0 \rbrace \quad \textit{if $h$ is even}\\
\Bscr_{h_-} \cup \Bscr_{h_+} \quad \textit{if $h$ is odd}
\end{cases} \, ,
$$
where $\Bscr_{h_-}=\Bscr_{h}\cap \interi_{<0}$ and $\Bscr_{h_+}=\Bscr_{h}\cap \interi_{>0}\, $.
With the following notation, the sum appearing in Eq.~\eqref{final0} can be decomposed as follows: 
\begin{align*}
&\sum_{h=1}^{+\infty}\,\sum_{j\in\Bscr_h }=\sum_{\substack{h=1\\ h \textit{odd}}}^{+\infty}\,\left(\sum_{j\in \Bscr_{h_-}} + \sum_{j\in \Bscr_{h_+}}\right)+\sum_{\substack{h=2\\ h \textit{even}}}^{+\infty}\,\left(\sum_{j\in \Bscr_{h_-}} + \sum_{j\in \Bscr_{h_+}}+\sum_{j=0}\right)\, .
\end{align*}
Thus, observing that
$$
\sum_{j\in \Bscr_{h_-}} \frac{D_{h}}{2^h}\frac{h!}{\left(\frac{h-j}{2}\right)!\left(\frac{h+j}{2}\right)!}\cos(j\sigma)=\sum_{j\in \Bscr_{h_+}} \frac{D_{h}}{2^h}\frac{h!}{\left(\frac{h-j}{2}\right)!\left(\frac{h+j}{2}\right)!}\cos(j\sigma)\, ,
$$
we can write Eq.~\eqref{final0} as:
\begin{align*}
\frac{1}{\left(1+\alpha^2-2\alpha \cos(\sigma)\right)^{\frac{l}{2}}}=&\sum_{\substack{h=0\\h \textit{even}}}^{+\infty}\,\frac{D_{h}}{2^h}\frac{h!}{\left(\frac{h}{2}\right)!\left(\frac{h}{2}\right)!}+\sum_{\substack{h=2\\h \textit{even}}}^{+\infty}\,\sum_{j\in\Bscr_{h_+} } \frac{D_{h}}{2^{h-1}}\frac{h!}{\left(\frac{h-j}{2}\right)!\left(\frac{h+j}{2}\right)!}\cos(j\sigma)\notag\\
&+\sum_{\substack{h=1\\h \textit{odd}}}^{+\infty}\,\sum_{j\in\Bscr_{h_+} }\frac{D_h}{2^{h-1}}\frac{h!}{\left(\frac{h-j}{2}\right)!\left(\frac{h+j}{2}\right)!}\cos(j\sigma)\, .
\end{align*}
Finally, reversing the order of the sums (over $j$ and $h\,$), we arrive at
\begin{equation}
\label{final1}
\begin{aligned}
\frac{1}{\left(1+\alpha^2-2\alpha \cos(\sigma)\right)^{\frac{l}{2}}}=&\sum_{\substack{h=0\\h \textit{even}}}^{+\infty}\,\frac{D_{h}}{2^h}\frac{h!}{\left(\frac{h}{2}\right)!\left(\frac{h}{2}\right)!}+\sum_{\substack{j=2\\j \textit{even}}}^{+\infty}\,\sum_{h\in\Ascr_{j} } \frac{D_{h}}{2^{h-1}}\frac{h!}{\left(\frac{h-j}{2}\right)!\left(\frac{h+j}{2}\right)!}\cos(j\sigma)\\
&+\sum_{\substack{j=1\\j \textit{odd}}}^{+\infty}\,\sum_{h\in\Ascr_{j} }\frac{D_h}{2^{h-1}}\frac{h!}{\left(\frac{h-j}{2}\right)!\left(\frac{h+j}{2}\right)!}\cos(j\sigma)\, , \\
=&\sum_{\substack{h=0\\h \textit{even}}}^{+\infty}\,\frac{D_{h}}{2^h}\frac{h!}{\left(\frac{h}{2}\right)!\left(\frac{h}{2}\right)!}+\sum_{j=1}^{+\infty}\,\sum_{h\in\Ascr_{j} } \frac{D_{h}}{2^{h-1}}\frac{h!}{\left(\frac{h-j}{2}\right)!\left(\frac{h+j}{2}\right)!}\cos(j\sigma)\, ,
\end{aligned}
\end{equation}
where
$$
\Ascr_j=\left\lbrace h\in\naturali \, : \, h=
\begin{cases}
2i \quad \textit{if $j$ is even}\,, \, i\in\naturali\,,\,i\geq  \frac{j}{2}\\
2i+1 \quad \textit{if $j$ is odd}\,, \, i\in\naturali\,,\,i\geq \lfloor \frac{j}{2}\rfloor
\end{cases} \right\rbrace\, , \qquad j\geq 1\, .
$$
Now we can finally put Eq.~\eqref{final1} in the expression~\eqref{start} and compare the obtained quantity with~\eqref{espans.flam}, having
\begin{align}
\label{coeff.Lapl.new}
&b_{s+\frac{1}{2}}^{(0)}(\alpha)=D_0+\sum_{h\in\Ascr_2}\frac{D_h}{2^{h}}\frac{h!}{\left(\frac{h}{2}\right)!\left(\frac{h}{2}\right)!}\, , & &b_{s+\frac{1}{2}}^{(j)}(\alpha)=\sum_{h\in\Ascr_j}\frac{D_h}{2^{h-1}}\frac{h!}{\left(\frac{h-j}{2}\right)!\left(\frac{h+j}{2}\right)!}\, ,
\end{align}
with $\Ascr_2=\left\lbrace h\in\naturali\, : \, h=2 i\,, \, i\in\naturali\,,\,i\geq 1 \right\rbrace\, $.
Finally, substituting in the previous expression~\eqref{coeff.Lapl.new} the definition of $D_h$ (given by~\eqref{Dh}) and remembering that $l=2s+1\,$ and $\sigma=\lambda_2-\lambda_3\,$, we obtain~\eqref{Lapl.coeff.analitic}. This concludes the proof.\\
\end{proof}

%--------------------------------------------------------------------
\section{Normal form around the periodic orbits A or B} 
%--------------------------------------------------------------------
The normal form function $Z_{sec}^{(r)}$ and the normalizing transformation $\Phi^{(r)}$ of Eqs.~\eqref{hamnophi} and (\ref{phiper}) are computed by the following steps:

\begin{enumerate}

\item \textit{Development of the Hamiltonian around the apsidal solution of the integrable part $\Hscr_{int}$}: Let $(\psi_{\ast},\,\Gamma_{\ast},\,J_{\ast})$ be one of the two fixed points (apsidal corotation solutions) of the integrable Hamiltonian $\Hscr_{int}(\psi,\,\Gamma\,;J)\,$. We consider the translations
\begin{equation}
\label{trasla.apsidal.integ}
\psi=\psi_{\ast}+\varepsilon_*\delta\psi\, , \qquad \Gamma=\Gamma_{\ast}+\varepsilon_*\,\delta\Gamma\,,\qquad 
J=J_{\ast}+\varepsilon_{J_*}\,\delta J\, ,
\end{equation}
where $\varepsilon_*$, $\varepsilon_{J_*}$ are both book-keeping symbols with numerical value equal to one. We then perform the following algebraic operations:
\begin{itemize}
\item
substitute (\ref{trasla.apsidal.integ}) into the Hamiltonian $\Hscr_{sec}(\psi,\varphi,\Gamma,J)$ and expand the Hamiltonian in powers of the symbols $\varepsilon_*$, $\varepsilon_{J_*}$; 
\item 
in the resulting expression, replace $\cos(\varphi)$ and $\sin(\varphi)$ with $\varepsilon_*\cos(\varphi)$ and $\varepsilon_*\sin(\varphi)$ respectively;
\item assign a unique book-keeping symbol to the expanded Hamiltonian $\Hscr_{sec}(\delta\psi,\,\varphi,\,\delta\Gamma,\,\delta J\,)$, by setting $\varepsilon_{J_*}^{a}\rightsquigarrow\,\lambda^{a-1}\,\, a\geq 1\,$, $\varepsilon_*\rightsquigarrow\,\lambda\,$;
\item
truncate the resulting expression at a maximum order $N_t$ in the book-keeping symbol $\lambda$.
\end{itemize}
After performing the above algebraic steps, the Hamiltonian resumes the form
\begin{equation}\label{ham0}
\begin{aligned}
\Hscr^{(0)}=\Hscr_{sec} &= c_1\,+Z_0
+\sum_{s=1}^{N_t}\lambda^s h_{s}^{(0)}(\delta\psi,\,\varphi,\,\delta\Gamma,\,\delta J\,) \\
&= c_1\,+\nu_*^{(0)}\delta J +\sum_{s=1}^{N_t}\lambda^s 
\left(\sum_{m,\,k,\,n,\,l}\theta^{(0)}_{m,k,n,l}\,\delta\psi^{m}\delta\Gamma^{n} \delta J^{l}\,e^{i\,k\,\varphi} \right)
\end{aligned}
\end{equation}
where $c_1$ is a constant term, $\theta^{(0)}_{m,k,n,l}$ are constant coefficients, and $\nu_*^{(0)}<0$ is the (also constant) unperturbed frequency of the apsidal periodic orbit in the integrable approximation. 

\item \textit{Normal form to eliminate $\varphi$}: The Hamiltonian (\ref{ham0}) is normalized by iterative steps using the method of composition of Lie series. The normalization algorithm is defined recursively, for $r=1,2,...$ by the relations: 
\begin{equation}\label{hamrm1}
\Hscr^{(r-1)}=c_1\,+Z_0
+\sum_{s=1}^{r-1}\lambda^s Z_s(\delta\psi,\delta\Gamma,\delta J)
+\sum_{s=r}^{N_t}\lambda^s 
\left(\sum_{m,\,k,\,n,\,l}
\theta_{m,k,n,l}^{(r-1)}\,\delta\psi^{m}\delta\Gamma^{n} \delta J^{l}\,e^{i\,k\,\varphi} \right) ,
\end{equation}
\begin{equation}\label{chir}
\chi^{(r)}=\lambda^{r}\!\!\!\sum_{\substack{m,\,k,\,n,\,l\\ k\neq 0}}\frac{\theta^{(r-1)}_{m,k,n,l}}{i\,\nu_*^{(0)}\,k}\,
\delta\psi^{m}\delta\Gamma^{n} \delta J^{l}\,e^{i\,k\,\varphi}\, ,
\end{equation}
\begin{equation}\label{hamr}
\Hscr^{(r)}=\Bigg[\exp(L_{\chi^{(r)}})\Hscr^{(r-1)}\Bigg]^{\leq N_t} \, ,
\end{equation}
where $L_{\chi^{(r)}}$ denotes the Poisson bracket operator $L_{\chi^{(r)}}\cdot=\{\cdot,\chi^{(r)}\}$, and $[\cdot]^{\leq N_t}$ means truncation at the order $N_t$ in the book-keeping parameter $\lambda$. 

\item \textit{Final Hamiltonian and normalizing transformation}: the normalizing transformation $\Phi^{(r)}$ is defined by:
\begin{eqnarray}\label{phifinal}
(\delta\psi,\varphi,\delta\Gamma,\delta J)&=&
\Phi^{(r)}(\delta\t\psi,\t\varphi,\delta\t\Gamma,\delta\t J) \\
&=&
\Bigg[
\left(\exp\left(L_{\chi^{(r)}}\right)\exp\left(L_{\chi^{(r-1)}}\right)\ldots\exp\left(L_{\chi^{(1)}}\right)(\delta\psi,\varphi,\delta\Gamma,\delta J)\right)
\Bigg|_{\substack{
\delta\psi={\delta\t\psi}\\ 
\varphi=\t{\varphi}\\ 
\delta\Gamma={\delta\t\Gamma}\\
\delta J={\delta\t{J}}}}
\Bigg]^{\leq N_t}~ \nonumber
\end{eqnarray}
and the final Hamiltonian
\begin{equation}\label{hamfinal}
\begin{aligned}
\Hscr^{(r)}&=
\Bigg[
\left(\exp\left(L_{\chi^{(r)}}\right)\exp\left(L_{\chi^{(r-1)}}\right)\ldots\exp\left(L_{\chi^{(1)}}\right)
\Hscr^{(0)}\right)
\Bigg|_{\substack{
\delta\psi={\delta\t\psi}\\ 
\varphi=\t{\varphi}\\ 
\delta\Gamma={\delta\t\Gamma}\\
\delta J={\delta\t{J}}}}
\Bigg]^{\leq N_t}~ \\
&=
Z^{(r)}(\delta\t{\psi},\,\delta\t{\Gamma},\,\delta\t{J}\,)
+R^{(r)}(\delta\t{\psi},\,\t{\varphi},\,\delta\t{\Gamma},\,\delta\t{J}\,)\, ,
\end{aligned}
\end{equation}
where $1\leq r\leq N_t$ is chosen so that the normal form term $Z^{(r)}(\delta\t{\psi},\,\delta\t{\Gamma},\,\delta\t{J}\,)$
yields the best possible approximation to the dynamics. 

Note that the constant term $c_1$ produced at step 1 above has to be carried along all the successive Hamiltonians $\Hscr^{(r)}$ produced through the normalization process, being eventually included in the normal form $Z^{(r)}$. This is necessary, since this constant appears in the third of the algebraic equations (\ref{compnfper}) by which the fixed point corresponding to the apsidal periodic orbit is computed in the new canonical variables. 

\end{enumerate}

%--------------------------------------------------------------------
\section{Proof of Proposition~\ref{prop.Kozai}}
\label{subappendix:proof Kozai}
%--------------------------------------------------------------------
\begin{proof}
The eigenvalues of the matrix M~\eqref{matrix.M} are $\lambda_{1,2}=\pm \sqrt{\mathfrak{a}\mathfrak{b}}\,$; then if $\mathfrak{a}\mathfrak{b}>0\,$, then the eigenvalues are real and opposite, instead if $\mathfrak{a}\mathfrak{b}<0\,$, then the eigenvalues are complex and conjugate. We have to analyze all the possible cases. We can have
\begin{subequations}
\begin{align}
&\begin{cases}
\mathfrak{a}>0\\
\mathfrak{b}>0
\end{cases}\, , &
&\begin{cases}
\mathfrak{a}<0\\
\mathfrak{b}<0
\end{cases} \quad \Longrightarrow\quad \lambda_{1,2}\quad \mathit{reals} \label{autoval.real}
\\
&\begin{cases}
\mathfrak{a}>0\\
\mathfrak{b}<0
\end{cases}\, , &
&\begin{cases}
\mathfrak{a}<0\\
\mathfrak{b}>0
\end{cases} \quad \Longrightarrow\quad \lambda_{1,2}\quad \mathit{immaginary}\, . \label{autoval.imm}
\end{align}
\end{subequations}
Let us start to understand the change of sign of $\mathfrak{b}\,$; remembering the definition~\eqref{ab}
\begin{equation*}
\mathfrak{b}=\frac{3 \mathcal{G}^2 L_2^3\, (3 G_3^2 - L_2^2 + L_z^2)\, m_0\, m_3^7}{8 \,G_3^5\, L_3^3\, m_2^3}>0 \quad\Longleftrightarrow\quad  \tilde{\mathfrak{b}}:= L_z^2 +3L_3^2(1-\e_3^2)-L_2^2>0\, ,
\end{equation*}
being $(3 \mathcal{G}^2 L_2^3 m_0 m_3^7)/(8 (1-\e_3^2)^{5/2} L_3^8 m_2^3)>0\,$. Now, let us observe that if 
\begin{align}
\label{segnob1}
C:=L_2^2-3L_3^2(1-\e_3^2)<0\quad\Longrightarrow\quad \e_3^2<1-\frac{L_2^2}{3 L_3^2}\, ,
\end{align}
then $\tilde{\mathfrak{b}}>0$ automatically; in order to have that, it is necessary that $1-\frac{L_2^2}{3 L_3^2}>0\,$, i.e. to have $0< L_2< \sqrt{3}L_3\,$. Thus, we can conclude that $\mathfrak{b}>0$ ( or equivalently $\tilde{\mathfrak{b}}>0\,$) if one of the two is fullfilled
\begin{itemize}
\item[i)] $0< L_2<\sqrt{3}L_3 \,\,\wedge\,\,0\leq \e_3\leq\sqrt{1-\frac{L_2^2}{3L_3^2}}\,$, with $L_z^2>0\,$ \textbf{or}\\
\item[ii)] $\left(0< L_2<\sqrt{3}L_3 \,\,\wedge\,\,\e_3>\sqrt{1-\frac{L_2^2}{3L_3^2}}\right)\,\,\vee\,\,L_2\geq\sqrt{3}L_3\,$, with $L_z^2>C\,$.
\end{itemize}
Now, let us study the sign of $\mathfrak{a}\,$; as before, from definition~\eqref{ab}
\begin{equation}
\begin{split}
&\mathfrak{a}=-\frac{3 \mathcal{G}^2 L_2 (5 G_3^4 - 4 G_3^2 L_2^2 + 3 L_2^4 - 10 G_3^2 L_z^2 - 
    8 L_2^2 L_z^2 + 5 L_z^4) m_0 m_3^7}{16 G_3^5 L_3^3 m_2^3}>0\quad\Longleftrightarrow\\
&\tilde{\mathfrak{a}}:= L_3^4(1-\e_3^2)^2 - \frac{4}{5} L_3^2 L_2^2(1-\e_3^2) + \frac{3}{5} L_2^4 - 2 L_3^2(1-\e_3^2) L_z^2 - 
    \frac{8}{5} L_2^2 L_z^2 +  L_z^4<0\, ,
\end{split}
\end{equation}
being $-(3 \mathcal{G}^2 L_2\, m_0 \,m_3^7)/(16\, L_3^8(1-\e_3)^{5/2} m_2^3)<0\,$. Then $\mathfrak{a}>0$ (or equivalently $\tilde{\mathfrak{a}}<0\,$) iff $A<L_z^2<B\,$,
where $A$ and $B$ are defined as in~\eqref{critical.points}, i.e.
\begin{align*}
&A=\frac{1}{5}\left(4L_2^2+5 L_3^2(1-\e_3^2)-L_2\sqrt{L_2^2+60 L_3^2(1-\e_3^2)}\right)\, ,\\
&B=\frac{1}{5}\left(4L_2^2+5 L_3^2(1-\e_3^2)+L_2\sqrt{L_2^2+60 L_3^2(1-\e_3^2)}\right)\, .
\end{align*} Let us observe that $A$ and $B$ are both greater than zero. Infact, being $0\leq \e_3 < 1\,$, it is obvious that $B>0\,$; on the other hand $A>0$ iff
\begin{align*}
&\sqrt{L_2^2+60 L_3^2(1-\e_3^2)}<4L_2+\frac{5 L_3^2}{L_2}(1-\e_3^2) & &\Longleftrightarrow\\
&L_2^2+60 L_3^2(1-\e_3^2)<16 L_2^2 + 25 \frac{L_3^4}{L_2^2}(1-\e_3^2)^2+40 L_3^2(1-\e_3^2) & &\Longleftrightarrow \\
& 3L_2^4-4 L_3^2 (1-\e_3^2) L_2^2 + 5 (1-\e_3^2)^2 L_3^4 > 0 
\end{align*}
that holds $\forall\,\, L_2\, ,L_3\,, 0\leq \e_3< 1\,$, being $\Delta=-44 L_3^4 (1-\e_3^2)^2 <0\,$. Then, we can conclude that if we are in the case i), (i.e. $\displaystyle{0< L_2<\sqrt{3}L_3 \,\,\wedge\,\,0\leq \e_3\leq\sqrt{1-\frac{L_2^2}{3L_3^2}}}\,$) and $A< L_z^2< B\,$, then $\mathfrak{b}>0\,$ and $\mathfrak{a}>0\,$, then we have reals eigenvalues; instead if, in the same case i), $0<L_z^2< A\,$ or $L_z^2> B\,$, then $\mathfrak{b}>0\,$, $\mathfrak{a}<0\,$, having complex immaginary eigenvalues. This proves the {\bf{Case i)}} of the Proposition.

In the case ii), i.e. $\left(0< L_2<\sqrt{3}L_3 \,\,\wedge\,\,\e_3>\sqrt{1-\frac{L_2^2}{3L_3^2}}\right)\,\,\vee\,\,L_2\geq\sqrt{3}L_3\,$, we have to solve~\eqref{autoval.imm} and~\eqref{autoval.imm}, that become, respectively
\begin{subequations}
\begin{align}
&\begin{cases}
\mathfrak{a}>0\\
\mathfrak{b}>0
\end{cases}
\, =\,
\begin{cases}
A<L_z^2< B\\
L_z^2>C
\end{cases}\, , &
&\begin{cases}
\mathfrak{a}<0\\
\mathfrak{b}<0
\end{cases} \, =\,
\begin{cases}
0<L_z^2< A \, \vee\, L_z^2>B\\
0<L_z^2<C
\end{cases}\, ,\label{autoval.real2}
\\
&\begin{cases}
\mathfrak{a}>0\\
\mathfrak{b}<0
\end{cases}\, =\,
\begin{cases}
A<L_z^2< B\\
0<L_z^2<C
\end{cases}\, , &
&\begin{cases}
\mathfrak{a}<0\\
\mathfrak{b}>0
\end{cases}\, =\,
\begin{cases}
0<L_z^2< A \, \vee\, L_z^2>B\\
L_z^2>C
\end{cases}\, ;\label{autoval.imm2}
\end{align}
\end{subequations}
then, we have to understand the position of the value $C$ with respect to $A$ and $B\,$.
\begin{description}
\item[Case 1 :] $C< A$
\end{description}
$$C< A\quad\Longleftrightarrow\quad \sqrt{L_2^2+60 L_3^2(1-\e_3^2)}< -L_2+ \frac{20 L_3^2}{L_2}(1-\e_3^2)$$
i.e.
\begin{align*}
\begin{cases}
\displaystyle{-L_2+ \frac{20 L_3^2}{L_2}(1-\e_3^2)\geq 0}\\
\displaystyle{L_2^2+60 L_3^2(1-\e_3^2)< L_2^2+ \frac{400 L_3^4}{L_2^2}(1-\e_3^2)^2- 40 L_3^2 (1-\e_3^2)}\,
\end{cases}
\quad\Longleftrightarrow\quad
 \begin{cases}
\displaystyle{\e_3^2\leq 1-\frac{L_2^2}{20 L_3^2} }\\
\displaystyle{\e_3^2< 1-\frac{L_2^2}{4 L_3^2} }\,,
\end{cases}
\end{align*}
then if $\e_3^2 < 1-\frac{L_2^2}{4 L_3^2}\, $. The last inequality makes sense if $1-\frac{L_2^2}{4 L_3^2}> 0\,$, that is $0\leq L_2< 2 L_3\,$. Remembering that we are in the case ii), it follows that 
\begin{align*}
C<A \quad\Longleftrightarrow\quad &\left(0< L_2<\sqrt{3}L_3 \,\,\wedge\,\,\sqrt{1-\frac{L_2^2}{3L_3^2}}< \e_3 < \sqrt{1-\frac{L_2^2}{ 4 L_3^2}} \right)\, \, \vee \, \,\\
& \left(\sqrt{3}L_3\leq L_2 < 2 L_3 \,\,\wedge\,\,0\leq \e_3 < \sqrt{1-\frac{L_2^2}{ 4 L_3^2}}\right)\, .
\end{align*}
Therefore, studying~\eqref{autoval.real2} and~\eqref{autoval.imm2}, we can conclude that if\\ $\left(0< L_2<\sqrt{3}L_3 \,\,\wedge\,\,\sqrt{1-\frac{L_2^2}{3L_3^2}}< \e_3 < \sqrt{1-\frac{L_2^2}{ 4 L_3^2}} \right)$  $\vee$ $ \left(\sqrt{3}L_3\leq L_2 < 2 L_3 \,\,\wedge\,\,0\leq \e_3 < \sqrt{1-\frac{L_2^2}{ 4 L_3^2}}\right)\,$,
then the eigenvalues are reals and opposite if $0<L_z^2<C\,$ (see the second of~\eqref{autoval.real2}), purely immaginary and coniugates if $C<L_z^2<A\,$ (see the second of~\eqref{autoval.imm2}), reals and opposite if $A<L_z^2<B$ (see the first of~\eqref{autoval.real2}) and again purely immaginary and coniugates if $L_z^2>B\,$ (see the second of~\eqref{autoval.imm2}), proving the {\bf{Case ii)}} of the Proposition.

\begin{description}
\item[Case 2 :] $C= A$
\end{description}
Following the same calculations done for the previous case ($C<A\,$), it easily follows that
\begin{align*}
C= A\quad\Longleftrightarrow\quad &\left(0< L_2\leq 2 L_3 \,\,\wedge\,\, \e_3 = \sqrt{1-\frac{L_2^2}{ 4 L_3^2}} \right) \,\, ;
\end{align*}
in this case, studying~\eqref{autoval.real2} and~\eqref{autoval.imm2}, we can conclude that the eigenvalues are reals and opposite if $0<L_z^2<C=A\,$ (see the second of~\eqref{autoval.real2}), again reals and opposite if $C=A<L_z^2<B$ (see the first of~\eqref{autoval.real2}) (i.e. are reals and opposite if $0<L_z^2<B\,$, passing from $0$ if $L_z^2=A=C\,$) and purely immaginary and coniugates if $L_z^2>B\,$ (see the second of~\eqref{autoval.imm2}). This proves the {\bf{Case iii)}} of the Proposition.

\begin{description}
\item[Case 3 :] $A< C< B$
\end{description}
\begin{align}
\label{prima}
&C> A\quad&\Longleftrightarrow&\quad \sqrt{L_2^2+60 L_3^2(1-\e_3^2)}> -L_2+ \frac{20 L_3^2}{L_2}(1-\e_3^2)\\
\label{seconda}
&C< B\quad&\Longleftrightarrow&\quad \sqrt{L_2^2+60 L_3^2(1-\e_3^2)}> L_2- \frac{20 L_3^2}{L_2}(1-\e_3^2)\, .
\end{align}
Let us start from~\eqref{prima}; we have to solve the following
\begin{align*}
&\begin{cases}
\displaystyle{-L_2+ \frac{20 L_3^2}{L_2}(1-\e_3^2)\geq 0}\\
\displaystyle{L_2^2+60 L_3^2(1-\e_3^2)> L_2^2+ \frac{400 L_3^4}{L_2^2}(1-\e_3^2)^2- 40 L_3^2 (1-\e_3^2)}\,
\end{cases}
\quad \wedge\quad
\begin{cases}
\displaystyle{-L_2+ \frac{20 L_3^2}{L_2}(1-\e_3^2)< 0}\\
\phantom{\displaystyle{L_2^2+60 L_3^2(1-e_3^2)> L_2^2+ \frac{400 L_3^4}{L_2^2}(1-\e_3^2)^2- 40 L_3^2 (1-\e_3^2)}\,}
\end{cases}
\end{align*}
that gives
\begin{equation}
\label{e3.prima}
\left(\begin{cases}
\displaystyle{\e_3^2\leq 1-\frac{L_2^2}{20 L_3^2} }\\
\displaystyle{\e_3^2> 1-\frac{L_2^2}{4 L_3^2} }\,
\end{cases}
\quad \wedge\quad
\begin{cases}
\displaystyle{\e_3^2> 1-\frac{L_2^2}{20 L_3^2} }\\
\phantom{\displaystyle{\e_3^2> 1-\frac{L_2^2}{4 L_3^2} }\,,}
\end{cases}\right)\qquad\Longrightarrow\qquad \e_3^2> 1-\frac{L_2^2}{4 L_3^2}\, .
\end{equation}
Similarly, we can solve~\eqref{seconda}, having the following 
\begin{align*}
&\begin{cases}
\displaystyle{L_2- \frac{20 L_3^2}{L_2}(1-\e_3^2)\geq 0}\\
\displaystyle{L_2^2+60 L_3^2(1-\e_3^2)> L_2^2+ \frac{400 L_3^4}{L_2^2}(1-\e_3^2)^2- 40 L_3^2 (1-\e_3^2)}\,
\end{cases}
\quad \wedge\quad
\begin{cases}
\displaystyle{L_2- \frac{20 L_3^2}{L_2}(1-\e_3^2)< 0}\\
\phantom{\displaystyle{L_2^2+60 L_3^2(1-\e_3^2)> L_2^2+ \frac{400 L_3^4}{L_2^2}(1-\e_3^2)^2- 40 L_3^2 (1-\e_3^2)}\,}
\end{cases}
\end{align*}
that gives 
\begin{equation}
\label{e3.seconda}
\left(\begin{cases}
\displaystyle{\e_3^2\geq 1-\frac{L_2^2}{20 L_3^2} }\\
\displaystyle{\e_3^2> 1-\frac{L_2^2}{4 L_3^2} }\,
\end{cases}
\quad \wedge\quad
\begin{cases}
\displaystyle{\e_3^2< 1-\frac{L_2^2}{20 L_3^2} }\\
\phantom{\displaystyle{\e_3^2> 1-\frac{L_2^2}{4 L_3^2} }\,,}
\end{cases}\right)\qquad\Longrightarrow\qquad \forall\,\e_3 .
\end{equation}
Finally, putting together~\eqref{e3.prima} and~\eqref{e3.seconda}, we get $\e_3^2> 1-\frac{L_2^2}{4 L_3^2}\,$. Let us observe that if $1-\frac{L_2^2}{4 L_3^2}< 0\,$, i.e. $ L_2> 2 L_3\,$,
then the last inequality for $\e_3$ is automatically satisfy. Then, remembering that we are in the case ii) and putting togheter the results, it follows that 
\begin{align*}
A<C<B \quad\Longleftrightarrow\quad &\left(0< L_2\leq 2 L_3 \,\,\wedge\,\, \e_3 > \sqrt{1-\frac{L_2^2}{ 4 L_3^2}} \right)\, \, \vee \, \,\\
& \quad L_2 > 2 L_3\, .
\end{align*}
Again, studying~\eqref{autoval.real2} and~\eqref{autoval.imm2}, we can conclude that if $\displaystyle{\left(0< L_2\leq 2 L_3 \,\,\wedge\,\, \e_3 > \sqrt{1-\frac{L_2^2}{ 4 L_3^2}} \right)}$  $\vee$ $ L_2 > 2 L_3\,$, then the eigenvalues are reals and opposite if $0<L_z^2<A\,$ (see the second of~\eqref{autoval.real2}), purely immaginary and coniugates if $A<L_z^2<C\,$ (see the first of~\eqref{autoval.imm2}), reals and opposite if $C<L_z^2<B$ (see the first of~\eqref{autoval.real2}) and again purely immaginary and coniugates if $L_z^2>B\,$ (see the second of~\eqref{autoval.imm2}). This proves the last {\bf{Case \textit{iv})}} of the Proposition. 

Actually, there is a last case to be analyzed, that will not give any contribution, i.e.
\begin{description}
\item[Case 4 :] $ C\geq  B$
\end{description}
$$C\geq B\quad\Longleftrightarrow\quad \sqrt{L_2^2+60 L_3^2(1-\e_3^2)}\leq L_2- \frac{20 L_3^2}{L_2}(1-\e_3^2)$$
i.e.
\begin{align*}
\begin{cases}
\displaystyle{L_2- \frac{20 L_3^2}{L_2}(1-\e_3^2)\geq 0}\\
\displaystyle{L_2^2+60 L_3^2(1-\e_3^2)\leq L_2^2+ \frac{400 L_3^4}{L_2^2}(1-\e_3^2)^2- 40 L_3^2 (1-\e_3^2)}\,
\end{cases}
\quad\Longleftrightarrow\quad
 \begin{cases}
\displaystyle{\e_3^2\geq 1-\frac{L_2^2}{20 L_3^2} }\\
\displaystyle{\e_3^2\leq 1-\frac{L_2^2}{4 L_3^2} }
\end{cases}
\end{align*}
that has no solutions. This concludes the proof of the Proposition.

\end{proof}

\end{document}